\documentclass[12pt]{article}

\usepackage{epsfig,latexsym,amsfonts,amsmath,amsthm,amssymb,amsbsy,multirow,slashed,wasysym,textcomp,dsfont,comment,mathtools,cancel,slashed,cite,datetime,appendix,BOONDOX-calo}
\usepackage{tocloft}
\usepackage{bbold}

\usepackage{graphicx}
\usepackage{float}
\usepackage{tikz}
\usetikzlibrary{backgrounds}
\usetikzlibrary{patterns}
\usetikzlibrary{arrows.meta}
\usetikzlibrary{shapes,snakes}
\tikzstyle{bag} = [align=center]
\usetikzlibrary{decorations.pathmorphing}
\usetikzlibrary{decorations.markings}
\usetikzlibrary{patterns, arrows.meta}
\usetikzlibrary{fadings}
\usetikzlibrary{shapes.misc}
\tikzset{cross/.style={cross out, draw=blue, minimum size=2*(#1-\pgflinewidth), inner sep=0pt, outer sep=0pt},
cross/.default={1.7pt}}
\definecolor{newred}{rgb}{0.65, 0.16, 0.16}
\definecolor{brg}{rgb}{0.0, 0.26, 0.45}

\usepackage{xspace}
\usepackage{hyperref}
\usepackage{subcaption}
\usepackage{ytableau}
\usepackage[vcentermath]{youngtab}

 \newcommand{\badat}{\begin{alignedat}}
 \newcommand{\eadat}{\end{alignedat}}
 
 \def\be{\begin{equation}}
\def\ee{\end{equation}}

\def\p{\partial}

\usepackage{xcolor}

\newcommand{\pink}[1]{\textcolor{\pink}{#1}}

\definecolor{dblue}{rgb}{0.2,0.50,0.80}

 \definecolor{dred}{rgb}{0.65,0.10,0.20} 
\definecolor{dblue}{rgb}{0.2,0.50,0.80}

\newcommand{\I}{\textrm{I}\xspace}
\newcommand{\Ia}{\textrm{Ia}\xspace}
\newcommand{\Ib}{\textrm{Ib}\xspace}
\newcommand{\II}{\textrm{II}\xspace}
\newcommand{\III}{\textrm{III}\xspace}

\def\Hcal{\mathcal{H}}

\def\Jcal{\mathcal{J}}

\def\Ocal{\mathcal{O}}
\def\Pcal{\mathcal{P}}
\def\Vcal{\mathcal{V}}

\def\ecal{\mathcal{e}}
\def\fcal{\mathcal{f}}

\def\kcal{\mathcal{n}}

\def\a{{\alpha}}
\def\b{{\beta}}

\def\S{{\rm{S}}}
\def\P{{\rm{P}}}
\def\Wcal{{\mathcal{W}}}

\def\bw{{\bar w}}

\def\bz{{\bar z}}

\def\hd{\frac{d}{2}}

\usepackage{comment}
\usepackage{todonotes}

\numberwithin{equation}{section} 

\begin{document}

 \begin{titlepage}
  \thispagestyle{empty}
  \begin{flushright}
  CPHT-RR071.122022
  \end{flushright}
  
  \bigskip \bigskip
   
  \begin{center}

        \baselineskip=13pt {\Huge \scshape{
       Symmetries in Celestial CFT$_d$
        }}
  
      \vskip1cm 

   \centerline{\large Yorgo Pano${}^\diamond$,
   {Andrea Puhm}${}^\diamond$,
   {and Emilio Trevisani}${}^{\diamond,\circ}$
   }

\bigskip \bigskip\bigskip
 
\centerline{\em ${}^\diamond$CPHT, CNRS, Ecole Polytechnique, IP Paris, F-91128 Palaiseau, France
}
\smallskip
\vspace{0.1cm}
\centerline{\em
${}^\circ$Université Paris-Saclay, CNRS, CEA, Institut de Physique Théorique,
}
\smallskip
\centerline{\em
91191, Gif-sur-Yvette, France
}

\bigskip \bigskip
  
\end{center}

\begin{abstract}
  \noindent

We use tools from conformal representation theory to classify the symmetries associated to conformally
soft operators in 
celestial CFT (CCFT) in general dimensions $d$. 
The conformal multiplets in $d>2$ take the form of {\it celestial necklaces} whose structure is much richer than the celestial diamonds in $d=2$, it depends on whether $d$ is even or odd and involves mixed-symmetric tensor representations of $SO(d)$.
The existence of primary descendants in CCFT multiplets corresponds to (higher derivative) conservation equations for conformally soft operators.
We lay out  a unified method for constructing the conserved charges associated to operators with primary descendants.
In contrast to the infinite local symmetry enhancement in CCFT${}_2$, we find the soft symmetries in CCFT${}_{d>2}$  to be finite-dimensional. 
The conserved charges that follow directly from soft theorems are trivial in $d>2$, while non-trivial charges associated to (generalized) currents and stress tensor are obtained from the shadow transform of soft operators which we relate to (an analytic continuation of) a specific type of primary descendants. 
We aim at a pedagogical discussion synthesizing various results in the literature.

\end{abstract}

\end{titlepage}

\setcounter{tocdepth}{2}
\tableofcontents

\newpage
\section{Introduction}

The S-matrix in $d+2$ spacetime dimensions exhibits conformal properties akin to those of correlation functions in a $d$ dimensional conformal field theory (CFT) when expressed in a basis of boost eigenstates~\cite{Pasterski:2016qvg}. 
Underpinning this relation is the well-known fact that the $d+2$ dimensional Lorentz group acts as the Euclidean global conformal group on the $d$ dimensional celestial sphere. 
This presents an opportunity to harness the power of CFT to make headway in the attempt of a holographic description of bulk quantum gravitational physics in terms of a lower-dimensional theory on the boundary of asymptotically flat spacetimes.

This is the goal of the celestial holography program. The identification of the symmetries which must govern both sides of any holographically dual pair is an essential first step.
A concerted effort over the past few years has revealed that the infrared structure of gauge theory and gravity encodes the symmetries of a ``celestial conformal field theory" (CCFT) that lives on the sphere at the null boundary of asymptotically flat spacetimes.
This builds on the realization that the Ward identities of four-dimensional asymptotic symmetries, such as large gauge transformations, BMS supertranslations and superrotations, are equivalent to quantum field theory (QFT) soft theorems ~\cite{He:2014laa,Kapec:2014opa,Lysov:2014csa,Campiglia:2014yka,He:2014cra,Kapec:2015vwa,Campiglia:2015yka,Campiglia:2015qka,Kapec:2015ena,Campiglia:2015kxa,Campiglia:2016jdj,Campiglia:2016hvg}.
When recast in the conformal primary basis of boost eigenstates $d+2$ dimensional soft theorems take the form of correlation functions of conserved operators in CCFT$_{d}$~\cite{Donnay:2018neh,Fan:2019emx,Nandan:2019jas,Pate:2019mfs,Adamo:2019ipt,Puhm:2019zbl,Guevara:2019ypd,Kapec:2017gsg,Kapec:2021eug}.
These {\it conformally soft} operators give rise to  $d$~dimensional generators of the asymptotic symmetries in the $d+2$~dimensional asymptotically flat bulk spacetime~\cite{He:2014laa,Kapec:2016jld,Nande:2017dba,Donnay:2018neh,Donnay:2020guq,Pasterski:2021fjn,Donnay:2022sdg}. Soft theorems and conformal representation theory then determine all celestial symmetries.

This approach was taken in~\cite{Pasterski:2021fjn} to classify the symmetries of $d=2$ CCFTs and construct all $SL(2,\mathbb C)$ primary descendants. They are organized into conformal multiplets that take the form of diamond-shaped descendancy relations -- or {\it celestial diamonds} -- and encode the conformally soft operators and their conservation equations relevant to construct conserved charges, as well as conformal Faddeev-Kulish dressings that render celestial scattering amplitudes infrared finite.

In this work we tackle the more challenging problem of classifying the symmetries of $d>2$ dimensional CCFTs. The global conformal multiplets now take the form of {\it celestial necklaces} and their precise structure, e.g. whether the necklace is plain or contains a diamond, depends on the dimensionality $d$ of the CCFT and the types of conserved operators.
From a CFT$_d$ perspective the main new difficulty lies in the much richer structure of $SO(d)$ representations. Indeed, to capture all symmetries imposed by soft theorems requires an important refinement of the naive extension of the $d=2$ classification. Moreover, there are different types of conserved operators depending on whether $d$ is even or odd. Another major difference is that soft theorems in $d+2>4$ dimensions do not correspond on the nose to correlation functions of familiar conserved CFT operators such as currents and the stress tensor. These operators arise only after applying a shadow transform to the soft theorems~\cite{Kapec:2017gsg}. 

We classify the primary descendants of CCFT$_{d>2}$ with particular focus on those associated to soft theorems. The conformally soft operators are conserved in the sense that they have primary descendants which give either exactly zero or vanish up to contact terms in correlation functions. In even $d$ the independent types of descendant operators are I, II, III (as in CCFT$_2$) while in odd $d$ they are I, II, S (for shadow), P (for parity). Roughly speaking, the different types refer to whether the spin of a descendant is larger, smaller or equal to that of its parent primary, albeit with refinements. We focus on traceless and symmetric primaries whose primary descendants may or may not be traceless and symmetric. Our results build on and extend the work of~\cite{Penedones:2015aga,Costa:2016xah}, where the conditions for descendants in CFT$_d$ to become primaries were studied and several (but not all) primary descendant operators showing up in our work were constructed. 

The most important operators are of type I and II as they are related to universal soft theorems including the leading soft theorem for photons, gluons or gravitons and the subleading soft graviton theorem as well as their shadow transforms. In contrast to CCFT$_2$, we need a more refined definition of type I and II operators for CCFT$_{d>2}$ owing to the fact that in general dimensions the representations of the $SO(d)$ spin of the operator can be more complicated than the traceless and symmetric one. Type III operators are expected to play a role for the subleading soft photon and gluon theorems and the subsubleading soft graviton theorem when $d$ is even (as in CCFT$_2$). Their form, however, is not yet known. Moreover, in odd $d$ the apparent absence of primary descendant operators with the correct conformal dimensions to match those soft theorems presents an interesting puzzle. Primary descendants of type P  have opposite parity than the parent primary they descend from (hence the name P), and are relevant in our work only for $d=3$ where they replace certain type I operators which do not exist in $d=3$.

An interesting operator is that of type S. It is an independent primary in odd $d$ while in even~$d$ we prove that a suitable analytic continuation  acts  as the shadow transform for soft operators in CCFT$_d$ (hence the name S). A relation between the shadow integral and a certain differential operator was already pointed out in~\cite{Kapec:2017gsg,Kapec:2021eug}. In our work not only do we explain the form of such differential operators in terms of the type S operators but we also generalize this result to any spin and any  subleading order of soft theorems.

Armed with this primary descendant classification and the knowledge of what types of conserved operators follow from soft theorems we turn to the construction of conserved charges. 
In quantum field theory symmetries are associated to Ward identities for Noether currents. Integrating them over regions  defines conserved charges in the form of topological surface operators. They are conserved in the sense that they do not change upon a deformation of the region of integration so long as no other operator insertions are crossed in the correlation function.
We explain a general framework  to define Noether currents from operators with primary descendants in CFT. This in turn can be used to classify Noether currents and  conserved charges of CCFTs.

While our main focus is on CCFTs in $d>2$ dimensions, we take the opportunity to review some of the literature on the symmetries of $d=2$ CCFTs with a two-fold purpose. The first is to add to the results of~\cite{Pasterski:2021fjn} a general expression for the CCFT${_2}$ Noether currents associated to soft symmetries and a construction of infinite towers of conserved charges. The second is to provide a contrast for the $d>2$ dimensional case which has a much richer structure. For example, while the CCFT$_2$ Noether current construction is identical for all types of primary descendants, in CCFT$_{d>2}$ they take different forms for the different types of conserved operators.

The natural language for correlation functions in CCFT$_d$ is the embedding space formalism.
Indeed, the defining properties of celestial amplitudes are exactly the transformation properties of correlation functions of operators lifted to the embedding space~\cite{Costa:2011mg}. 
Every light ray passing through the origin in embedding space corresponds to a point in CFT$_d$ space while different sections of the lightcone correspond to different conformally flat spaces where the CFT$_d$ is defined.
This ensures Weyl covariance of celestial amplitudes.
Using the all-{\it out} representation of the scattering amplitude, the {\it in} and {\it out} momenta are only distinguished by an overall sign, thus null momenta live on the same set of null rays (at antipodally related points) and the resulting CCFT$_d$ lives on a single celestial sphere.

The antipodal map is a crucial ingredient in showing the equivalence between asymptotic symmetries and soft theorems~\cite{Strominger:2013jfa}. In proving this connection one typically starts from the conservation law for the charge associated to an asymptotic symmetry which is split into ``soft" and ``hard" charges, and one lands on the soft theorem by a suitable (singular) choice of the symmetry parameter. We find it, instead, more natural to start from soft and hard operators defined as the ``unsmeared charges" -- that is after integrating over the null coordinate on the conformal boundary of the spacetime but before integrating over the celestial sphere. Indeed, it is this unsmeared conservation law for the combination of soft and hard operators on the past and future null boundaries that is equivalent to the Ward identity of the conserved operator showing up in the corresponding soft theorem. We flesh this out for $d=2$, but this perspective also applies to $d>2$ where it is not known how to recover soft theorems from conservation laws of asymptotic charges. Indeed the role of infinite symmetry enhancements of the $d+2>4$ dimensional Poincaré group are more mysterious. Interesting recent work related to this point appeared in \cite{Kapec:2022hih}, see also~\cite{Cheung:2021yog,Cheung:2022vnd,Kapec:2022axw,Kapec:2021eug}.

This paper is organized as follows. We begin in section~\ref{sec:QFTandCCFT} with a general discussion of symmetries in quantum field theory which we then apply to celestial conformal field theory: 
we review in section~\ref{sec:TopoCharge} general statements about QFT Ward identities and conservation laws, introduce celestial amplitudes via the embedding space formalism in section~\ref{sec:CelestialAmplitudes}, and discuss in section~\ref{sec:CCFTWardID} conformally soft theorems and when they give rise to CCFT Ward identities.
In section~\ref{sec:CCFT2} we recast the symmetries of $d=2$ dimensional CCFTs in this language.
We briefly review the conformal multiplet structure in section~\ref{sec:CelestialDiamonds}, give a general expression for the Noether currents and infinite towers of conserved charges in section~\ref{sec:CCFT2Charges}, use this language to discuss the Ward identities associated to soft theorems, and conclude in section~\ref{sec:softhardtopo} with some remarks on the soft theorem = asymptotic symmetry connection and the relation to the CFT framework used here.
In section~\ref{sec:CCFTd} we discuss the symmetries of $d>2$ dimensional CCFTs. 
We review in section~\ref{sec:SOd_tensors} efficient technology to deal with $SO(d)$ tensors. We classify and construct in section~\ref{sec:CelestialNecklaces} the primary descendants that encode the celestial symmetries, in both even $d$ and odd $d$, for which we compute in section~\ref{sec:CCFTdCharges} the associated conserved charges. In section~\ref{sec:CCFTdCWardId} we discuss $d+2$ dimensional soft theorems and in section~\ref{sec:CCFTdSHCWardId} their shadow transforms highlighting the differences between $d>2$ and $d=2$.
We end with a discussion of open problems in section~\ref{sec:Conclusions}. The appendices~\ref{app:projectors},~\ref{app:typeI},~\ref{app:deltadist},~\ref{app:2pt},~\ref{app:Shadow} and \ref{app:typeSeven} collect various details of the computations.

\paragraph{Notation} We work in mostly plus signature $(-,+,...,+)$, $d+2$ dimensional spacetime indices are denoted by Greek letters $\mu,\nu\in \mathbb R^{1,d+1}$ while Latin letters $a,b\in \mathbb R^d$ are reserved for indices in CCFT${}_d$ which by our signature choice is Euclidean. Note that we use $\cdot$ for contraction of both $a$ and $\mu$ indices but from the context no confusion should arise.

\section{
Ward Identities and Charges in QFT and CCFT
}
\label{sec:QFTandCCFT}
The aim of this section is to review basic facts about symmetries in QFT and CCFT. 
In quantum field theory symmetries are associated to Ward identities and conserved charges are associated to topological surface operators. In section~\ref{sec:TopoCharge} we review  the construction of conserved charges from operators satisfying conservation equations that follow from Ward identities. Of great interest for understanding quantum gravity in asymptotically flat spacetimes are asymptotic symmetries whose associated Ward identities can be understood from soft factorization theorems of QFT scattering amplitudes. Upon a basis change from plane wave asymptotic states to conformal boost eigenstates momentum space amplitudes can be recast as celestial amplitudes which take the form of correlation functions on the co-dimension two celestial sphere. We introduce celestial amplitudes in section~\ref{sec:CelestialAmplitudes} using the embedding space formalism.

\subsection{Topological Operators in QFT}
\label{sec:TopoCharge}
We start with an elementary review on 
symmetries in QFTs (see e.g. \cite{Simmons-Duffin:2016gjk} for more details).
Let us assume  the existence of a conserved current 
 $\Jcal^{a}$ satisfying the following condition
\be
\partial_a \Jcal^{a} = 0 \, ,
\ee 
which must be understood as an operator equation valid inside correlation functions away from other insertions.
A better definition for the conservation equation is provided by the Ward identity which also defines the contribution of $\partial_a \Jcal^{a} (x)$ when $x$ coincides with another operator insertion.
In general a Ward identity takes the form
\be
\label{Ward_id}
\langle \partial_a \Jcal^{a} (x) \Ocal_1(x_1) \dots \Ocal_N(x_N) \rangle = \sum_{i=1}^N  \delta^{(d)} (x-x_i) \langle  \Ocal_1(x_1) \dots  \delta \Ocal_i(x_i) \dots\Ocal_N(x_N) \rangle \,,
\ee
where the variation $\delta  \Ocal(x)$ is some explicit operation on $\Ocal(x)$.

Using the current $\Jcal$ it is possible to define a topological surface operator
\be
\label{top_op}
Q_\Sigma = \int_{\Sigma} d S^{a}  \Jcal_ a\, ,
\ee
where $\Sigma$ is a $d-1$ dimensional (hyper)surface in $\mathbb{R}^d$ and $d S^{a}$ is the surface element multiplied by the unit vector $n^a$ normal to the surface $\Sigma$ at each point.

The fact that  $\Jcal_a$ is conserved is crucial to ensure that the operator $Q_\Sigma $ is topological, namely it does not depend on the choice of $\Sigma$. 
To be precise  $Q_\Sigma$ in \eqref{top_op} is understood as an operator that should be inserted in a correlation function and $Q_\Sigma =Q_{\Sigma' }$ as long as there exists a way to deform $\Sigma$ to $\Sigma'$ without crossing any other insertion in the correlation function.
If $\Sigma=\Sigma_i$ is a closed surface that contains a single point $x_i$ with $i\in[1,N]$, from the definition \eqref{Ward_id} and \eqref{top_op} it is easy to see that
\be
\label{QOi}
\langle Q_{\Sigma_i} \Ocal_1(x_1) \dots \Ocal_N(x_N) \rangle = \langle  \Ocal_1(x_1) \dots   \delta\Ocal_i(x_i) \dots  \Ocal_N(x_N) \rangle \, .
\ee
Here we used  Gauss's law to express~\eqref{top_op} as the divergence of the current 
\begin{equation}
  Q_\Sigma = \int_{R}d^dx \partial_a \Jcal^a\
\end{equation}
integrated over a region $R\in \mathbb R^d$ that is bounded by $\Sigma=\partial R$.
Similarly, when $\Sigma$ contains a set of points the right-hand-side of \eqref{QOi}  would be a sum of terms with $\delta$ acting on each operator insertion inside $\Sigma$. From the (integrated) Ward identity~\eqref{QOi} we see that
in practice $Q_\Sigma$ just acts by taking a variation of the operators enclosed by $\Sigma$ 
\begin{equation}
\label{charge_action}
  Q_\Sigma \Ocal(x)=\delta \Ocal(x)\,.
\end{equation}
When $\Sigma$ contains all insertions and the Noether current is smooth in $\mathbb R^d$ we can deform the integral to infinity and get zero. We find that the sum of the variations of each insertion gives zero
\be
\label{generic_symmetry}
0= \sum_{i=1}^N  \langle  \Ocal_1(x_1) \dots   \delta\Ocal_i(x_i) \dots  \Ocal_N(x_N) \rangle \,,
\ee
which defines a symmetry transformation.

In the quantization picture $Q_\Sigma$ is what defines a conserved charge. Let us see how this works.
In order to quantize a theory we specify a foliation of the spacetime in hypersurfaces. 
The Hamiltonian allows us to evolve from one slice of the foliation to the others. 
The direction of the evolution is typically referred to as the quantization ``time'' $t$ even if this does not need to be a time direction, e.g. in Euclidean signature we can pick any direction to be the quantization time. In CFTs it is often convenient to use the radial direction as a quantization time.
On each slice, which will be labelled by coordinates $\bf x$, one can define a Hilbert space $\Hcal$ of states and Hilbert spaces at different times are all isomorphic.
Correlation functions are computed as vacuum expectation values of time ordered products of operators
\be
\langle \Ocal_1(x_1) \dots  \Ocal_N(x_N) \rangle = \langle 0|  T\{ \hat \Ocal_1(t_1, { \bf x_1}) \dots \hat \Ocal_N(t_N,{ \bf x_N}) \} |0  \rangle \, ,
\ee
where $T\{ \dots \}$ implements time ordering with respect to the quantization time, $ |0  \rangle \in \Hcal $ is the vacuum state and  $\hat \Ocal_i$ are quantum operators $\hat \Ocal_i(t_i,{ \bf x_i}): \Hcal \to \Hcal$. 

In order to have $Q_\Sigma$ as an operator acting on a single Hilbert space we pick $\Sigma=\Sigma_t$ to be a slice at some time $t$.
The fact that  $Q_{\Sigma_t}$  is topological means that it does not change if we consider it at a different time $t'$, or in other words that it is conserved in time $Q_{\Sigma_t}=Q_{\Sigma_{t'}}$.
This of course should be considered again inside correlation functions and it assumes that no other insertion appeared between time $t$ and $t'$.
Indeed, if an insertion $\Ocal(t_\star, {\bf x})$ is present with $t<t_\star<t'$, the difference $Q_{\Sigma_t}-Q_{\Sigma_{t'}}$ is not zero and is given by the commutator
\begin{equation}
\label{Q_commutator_quant}
\badat{2}
\langle 0| T\{  [\hat Q, \hat \Ocal(t_ \star,{\bf x})] \hat \Ocal_1(x_1) \dots \hat \Ocal_N(x_N) \} |0  \rangle &= \langle (Q_{\Sigma_t}-Q_{\Sigma_{t'}}) \Ocal(t_ \star,{\bf x})   \Ocal_1(x_1) \dots  \Ocal_N(x_N) \rangle \\
&=  \langle 0| T\{ \delta \hat \Ocal(t_ \star,{\bf x})  \hat  \Ocal_1(x_1) \dots \hat \Ocal_N(x_N) \} |0 \rangle \, .
\eadat
\end{equation}
Here we assumed that all points $x_i$ have time coordinates $t_i$ outside of the interval $[t,t']$ and we used that the difference $Q_{\Sigma_t}-Q_{\Sigma_{t'}}$ can be rewritten -- by opportunely deforming the surfaces of integration --  in terms of a single operator $Q_{\Sigma}$ where $\Sigma$ is a closed surface that surrounds the point $(t_ \star,{\bf x})$. 
In the following we will use less precise notation for the quantized picture and we will often write formulae like
\be
[Q,\Ocal(x)]=\delta \Ocal(x) \, ,
\ee
which should be understood as \eqref{Q_commutator_quant} or as \eqref{QOi}.

We thus find that given any operator that satisfies a generic Ward identity of the form~\eqref{Ward_id}, we can associate a topological operator $Q_\Sigma$ which in the quantization picture has the meaning of a conserved charge.
These statements are of course true for generic QFTs. 

 Ward identities of the form \eqref{Ward_id} appear by considering the soft theorems written in a boost basis. One of the main goals of this paper is to classify the charges  associated to such Ward identities. Before getting to this classification program, in the rest of this section, we review how the ``soft'' Ward identities arise.

\subsection{Celestial Amplitudes and Embedding Space}
\label{sec:CelestialAmplitudes}

Let us consider a scattering amplitude $\mathcal A(p_i)$ in $d+2$ spacetime dimensions dependent on the momenta $p_i \in \mathbb{R}^{1,d+1}$ and including the momentum conserving delta function.
For simplicity let us focus on the case where all particles are massless scalars (we will generalize to massive and spinning particles below).
The momenta are null $p_i^2=0$ and can be conveniently parametrized as $p^{\mu}_i= \omega_i q^\mu_i$  where $q_ i^2=0$. The celestial amplitude is defined by the following change of basis\footnote{The amplitudes in~\eqref{Acelestial} should be understood to depend on sets of momenta $\{p_i=\omega_i q_i\}$ or null vectors $\{q_i\}$ and labels $\{\Delta_i\}$; we omit the brackets to avoid notational clutter.
The incoming and outgoing labels are also suppressed here; more about that in subsection~\ref{subsec:in_out_states}. 
} 
\be
\mathcal M_{\Delta_i}(q_i) \equiv \int_0^\infty \left(\prod_{i=1}^N d\omega_i \omega_i^{\Delta_i -1}\right)  \mathcal A(\omega_i q_i) \, .
\label{Acelestial}
\ee
We notice that the function $\mathcal M_{\Delta_i}(q_i) $:
\begin{itemize}
\item  is defined on the null cones $q_i^2=0$,
\item is Lorentz invariant, namely it can only depend on scalar products $q_i \cdot q_j$,
\item is homogeneous of degree $-\Delta_i$ in $q_i$, namely $\mathcal M_{\Delta_i}(\lambda_i q_i) = \left(\prod_{i=1}^{N} \lambda_i^{-\Delta_i}\right)  \mathcal M_{\Delta_i}(q_i) $ for real coefficients $\lambda_i$.
\end{itemize}
These are exactly the transformation properties of correlation functions of operators in embedding space \cite{Costa:2011mg}. So we can formally identify 
\be
\mathcal M_{\Delta_i}(q_i) \equiv \langle \Ocal_{\Delta_1}(q_1) \dots \Ocal_{\Delta_N}(q_N) \rangle \, ,
\ee
where $\Ocal_{\Delta}(\lambda q)=\lambda^{-\Delta}  \Ocal_{\Delta}(q)$ are CFT$_d$ primary operators uplifted to embedding space -- which in the current context corresponds to the physical spacetime.
Every light ray passing through the origin in embedding space corresponds to a point in CFT$_d$ space.  The CFT$_d$ space is described by a section of the null cone. Different choices of sections correspond to different spaces where the CFT lives. One does not lose any information by restricting a correlation function to a section, in fact by homogeneity one can always uplift it to the full null cone.  There are a few important sections which are frequently used:
\begin{itemize}
\item the Poincaré section $q^0+q^{d+1}=1$, parametrized  by  $q^\mu_p=(\frac{1+x^2}{2},x^a, \frac{1-x^2}{2})$ where $x^{a}\in\mathbb{R}^{d}$,
\item the sphere section $q^0=1$, parametrized  by $q^\mu_s=\frac{2}{1+x^2}(\frac{1+x^2}{2},x^a, \frac{1-x^2}{2})$ where $x^{a}\in\mathbb{R}^{d}$,
\item the cylinder  section  $(q^0)^2-(q^{d+1})^{2}=1$, parametrized  by $q^\mu_c=(\cosh \tau, \Omega^a, \sinh \tau)$ with $\Omega^2=1, \Omega^{a}\in\mathbb{R}^{d}$ and $\tau \in \mathbb{R}$.
\end{itemize}

It is easy to show  that the induced metric on the Poincaré section is the flat space one, namely $ds^2=\delta_{ab} dx^a dx^b$. In general, by appropriately choosing a section, one can obtain a correlation function defined in any conformally flat space.
This is easy to see, indeed  the section $q^\mu=\Omega(x) q^\mu_{p}$ has induced metric given by the most generic conformally flat metric $ds^2=\Omega(x)^2 \delta_{ab} dx^a dx^b$. The correlation functions in embedding space are in fact automatically Weyl covariant, which implies that the celestial amplitudes are also Weyl covariant.

\subsubsection{Celestial Amplitudes for Particles with Mass and Spin}
The embedding space definition of celestial amplitudes can be generalized to the massive case.
The celestial amplitude for scattering of massive scalar particles is defined by~\cite{Pasterski:2016qvg}
\be\label{MassiveCelestial}
\mathcal M_{\Delta_i}(q_i) \equiv \int \left(\prod_{i=1}^N Dp_i \right)\frac{1}{(-p_i\cdot q_i)^{\Delta_i}}  \mathcal A(p_i) \, ,
\ee
where the measure is over the mass shell $Dp=d^{D}p \delta(p^2+1) \theta(p^0)$.
Also in this case $\mathcal M_{\Delta_i}(q_i)$  is a homogeneous function of $q_i$ with weight $-\Delta_i$. Therefore also massive celestial amplitudes transform as embedding space correlation functions. 
Massive celestial correlators have received far less attention in the literature than their massless counterparts and it would be desirable to study their properties more thoroughly. 
We leave this for future work.

Let us now consider a spin $\ell$ massless bosonic field $F_{\mu_1 \dots \mu_\ell}$, where e.g. the case $\ell=1$ corresponds to photons $A_\mu$ and $\ell=2$ to gravitons $h_{\mu \nu}$. 
Gauge invariance implies
\be
\label{GaugeSpinl}
F_{\mu_1 \dots \mu_\ell}(p) \to F_{\mu_1 \dots \mu_\ell}(p) + p_{(\mu_1} \Lambda_{\mu_2 \dots \mu_\ell)}(p) \, .
\ee

For convenience we contract the field $F_{\mu_1 \dots \mu_\ell}(p)$ with some polarization vectors $\varepsilon^{\mu}$ such that $\varepsilon^2=0$ and $p \cdot \varepsilon =0$. We then define $F(p,\varepsilon)\equiv F_{\mu_1 \dots \mu_\ell}(p) \varepsilon^{\mu_1} \cdots \varepsilon^{\mu_\ell}$. 
We can define celestial amplitudes associated to $\mathcal A_{\ell_i}(p_i, \varepsilon_i)$, where all indices of the particle $i$ are contracted with $\varepsilon_i$, as
\be
\mathcal M_{\Delta_i,\ell_i}(q_i,\varepsilon_i) \equiv \int_0^\infty \left(\prod_{i=1}^N d \omega_i \omega_i^{\Delta_i-1} \right) \mathcal A_{\ell_i}(\omega_i q_i,\varepsilon_i) \, .
\ee
The function $\mathcal M_{\Delta_i,\ell_i}(q_i,\varepsilon_i)$ satisfies
\be
\mathcal M_{\Delta_i,\ell_i}(\lambda_i q_i,\alpha_i \varepsilon_i+\beta_i q_i) =\left(\prod_{i=1}^N\lambda_i^{-\Delta_i} \alpha_i^{\ell_i}\right) \mathcal M_{\Delta_i,\ell_i}(q_i,\varepsilon_i) \, ,
\ee
and thus it again defines a correlation function of primary operators $\Ocal_{\Delta_i,\ell_i}$ with dimensions $\Delta_i$ and spin $\ell_i$ in embedding space,
\be
\mathcal M_{\Delta_i,\ell_i}(q_i,\varepsilon_i) \equiv \langle \Ocal_{\Delta_1,\ell_1}(q_1,\varepsilon_1) \dots \Ocal_{\Delta_N,\ell_N}(q_N,\varepsilon_N) \rangle \, .
\ee
Here we are using a notation where the spinning operators in embedding space are contracted with polarization vectors
 $\Ocal_{\Delta,\ell}(q,\varepsilon) \equiv \Ocal^{\mu_1 \dots \mu_\ell}_{\Delta,\ell}(q) \varepsilon_{\mu_1} \dots \varepsilon_{\mu_\ell}$. Notice that in embedding space the spinning operators $\Ocal^{\mu_1 \dots \mu_\ell}_{\Delta,\ell}(q)$ are required to obey the same gauge condition \eqref{GaugeSpinl}, which from this construction is automatically satisfied.
 
 To project the operator into CFT$_d$ space we  need to both set $q$ to a section while at the same time projecting the embedding indices $\mu_i=0, \dots, d+1$ to CFT$_d$ indices $a_i=1,\dots, d$.
 This is achieved by contracting the indices with the Jacobian of the immersion. E.g. for the Poincaré section
\cite{Costa:2011mg}
\begin{equation}
\Ocal^{\mu_1 \dots \mu_\ell}_{\Delta,\ell}(q) \to \Ocal^{a_1 \dots a_\ell}_{\Delta,\ell}(x)= \partial^{a_1}_{x}q_{\mu_1} \dots \partial^{a_\ell}_{x}q_{\mu_\ell} \Ocal^{\mu_1 \dots \mu_\ell}_{\Delta,\ell}(q)\,.
\end{equation}
Often it is convenient to work in an index-free notation where all indices are contracted with polarization vectors both in embedding and in CFT$_d$ space.
To do so we can directly project a polarization vector $\varepsilon$ associated to an operator inserted  at a point $q$ to CFT$_d$ space. 
E.g. if $q$ is projected to the Poincaré section (and similarly for other sections), then $\varepsilon$ is projected as follows 
\begin{equation}
\varepsilon^\mu \to   
{\mathcal e}_a \partial^a_{x} q^{\mu}
\, ,
\end{equation}
where ${\mathcal e}^a$ is a polarization vector in $\mathbb{R}^d$, generically not transverse to $x^a$, that satisfies ${\mathcal e}^a {\mathcal e}_a=0$ (see the discussion in section \ref{sec:SOd_tensors} to recover the indices of the tensor operators).

\subsubsection{Celestial {\it in} $\leftrightarrow$ {\it out} States and the Antipodal Map}
\label{subsec:in_out_states}
Let us comment on the role of {\it in} and {\it out} states for celestial amplitudes. 
We can adopt an all-{\it out} formalism where we write all momenta as outgoing and flip the sign of incoming momenta,
namely the 
momenta are written as $p^{\mu}=\eta \omega q^{\mu}$ where $\eta=+1$ ($\eta=-1$) for outgoing (incoming) particles.
This implies that we can read off if a particle is {\it in} or {\it out} from the sign of the energy component of $q^\mu$: $q^0<0$ for incoming and $q^0>0$ for outgoing particles.
We thus find that incoming  operators naturally live on the null cone $q^2=0$ with $q^0<0$ while outgoing  operators live on the flipped null cone with $q^0>0$.
However, it is not necessary to use two different cones because homogeneity relates operators insertions at  $-q$ to insertions at $q$ via $\Ocal_\Delta(-q)=(-1)^{-\Delta}\Ocal_\Delta(q)$. 
On the sphere section the map $q \to -q$ which takes {\it out} $\leftrightarrow$ {\it in} is achieved by flipping the sign of $q^0$ and by performing the antipodal\footnote{In $d=2$ using complex coordinates this map is written as $z\to-1/\bar z$, $\bar z\to-1/z$.} map $x^a \to - \frac{x^a}{x^2}$, namely
\be
-q_s^\mu=-\frac{2}{1+x^2}\left(\frac{1+x^2}{2},x^a, \frac{1-x^2}{2}\right) \underset{q^0 \to - q^0 \atop x^a \to - \frac{x^a}{x^2} }{\longrightarrow}
q_s^\mu=\frac{2}{1+x^2}\left(\frac{1+x^2}{2},x^a, \frac{1-x^2}{2}\right) \, .
\ee

This map was discussed in the literature~\cite{Strominger:2013jfa} as a continuity condition for the asymptotic data at past and future null infinities when connected across spatial infinity, and can be derived from the dynamics of the fields at spatial infinity~\cite{Campiglia:2017mua,Capone:2022gme}.
On the other hand in embedding space this comes about trivially. Indeed a single light ray passing through the origin defines a single point in CFT$_d$ space so independently of the sign of $q^0$ it is natural to consider {\it in} and {\it out} operators living in the same CFT$_d$ space. Moreover, since a light ray intersects  spherical sections with opposite values of $q^0$ at antipodal points, the antipodal matching condition is automatically imposed by the formalism.

\subsection{Soft Theorems and Celestial Ward Identities}
\label{sec:CCFTWardID}

Celestial amplitudes in gauge theory and gravity obey Ward identities~\cite{Fan:2019emx,Nandan:2019jas,Pate:2019mfs,Adamo:2019ipt,Puhm:2019zbl,Guevara:2019ypd,Fotopoulos:2020bqj,Kapec:2017gsg,Kapec:2021eug}.
Their origin in momentum-space amplitudes is the soft, or zero-energy, limit of massless particles. The scattering amplitude of $N$ hard particles and one soft particle factorizes into the amplitude for the hard particles times a soft factor\footnote{
We restrict to tree-level scattering and express soft theorems in the $SO(d)$ index-free notation.} 
\begin{equation}
\label{softtheorem}
    \lim_{\omega\to 0} \mathcal A_{N+1}(\omega q,\varepsilon)=\sum_k S^{(1-k)}(\omega q,\varepsilon) \mathcal A_N\,,
\end{equation}
where the soft factors scale with powers of the soft particle's energy, $S^{(1-k)}\sim \omega^{-k}$ 
with the leading term given by Weinberg's soft pole $\sim 1/\omega$,
and we have suppressed the dependence of the amplitude on the $N$ hard particles to avoid clutter. In QED the leading soft photon factor is given by
~\cite{Weinberg:1965nx}
\begin{equation}
\label{SoftPhoton}
S^{(0)}(\omega q,\varepsilon)=e\sum_{i=1}^N   \mathcal Q_i\frac{1}{\omega} \frac{q_i \cdot \varepsilon}{ q_i \cdot q}\, ,
\end{equation}
where $p_i^\mu=\eta_i\omega_i q_i^\mu$ are the momenta of the hard particles and $\mathcal Q_i$ are their charges. In gravity the leading and subleading soft graviton factors are~\cite{Weinberg:1965nx,Gross:1968in,Jackiw:1968zza,White:2011yy}
\begin{equation}
\label{SoftGraviton}
S^{(0)}(\omega q,\varepsilon)=\frac{\kappa}{2}  \sum_{i=1}^N \eta_i\frac{\omega_i}{\omega} \frac{(q_i \cdot \varepsilon)^2}{ q_i \cdot q} \, ,
\quad 
S^{(1)}(\omega q,\varepsilon)=-i\frac{\kappa}{2}\sum_{i=1}^N \frac{(q_i\cdot \varepsilon)(q\cdot J_i\cdot \varepsilon)}{q_i\cdot q}\, ,
\end{equation}
where $\eta_i=+1$ for \textit{out} particles and $\eta_i=-1$ for \textit{in} particles. Here $J_i^{\mu\nu}$ is the total angular momentum of particle $i$ which can be decomposed as $J_i^{\mu\nu}=\mathcal{L}_i^{\mu\nu}+\mathcal{S}_i^{\mu\nu}$, where $\mathcal{L}_i^{\mu\nu}=-i\left(p_{i \, \mu}\frac{\partial}{\partial p_i^\nu}-p_{i \, \nu}\frac{\partial}{\partial p_i^\mu}\right)$
is the orbital momentum and $\mathcal{S}_i^{\mu\nu}$ is the spin representation of particle $i$.
The leading soft factors in gauge theory and gravity are thus universal, in the sense that they only depend on the momenta, angular momenta and electromagnetic charges of the hard particles, but not on the details of the theory.
The factorization property of the S-matrix persists to more subleading orders. The subleading soft photon~\cite{Low:1954kd,Low:1958sn,Burnett:1967km,Gell-Mann:1954wra} 
and subsubleading soft graviton~\cite{Cachazo:2014fwa} 
factors are respectively
\begin{equation}
S^{(1)}(\omega q,\varepsilon)=-ie \sum_{i=1}^N  \eta_i\frac{\mathcal Q_i}{\omega_i} \frac{q\cdot J_i\cdot \varepsilon}{q_i \cdot q}\,, \quad 
    S^{(2)}(\omega q,\varepsilon)=-\frac{\kappa}{4} \sum_{i=1}^N \eta_i\frac{\omega}{\omega_i} \frac{(q\cdot J_i\cdot \varepsilon)^2}{q_i\cdot q}\,,
\end{equation}
but are subject to (non-universal) modifications in effective field theory~\cite{Elvang:2016qvq}. 
More subleading soft theorems can also be studied but the expressions for the corresponding soft factors $S^{(1-k)}$ are known to be non-universal~\cite{Li:2018gnc,Hamada:2018vrw}. Let us also mention that the leading soft factors $S^{(0)}$ are tree-level exact while $S^{(1)}$ receives a one-loop-exact correction in gravity~\cite{He:2014bga,Bern:2014oka}. All other soft theorems are corrected by further quantum corrections. 
In this paper we will focus on the tree level contributions.

We can recast soft theorems as Ward identities for celestial amplitudes. The individual terms in the energetically soft expansion of momentum-space amplitudes get mapped under the Mellin transform to poles in the conformal dimension of massless operators in the celestial correlation function,
\begin{equation}
 \frac{1}{\omega^{k}} \quad \stackrel{\int_0^\infty d\omega \omega^{\Delta-1}}{\longrightarrow
 } \quad \frac{1}{\Delta-k}
\end{equation}
with $k=1,0,-1...$, and the lower-point amplitudes are extracted by the residues at $\Delta=k$.\footnote{Here we focus on soft photons/gluons and gravitons. Conformally soft theorems for photinos and gravitinos in $d=2$ were discussed in\cite{Fotopoulos:2020bqj} and their associated conformally soft charges in~\cite{Pano:2021ewd}; see also~\cite{Iacobacci:2020por,Narayanan:2020amh}.}
The operators with these conformal dimensions obey conservation equations that take the form of CCFT Ward identities. Our interest lies in identifying all the conserved operators associated to conformally soft theorems in gauge theory and gravity. 
The (conformally) soft limit will capture the 
leading  contributions -- namely the contributions that determine the Ward identities -- to the operator product expansion~\cite{Fan:2019emx,Pate:2019lpp} (while the full tower of descendants can be accessed from soft-collinear limits\cite{Himwich:2021dau}).  
This limit can be expressed as 
\begin{equation}\label{Confsofttheorem}
    \lim_{\Delta \to k}(\Delta-k)\langle \Ocal_{\Delta,\ell}(q,\varepsilon) \Ocal_{\Delta_1,\ell_1}\cdots \Ocal_{\Delta_N,\ell_N}\rangle = \sum_{i=1}^N\hat S^{(1-k)}_i(q,\varepsilon) \langle  \Ocal_{\Delta_1,\ell_1}\cdots \delta \Ocal_{\Delta_i,\ell_i}\cdots \Ocal_{\Delta_N,\ell_N}\rangle\,,
\end{equation}
where the conformally soft factor $\hat S^{(1-k)}_i$ is defined as the $i$th soft factor in~\eqref{softtheorem} stripped off its $\omega$~dependence while the $\omega_i$~dependence of~\eqref{softtheorem} determines the operation $\delta$ on $\Ocal_{\Delta_i,\ell_i}$. As in~\eqref{softtheorem} we have suppressed the arguments $(q_i,\varepsilon_i)$ of the hard particles to avoid clutter. 

In what follows it will be most convenient to work in the Poincaré section where $q^0+q^{d+1}=1$ such that the null vector determining the propagation direction of the soft particles is parameterized by 
\begin{equation}\label{qPoincare}
    q^\mu(x)=\frac{1}{2}\left(1+x^2,2x^a,1-x^2\right)\,,
\end{equation}
and similarly for the massless hard particles.
The spacetime polarization vector projected in the Poincaré section is
\begin{equation}\label{epsPoincare}
    \varepsilon^\mu(x,\ecal)=\ecal^a \partial_a q^\mu(x) =(x \cdot \ecal, \ecal^a,-x\cdot \ecal)\,,
\end{equation}
where the CCFT$_d$ polarization vectors $\ecal_a$ are normalized such that $\ecal \cdot \ecal=0$.
Note the following simple rules to project scalar products of vectors $q^{\mu}_i$ and $\varepsilon^{\mu}_i$ into the Poincaré section 
\begin{equation}
\label{prods_poinc}
q_i \cdot q_j \to - \frac{x_{i j}^2}{2} \, ,\qquad 
q_i \cdot \varepsilon_j  \to {\mathcal e}_j \cdot x_{i j} \, , \qquad
\varepsilon_i \cdot \varepsilon_j  \to {\mathcal e}_i \cdot {\mathcal e}_j \, ,
\end{equation}
where $x^a_{ij}=x^a_i-x^a_j$. With that we can now easily express the $d+2$ dimensional soft theorems in CCFT$_d$ space in terms of operators $\Ocal_{\Delta,\ell}(x,\ecal)$, or in terms of operators $\Ocal_{\Delta,\ell}^{a_1\dots a_\ell}(x)$ with explicit $SO(d)$ indices that we may contract with polarization vectors $\ecal_{a_1}\dots \ecal_{a_\ell}$.

We define the following conformally soft operators~\cite{Donnay:2018neh,Guevara:2021abz}
\begin{equation}
\label{conf_soft_ops}
   R^a_k(x) :=  \lim_{\Delta \to k}(\Delta-k) \Ocal^a_{\Delta, \ell=1}(x) \,, \quad H^{a b}_{k}(x) := \lim_{\Delta \to k}(\Delta-k) \Ocal^{a b}_{\Delta, \ell=2}(x) \, ,
\end{equation}
where $a,b=1,\dots,d$ are the tensor indices for the spin $1$ and spin $2$ operators. 
Alternatively, we can define them via the residues of the operators $\Ocal^a_{\Delta, \ell=1,2}$ at $\Delta=k$.
Contracting them with the polarization vectors in CFT$_d$ space we get in index-free notation
\begin{equation}\label{conf_soft_ops_indexfree}
    R_k(x,\ecal)=R^a_k(x) \ecal_a\,, \quad H_k(x,\ecal)=H^{ab}_k(x) \ecal_a \ecal_b\,,
\end{equation}
which are the operators appearing in the CCFT$_d$ Ward identities~\eqref{Confsofttheorem} for $\ell=1$ and $\ell=2$.

Besides the conformally soft operators~\eqref{conf_soft_ops} it can be useful in CCFT to consider the associated shadow operators.
The shadow transform (see expression~\eqref{2dshadow} for CCFT${}_2$ and~\eqref{I_Shadow_indexfree} for CCFT${}_d$) maps a primary operator of dimension $\Delta$ to a primary operator of dimension $d-\Delta$.
In celestial CFT$_d$ we will see that conformally soft shadow operators are natural and physically important operators~\cite{Kapec:2017gsg,Kapec:2021eug} as they directly give rise to the standard Ward identities for the global (local in $d=2$) symmetries of  CFT$_d$.

In summary, by going to the conformal basis, soft theorems can be recast  as Ward identities of some operators with given integer dimensions $\Delta$ (fixed by the order of the expansion in $\omega$) and spin $\ell$ (equal to the spin of the soft particle).
The important observation is that these operators are special in the CFT description because their dimension must be protected. Indeed if $\Delta$  acquired an anomalous dimension, it would no longer be possible to associate to it an integer power of the  $\omega$ expansion. We can thus conclude that in order to understand the soft operators one should look at the protected sector of CFT operators. Luckily this is already classified. Indeed an operator is protected when there exists a shortening condition of its conformal multiplet, which means that the operator must have a descendant that becomes a primary. This happens for special values of the labels $\Delta, \ell$ of the primary. 
Our strategy is to use this classification to understand the properties of all soft operators relevant for celestial CFT$_d$ and to build their associated charges.

\section{Symmetries in Celestial CFT${}_{d=2}$}
\label{sec:CCFT2}
The symmetries of $d=2$ dimensional celestial CFTs implied by $d+2=4$ dimensional soft theorems is the topic of a fairly large body of literature. Here we review some of the salient features within the unified framework of section~\ref{sec:CelestialAmplitudes} which serves two objectives: it allows us to write down a single formula for the Noether currents and conserved charges for all soft theorems and it contrasts the discussion of celestial symmetries in higher dimensions.
We review in section~\ref{sec:CelestialDiamonds} the global conformal multiplets that encode the operators responsible for conformally soft factorization theorems in gauge theory and gravity -- the {\it celestial diamonds} introduced in in~\cite{Pasterski:2021fjn}. 
These operators are conserved and when inserted in correlation functions they give rise to celestial Ward identities. We give a general expression for the associated Noether currents and infinite towers of topological charges in section~\ref{sec:CCFT2Charges} and discuss their explicit form in examples in section~\ref{sec:CCFT2CWardId}. 
We contrast this CFT approach with the equivalence between soft theorems and conservation laws for asymptotic symmetry charges in section~\ref{sec:softhardtopo}.

\subsection{Celestial Diamonds}
\label{sec:CelestialDiamonds}
Conformally soft operators  in CCFT$_2$ have protected integer dimensions. The strategy in~\cite{Pasterski:2021fjn} was to compare these protected operators to global primary descendants in CFT$_2$ which also have protected integer dimensions and are associated to reducible global conformal multiplets that are classified by representation theory.\footnote{While in standard CFT primary descendants are typically associated to null states, in CCFT for a conventional choice of inner product this is not automatically true -- see the discussion in~\cite{Pasterski:2021fjn}. In this work we are interested in the classification of reducible conformal multiplets in CCFT${}_d$ which is independent of the choice of inner product.}

To review the primary descendant classification it is  convenient to work in complex coordinates $z=x_1+i x_2$ and $\bar z=x_1 -i x_2$, define holomorphic an antiholomorphic derivatives $\partial\equiv \partial_z$ and $\bar \partial \equiv \partial_{\bar z}$, and note that they implement the action of the Virasoro generators $L_{-1}$ and $\bar L_{-1}$. 
The classification is then straightforward: we write the most general descendant operator $\p^{n} \bar \p^{\bar n} \Ocal_{\Delta,\ell}$, and demand that it obeys the primary condition of being annihilated by the Virasoro generators $L_1$ and $\bar L_1$. 
The result of the computation is schematically presented in Figure~\ref{fig:CelestialDiamond} where the nodes correspond to primary operators and the arrows to the right (left) denote the action of $\partial\leftrightarrow L_{-1}$ ($\bar \partial\leftrightarrow \bar L_{-1}$).
\tikzset{->-/.style={decoration={
  markings,
  mark=at position #1 with {\arrow{>}}},postaction={decorate}}}
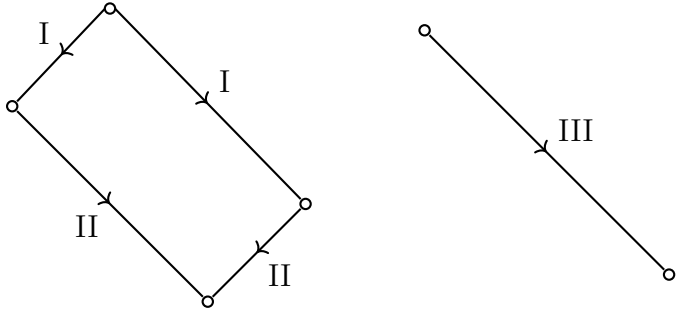
\begin{figure}[h!]
\begin{center}
\begin{tikzpicture}[scale=1.3]
\draw[thick,->-=0.5](0-.05,2)--node[above left]{$\rm I$} (-1+.05,1+.05);
\draw[thick,->-=0.5](0+.05,2)--node[above right]{$\rm I$} (2-.05,0+.05);
\draw[thick,->-=0.5] (-1+.05,1-.05)node[left]{} --node[below left]{$\rm II$} (1-.05,-1+.05) ;
\draw[thick,->-=0.5] (2-.05,0-.05)node[right]{}--node[below right]{${\rm II}$} (1.05,-1+.05) ;
\draw[thick](1,-1.2)node[below]{};
\draw[thick] (0,2) circle (1.5pt) ;
\draw[thick] (2,0) circle (1.5pt) ;
\draw[thick] (-1,1) circle (1.5pt) ;
\draw[thick] (1,-1) circle (1.5pt) ;
\end{tikzpicture}
\hspace{1cm}
\raisebox{2em}{\begin{tikzpicture}[scale=1.3]
\draw[thick,->-=0.5](0+.05,2.5-.05)--node[above right]{$\rm III$} (2.5-.05,0+.05);
\draw[thick] (0,2.5) circle (1.5pt) ;
\draw[thick] (2.5,0) circle (1.5pt) ;
\end{tikzpicture}}
\caption{Celestial diamonds in $d=2$.
 The picture shows the $\Delta-\ell$ plane, where the nodes correspond to the position of primaries while the arrows pointing to the bottom right/left denote the action of derivatives $\partial$/$\bar \partial$  (longer arrows correspond to higher numbers of derivatives). 
 The diamond picture on the left  shows the existence of a nested structure of descendants that are themselves primaries.
The picture on the right is understood as a diamond with zero area.
}
\label{fig:CelestialDiamond}
\end{center}
\end{figure}
We distinguish three categories of primary descendants $\I,\II,\III$ (defined below) depending on whether the spin of the primary descendant is larger, smaller or equal in absolute value compared to the spin of the parent primary operator. 
The conformally soft operators of section \ref{sec:CCFTWardID} are  primary operators with primary descendants of one of these three types. \\
The universal soft theorems which scale as $\omega^{-k}$ for $k=1,0,\dots,1-\ell$ (where $\ell$ is the spin of the soft particle) are associated to operators with type II primary descendants. 
The subleading soft theorem scaling as $\omega^{-1+\ell}$ are associated to type III operators that arise in zero-area celestial diamonds. All the ever more subleading soft theorems are related to primary descendants of type~I.

\paragraph{Type I}
A descendant of type I has spin ${\mathcal n}$ units larger in absolute value than its parent primary. Type I descendant operators are further divided in Type Ia and Ib, where the former is a descendant at level ${\mathcal n}$ of the form 
\begin{equation}\label{typeIa}
    O_{\Ia,{\mathcal n}}(z,\bar z)\equiv \partial^{\mathcal n} \Ocal_{\Delta,+|\ell|}\,, \quad\text{or}\quad  \Ocal_{\Ia,{\mathcal n}}(z,\bar z)\equiv \bar{\partial}^{\mathcal n} \Ocal_{\Delta,-|\ell|}\,,
\end{equation}
while the latter is a descendant at level $2|\ell|+{\mathcal n}$ of the form
\begin{equation}\label{typeIb}
    \Ocal_{\Ib,2|\ell|+{\mathcal n}}(z,\bar z)\equiv \partial^{2|\ell|+{\mathcal n}} \Ocal_{\Delta,-|\ell|}\,,  \quad\text{or}\quad    \Ocal_{\Ib,2|\ell|+{\mathcal n}}(z,\bar z)\equiv \bar \partial^{2|\ell|+{\mathcal n}} \Ocal_{\Delta,+|\ell|}\,.
\end{equation}
The operators $\Ocal_{{\rm Ia},{\mathcal n}}$ and $\Ocal_{{\rm Ib},2|\ell|+{\mathcal n}}$ become primary when
\begin{equation}
    \Delta=1-|\ell|-{\mathcal n}\,, \quad {\mathcal n}\in \mathbb Z_{>0}\,.
\end{equation}

\paragraph{Type II}
A level ${\mathcal n}$ descendant of type II has spin ${\mathcal n}$ smaller in absolute value than its parent. Type II descendant operators are of the form
\begin{equation}
    O_{\II,{\mathcal n}}\equiv \bar\partial^{\mathcal n} \Ocal_{\Delta,+|\ell|}\,, \quad \text{or} \quad \Ocal_{\II,{\mathcal n}}\equiv  \partial^{\mathcal n} \Ocal_{\Delta,-|\ell|}\,,
\end{equation}
and become primary when
\begin{equation}
    \Delta=1+|\ell|-{\mathcal n}\,,\quad {\mathcal n}\in \{1,\dots,2|\ell|-1\}\,.
\end{equation}
These primary descendants only exist for $|\ell|\geq 1$ in the range $1-|\ell|<\Delta<1+|\ell|$. 

\paragraph{Type III}
A type III descendant has the same absolute value of spin as the parent primary and takes the form
\begin{equation}
    O_{\III,2|\ell|}\equiv \bar \partial^{2|\ell|}\Ocal_{\Delta,+|\ell|}\,, \quad \text{or} \quad \Ocal_{\III,2|\ell|}\equiv  \partial^{2|\ell|}\Ocal_{\Delta,-|\ell|}\,.
\end{equation}
It becomes primary when
\begin{equation}
    \Delta=1-|\ell|\,.
\end{equation}
\smallskip

Primary descendants of type I, II and III are associated to conserved operators from which we can construct Noether currents and topological charges. %

Before moving on to their construction, let us point out a notable feature of celestial diamonds: the primary operators at the left and right corners are related by the shadow transform in $d=2$ CFT defined by the integral transform 
\be \label{2dshadow}
S[\Ocal_{h, \bar h}](z,\bar z) \equiv \frac{K_{h, \bar h}}{2 \pi} \int d^2 w \frac{\Ocal_{h, \bar h}(w,\bar w)}{(z-w)^{2-2 h}(\bar z- \bar w)^{2-2 \bar h}} \, .
\ee
Here and below we may equivalently trade the labels $\Delta,\ell$ of the operators for $h\equiv \frac{1}{2}(\Delta+\ell)$ and $\bar h\equiv \frac{1}{2}(\Delta-\ell)$, and for the choice of normalization $K_{h, \bar h}=2 \max\{h,\bar h\}-1$ the shadow squares to one up to a sign $S[S[\Ocal_{h, \bar h}]]=(-1)^{2(h-\bar h)}$. 
As we will see, shadow transformed conformally soft primary operators are naturally associated to standard conserved operators in CCFT.

\subsection{Conserved Charges}
\label{sec:CCFT2Charges}
Having reviewed how to classify all primary descendants in CFT$_{d=2}$, we are now in a position to show how to get topological charges from the knowledge of these primary descendants.
We begin by rewriting formula  \eqref{top_op} in $d=2$ where it is convenient to use complex variables.
The topological charges obtained by integrating Noether currents can be expressed as\footnote{Replace the outward-directed differential $dS_a$ orthogonal to $\Sigma$ by a counterclockwise differential parallel to a choice of contour defined via $dS_a=\epsilon_{ab}ds^b$ where $\epsilon_{z\bar z}=-\epsilon_{\bar z z}=\frac{i}{2}$, whose holomorphic (antiholomorphic) component is $dz$ ($d\bar z$).}
\begin{equation}\label{QSigma2d}
    Q_{\Sigma}=\int_{\Sigma} dS^a \Jcal_a=-\frac{1}{2\pi i}\oint_{\Sigma} (dz  \Jcal - d\bar z  \bar \Jcal)\, ,
\end{equation}
where we defined $\Jcal \equiv -2\pi \Jcal_z$ and  $\bar \Jcal\equiv -2\pi \Jcal_{\bar z}$, and $\Sigma$ is just a contour in $d=2$. 

In the following we show how to build the Noether current $\Jcal$  ($\bar \Jcal$) by opportunely combining an operator $\Ocal$ which has a primary descendant, with a parameter $\epsilon$  ($\bar \epsilon$).
We start by considering a primary operator  $\mathcal O$  with dimension $\Delta$ and spin~$|\ell|$ which has a level $\kcal$ primary descendant and thus satisfies the following shortening condition 
\begin{equation}
\label{gen_short_cond_2d}
    \bar \partial^{\kcal} \Ocal(z, \bar z)=0\, .
\end{equation} 
At this level, the shortening condition \eqref{gen_short_cond_2d} (which should be considered as an operator equation valid away from contact points) can be of any type I, II or III depending on the value of  $\ell$ and~$\Delta$ (which we assume to be fixed to one of the values defined in the previous section). 
The rest of this section will hold regardless of the chosen type.

We can construct a Noether current $\Jcal$ by combining $\Ocal$ with a parameter $\epsilon$ as follows
\be\label{Noether2d}
\Jcal^\epsilon=\sum_{m=0}^{\kcal-1} (-1)^{m}  \bar\partial^{m}\epsilon(z,\bz)  \bar\partial^{\kcal-m-1} \Ocal(z,\bar z) \,.
\ee
Conservation of the Noether current
\begin{equation}
    \bar \partial \Jcal^\epsilon=0\,,
\end{equation}
requires that $\epsilon$ satisfies the same type of (higher-derivative) conservation equation as the operator $\Ocal$, namely
\be
\bar \partial^{\kcal} \epsilon(z, \bar z)=0 \, ,
\ee
which we shall call generalized Killing tensor equation.
These conservation equations are solved by making a polynomial ansatz for both $\Ocal$ and $\epsilon$,
\be\label{epssum} 
\Ocal(z, \bar z)=\sum_{m=0}^{\kcal-1} \bar z^m \Ocal^{(m)}(z)\,,\quad \epsilon(z, \bar z)=\sum_{m=0}^{\kcal-1} \bar z^m \epsilon^{(m)}(z) \,,
\ee
where each $\Ocal^{(m)}(z)$ and $\epsilon^{(m)}(z)$ is holomorphic, and we further expand $\epsilon^{(m)}(z)=\sum_{n \in \mathbb{Z}} \epsilon_{m,n} z^{n}$. 
Integrating the Noether current~\eqref{Noether2d} we obtain
\begin{equation}
 Q^{\epsilon}_{\Sigma}= \frac{1}{2\pi i}\oint_{\Sigma} dz \Jcal^\epsilon(z)\equiv \sum_{m=0}^{\kcal -1}\sum_{n\in \mathbb Z}   m!(\kcal-m-1)! \epsilon_{m,n} Q^{m,n}\,,
\end{equation}
with $\kcal$ towers ($m=0, \dots, \kcal-1$) of infinitely many charges ($n\in\mathbb Z $),  
\be
\label{charges2d_a}
Q^{m,n}=  \frac{(-1)^{m}}{2\pi i} \oint_{\mathcal C} dz z^{n} \Ocal^{(n-m-1)}(z)\,.
\end{equation}
Following the same arguments for the analogous conservation equations with $\bar \partial^\kcal \to \partial^\kcal$ we can construct a Noether current 
$\bar \Jcal^{\bar \epsilon}(\bar z)$ 
and associated topological charge $\bar Q^{m,n}$.
The infinite tower is a special feature of $d=2$.

While the construction of these charges is general for all types I, II and III of conserved operators,\footnote{To be precise, for type I operators there are two shortening conditions (nameley Ia and Ib) for both holomorphic and anti-holomorphic derivatives.
So the Noether current \eqref{Noether2d} for type I,  secretly satisfies extra conservation equations in  the $z$ variable. 
In principle one could also define a type I Noether current which is  annihilated by both $\p$ and $\bar \p$ by generalizing formula \eqref{Noether2d}. The resulting current would not satisfy extra relations, but a function annihilated by both these derivatives should be constant. This is yet another way to see  that type I operators should not be associated to non-trivial charges.} 
their insertion in correlation functions and the action of these charges on other operators depends on their type. 
Indeed, since the Ward identities associated to type I operators do not have contact terms (see appendix \ref{app:typeI}), the charges associated to type I operators are trivial.\footnote{In \cite{Himwich:2021dau}  non-trivial charges were defined, by using light transform of type I operators. A similar picture seems to hold in $d>2$, where type I operators have trivial charges but their shadow transform give rise to non-trivial ones as we will discuss in section~\ref{sec:CCFTd}.} In contrast, the charges for type II and III operators are non-trivial as we will exemplify in the following.

\subsection{$\text{CCFT}_{d=2}$ Ward Identities}
\label{sec:CCFT2CWardId}
In this section we shall see how the construction of the charges applies to celestial CFT$_{d=2}$.
To start we want to recast the soft theorems reviewed in section~\ref{sec:CCFTWardID} as  correlation functions of conserved operators $\Ocal_{\Delta,\ell}(q,\varepsilon)$ in celestial CFT$_{d=2}$. To do so, we review some simple properties of the $d=2$ kinematics. 
The Poincaré section, expressed in $z$ and $\bz$ coordinates, is given by
\be\label{qPoincare2d}
q^\mu(z,\bar z)=\frac{1}{2}\left(1+z\bz,z+\bz,i(\bz-z),1-z\bz\right)\,.
\ee
We take as CCFT$_2$ polarization vectors 
\be
\mathcal e_+^a=\frac{1}{\sqrt{2}}(1,-i)\qquad\text{ and }\qquad \mathcal e_-^a=\frac{1}{\sqrt{2}}(1,i)\, ,
\ee
which are normalized such that $\mathcal e_+\cdot\mathcal e_-=1 $ and verify $\mathcal e_\pm^2=0$. They satisfy $x\cdot \mathcal e_-=\frac{1}{\sqrt{2}}z$ and $x\cdot \mathcal e_+=\frac{1}{\sqrt{2}}\bz$.
In index-free notation, the embedding space polarization vectors of positive and negative helicity orthogonal to $q^\mu$ are
\begin{equation}
    \varepsilon^\mu_+= \ecal_+^a  \partial_a q^\mu
    =\frac{1}{\sqrt{2}}(\bz,1,-i,-\bz)\,,  \qquad \text{ and }\qquad
\varepsilon^\mu_-= \ecal_-^a  \partial_a q^\mu
=\frac{1}{\sqrt{2}}(z,1,i,-z)\,.
\ee
They are normalized such that $\varepsilon_+\cdot \varepsilon_{-}=1$ and satisfy $\varepsilon_+\cdot \varepsilon_{+}=0$, $\varepsilon_- \cdot \varepsilon_{-}=0$.
To express the $d+2=4$ dimensional soft theorems in the Poincaré section~\eqref{qPoincare2d} we use that $q\cdot q_i=-\frac{1}{2}(z-z_i)(\bz-\bz_i)$, as well as $q_i\cdot \varepsilon_+=\frac{1}{\sqrt{2}}(\bar z_i-\bar z)$ and  $q_i\cdot \varepsilon_-=\frac{1}{\sqrt{2}}(z_i-z)$, and note that the helicity of massless particles in $\mathbb R^{1,3}$ gets identified with the spin $\ell$ in CCFT$_2$.

Each soft theorem then corresponds to a spin~$\ell$ primary operator $\Ocal_{\Delta,\ell}(z,\bar z)$ with a special conformal dimension $\Delta$ 
which has a given level $\kcal$ primary descendant determined by conformal representation theory.
We apply our general expression for the Noether current~\eqref{Noether2d} to the soft theorems, and we show that our infinite towers of topological charges reproduce the results in the literature.
For the conformally soft operators~\eqref{conf_soft_ops}-\eqref{conf_soft_ops_indexfree} with $\Delta \to k=1,0,-1,...$ it will be useful to introduce the mode expansion~\cite{Guevara:2021abz,Strominger:2021lvk}\footnote{Note that in the literature~\cite{Guevara:2021abz,Strominger:2021lvk} the labels are shifted, e.g. the sum there ranges between $\frac{k-\ell}{2}\leq m \leq \frac{\ell-k}{2}$.}
\begin{equation}\label{RHexpansion}
   R^{+}_k(z,\bar z)=\sum_{m=0}^{\kcal-1} \bar z^{m} R_k^{(m)}(z)\,, \quad  H_k^{+}(z,\bar z)=\sum_{m=0}^{\kcal-1} \bar z^{m} H_k^{(m)}(z)\,,
\end{equation}
where we have added a $+$ label for the spin $\ell=+|\ell|$ compared to~\eqref{conf_soft_ops_indexfree}. The corresponding $\ell=-|\ell|$ modes have a $-$ label in~\eqref{RHexpansion} and $z\leftrightarrow \bar z$.
The leading soft photon theorem ($\Delta=1$) and the leading and subleading soft graviton theorems $(\Delta=1$ and $\Delta=0$) give rise to Ward identities for type II primary descendants. The subleading soft photon theorem ($\Delta=0$) and the subsubleading soft graviton theorem ($\Delta=-1$) instead give rise to Ward identities for type~III primary descendants. The conformally soft operators with ever more negative conformal dimensions give rise to type I primary descendants.
\subsubsection{Leading Soft Photon}
The leading positive-helicity soft photon theorem corresponds to the insertion of a $U(1)$ current $J(z)\equiv R_1^{+}(z)$  in the celestial correlation function, namely~\cite{He:2014cra}
\begin{equation}
    \langle  J(z) \Ocal_{\Delta_1,\ell_1}\cdots  \Ocal_{\Delta_N,\ell_N}\rangle =\sqrt{2} e\sum_{i=1}^N \mathcal Q_i \frac{1}{z-z_i}\langle  \Ocal_{\Delta_1,\ell_1}\cdots \Ocal_{\Delta_i,\ell_i}\cdots \Ocal_{\Delta_N,\ell_N}\rangle\,,
\end{equation}
which satisfies $\bar \p J(z)=0$ up to contact terms.
The spin $|\ell|=1$ operator $\bar \partial R_{1}( z)$ is a primary descendants of type $\II$ at level~$1$ and the Noether current~\eqref{Noether2d} is simply
\begin{equation}
    \Jcal^\epsilon(z)=\epsilon(z) J(z)\,.
\end{equation}
Its conservation implies that the associated symmetry parameter has to satisfy $\bar \partial \epsilon=0$ which can therefore be expanded as $\epsilon(z)=\sum_{n\in \mathbb{Z}}  \epsilon_{n} z^n$ and we get one tower of charges
\begin{equation}
    Q^{0,n}={\frac{1}{\sqrt{2}e}}\frac{1}{2\pi i}\oint_{\Sigma} dz z^n J(z)\,.
\end{equation}
Similar statements are obtained for the negative-helicity soft photon upon  $z\leftrightarrow \bar z$ which gives $\bar \Jcal^{\bar \epsilon} =\bar \epsilon \bar J$ and another tower of charges $\bar Q^n$ contributing to~\eqref{QSigma2d}.
Due to the symmetry of the celestial photon diamond~\cite{Pasterski:2021fjn} analogous statements hold for the shadow transformed soft photon operator. 
The charges act on an operator with conformal dimension $\Delta$, spin $\ell$ and charge $\mathcal Q$ as 
\begin{equation}
    [Q^{0,n},\Ocal_{\Delta,\ell}(z,z)]=\mathcal Q z^n  \Ocal_{\Delta,\ell}(z,z)\, .
\end{equation}
The enhancement from a global Noether current where  $\epsilon=\text{const}$ to a local one in CCFT$_2$ corresponds to the large gauge symmetry in the $D=4$ dimensional bulk. 

\subsubsection{Leading Soft Graviton}
The leading positive-helicity soft graviton theorem corresponds to the insertion of an operator $H_1^{{+}}(z,\bar z)$ in the celestial amplitude~\cite{He:2014laa}
\be
\label{eq:WI_H1}
 \langle H_1^{+}(z,\bar z) \Ocal_{\Delta_1,\ell_1}\cdots  \Ocal_{\Delta_N,\ell_N}\rangle =- \frac{\kappa}{2}\sum_{i=1}^N \eta_i\frac{\bar z-\bar z_i}{z-z_i}\langle  \Ocal_{\Delta_1,\ell_1}\cdots \Ocal_{\Delta_i+1,\ell_i} \cdots \Ocal_{\Delta_N,\ell_N}\rangle \,,
\ee
that satisfies a non-standard (since higher-derivative) Ward identity with $\bar \partial^2 H_1^+(z,\bar z)=0$ up to contact terms. 
Note that $P(z)\equiv \bar \partial  H_1^{+}(z,\bar z)$, which satisfies the more standard conservation equation $\bar \p P(z)=0$, is a descendant but not a primary. Instead, the spin $|\ell|=2$ operator $\bar \partial^2 H_1^{+}(z,\bar z)$ is a primary descendant of type $\II$ at level~$2$. The associated symmetry parameter also satisfies a higher-derivative conservation equation $\bar \partial^2  \epsilon=0$ and can be expanded as $\epsilon(z,\bz)=\epsilon^{(0)}(z)+\bz \, \epsilon^{(1)}(z) $  where $\epsilon^{(m)}(z)=\sum_{n\in \mathbb{Z}}\epsilon_{m,n}z^{n}$ for $m=0,1$. The Noether current is
\begin{equation}\label{JcalBMS}
  \Jcal^\epsilon( z) =  H_1^{(1)}(z) \epsilon^{(0)}(z)- H_1^{(0)}(z) \epsilon^{(1)}(z)\,,
\end{equation}
which yields two towers of charges
\begin{equation}\label{Qleadinggraviton}
  Q^{m,n}\equiv {-\frac{2}{\kappa}} \frac{(-1)^{m}}{2\pi i} \oint_{\Sigma} dz z^{n}  H_1^{(1-m)}(z)\,, \quad m=0,1\,.
\end{equation}
From this we see that the operator $P(z)= H_1^{(1)}(z)$ (which is usually called the supertranslation current) does not know about the infinite tower of conserved charges associated to $ H_1^{(0)}$. This was emphasized in~\cite{Fotopoulos:2019vac,Banerjee:2021cly} and is by now well-known in the celestial literature. It is crucial to use~\eqref{JcalBMS} so to take into account all possible generators. 
Similar statements hold for the negative-helicity soft graviton as well as for the shadow transform since the leading celestial graviton diamond is symmetric. 
Using \eqref{eq:WI_H1} it is easy to see that the charges act as
\begin{equation}\label{BMScommutator}
    [Q^{m,n},\Ocal_{\Delta,\ell}(z,\bar z)]=\eta z^{n} \bar z^{m}\Ocal_{\Delta+1,\ell}(z,\bar z).
\end{equation}
In the above commutator, $\eta=+1$ for \textit{out} operators and $\eta=-1$ for \textit{in} operators.
In some of the literature (see e.g.~\cite{Fotopoulos:2019vac}) the global charges are called $Q^{0,0}\equiv P_{-\frac{1}{2},- \frac{1}{2}}$, $Q^{0,1}\equiv P_{\frac{1}{2},-\frac{1}{2}}$, $Q^{1,0}\equiv P_{-\frac{1}{2},\frac{1}{2}}$, $Q^{1,1}\equiv P_{\frac{1}{2},\frac{1}{2}}$.
 The action~\eqref{BMScommutator} corresponds to the enhancement of the $d+2=4$ spacetime translations to BMS supertranslations.
 While $m=0,1$ in~\eqref{Qleadinggraviton}, all other $m\in \mathbb Z$ supertranslations can be obtained from the commutator with the Virasoro generators using the celestial OPE~\cite{Fotopoulos:2019vac}.

\subsubsection{Subleading Soft Graviton}
The subleading celestial graviton diamond is not symmetric~\cite{Pasterski:2021fjn} and so we get different conservation equations for the conformally soft primary and its shadow.
The subleading negative-helicity soft graviton theorem corresponds to the correlation function
\be
 \langle H_0^{-}(z,\bar z) \Ocal_{\Delta_1,\ell_1}\cdots  \Ocal_{\Delta_N,\ell_N}\rangle = \frac{\kappa}{2}\sum_{i=1}^N 
 \frac{(z-z_i)^2}{\bar z-\bar z_i}\left(\frac{2h_i}{z-z_i}-\partial_{z_i}\right)
 \langle  \Ocal_{\Delta_1,\ell_1}\cdots \Ocal_{\Delta_i,\ell_i} \cdots \Ocal_{\Delta_N,\ell_N}\rangle \,,
\ee
while we recognize its shadow transform as a correlator with a stress tensor 
insertion~\cite{Kapec:2016jld,Kapec:2014opa}
\be
 \langle T(z) \Ocal_{\Delta_1,\ell_1}\cdots  \Ocal_{\Delta_N,\ell_N}\rangle = {\kappa}\sum_{i=1}^N 
 \left(\frac{h_i}{(z-z_i)^2}+\frac{1}{z-z_i}\partial_{z_i}\right)
 \langle  \Ocal_{\Delta_1,\ell_1}\cdots \Ocal_{\Delta_i,\ell_i} \cdots \Ocal_{\Delta_N,\ell_N}\rangle \,.
\ee
The stress tensor $T(z)\equiv \S[ H_0^{-}](z)\equiv\tilde H_2^+(z)$ 
is conserved $\bar \partial T(z)=0$ up to contact terms, as is the operator $\tilde T(z,\bar z)\equiv H_0^{-}(z,\bar z)$ albeit in the form of a higher-derivative conservation equation $\partial^3 \tilde T(z,\bar z)=- \bar \partial T(z)$.
The spin $|\ell|=2$ operators $\partial^3 H_{0}^{-}(z,\bar z)$ and $\bar \partial \tilde H_{2}^{+}(z)$ are primary descendants of type $\II$ at, respectively, level~$3$ and level~$1$. The symmetry parameter for the stress tensor is required to be holomorphic $\bar \partial \epsilon=0$ in order for the Noether current
\begin{equation}
    \Jcal^\epsilon(z)=\epsilon(z) T(z)
\end{equation}
to be conserved. We obtain the tower of charges 
\begin{equation}
Q^{0,n}={\frac{1}{\kappa}}\frac{1}{2\pi i}\oint_{\Sigma} dz z^n  T(z)\,,
\end{equation}
which act as Virasoro generators $Q^{0,n+1}\equiv L_n$ 
\begin{equation}
    [Q^{0,n+1},\mathcal O_{\Delta,\ell}(z,\bar z)]=z^{n}[(n+1)h+z\partial_z]\mathcal O_{\Delta,\ell}(z,\bar z)\,.
\end{equation}
They enhance the global to the local conformal transformations, or equivalently, Lorentz transformations to Virasoro superrotations.
Meanwhile, the symmetry parameter for the shadow stress tensor satisfies the higher-derivative conservation equation $\partial^3 \tilde \epsilon=0$. Expanding  $\tilde \epsilon(z,\bar z)=\tilde \epsilon^{(0)}(\bar z)+z \tilde \epsilon^{(1)}(\bar z)+z^2 \tilde \epsilon^{(2)}(\bar z)$ we can write the Noether current as 
\begin{equation}
\badat{2}
    \quad \tilde \Jcal^{\tilde\epsilon}(\bar z)
    &=2H_0^{(2)}(\bar z) \tilde \epsilon^{(0)}(\bar z)-H_0^{(1)}(\bar z) \tilde \epsilon^{(1)}(\bar z)+2H_0^{(0)}(\bar z)\tilde \epsilon^{(2)}(\bar z)\,,
\eadat
\end{equation}
and get three towers of charges 
\begin{equation}
\tilde Q^{m,n}=
\frac{2}{\kappa}\frac{(-1)^m}{2\pi i} \oint_{\Sigma} d\bz \bz^{n} H_0^{(2-m)}(\bz) \,, \quad m=0,1,2\,.
\end{equation}
These charges act on generic operators as follows
\be
\begin{aligned}
[\tilde Q^{0,n},\Ocal_{\Delta,\ell}(z,\bz)]=&-\bz^n \p_z \Ocal_{\Delta,\ell}(z,\bz)\,,\\
[\tilde Q^{1,n},\Ocal_{\Delta,\ell}(z,\bz)]=&-\bz^n(2h+2z\p_z)\Ocal_{\Delta,\ell}(z,\bz)\,,\\
[\tilde Q^{2,n},\Ocal_{\Delta,\ell}(z,\bz)]=&-\bz^n(2h z+z^2 \p_z)\Ocal_{\Delta,\ell}(z,\bz)\, .
\end{aligned}
\ee
Similar expressions are obtained for the opposite helicity charges.
%

\subsubsection{Subleading Soft Photon and Subsubleading Soft Graviton}
There are further subleading soft theorems which correspond to correlation functions of conserved operators albeit with primary descendants of a different type.
The positive-helicity subleading soft photon theorem can be expressed as~\cite{Lysov:2014csa}
\be
 \langle R_0^{+}(z,\bar z) \Ocal_{\Delta_1,\ell_1}\cdots  \Ocal_{\Delta_N,\ell_N}\rangle = -\sqrt{2}e\sum_{i=1}^N \eta_i\frac{\mathcal Q_i}{z-z_i}
 \left(2\bar h_i-1-( \bar z- \bar z_i)\partial_{ \bar z_i}\right)
 \langle  \Ocal_{\Delta_1,\ell_1}\cdots \Ocal_{\Delta_i-1,\ell_i} \cdots \Ocal_{\Delta_N,\ell_N}\rangle \,,
\ee
where the spin $|\ell|=1$ operator $R_0^{+}(z,\bar z)$ is conserved as $\bar \partial^2 R_0^{+}(z,\bar z)=0$ up to contact terms. This corresponds to a type III primary descendant at level~2.
For a symmetry parameter satisfying $\bar \partial^2 \epsilon=0$ we can write the Noether current
\begin{equation}\label{Noethersubphoton}
    \Jcal^\epsilon(z)= R_0^{(0)}(z) \epsilon^{(1)}(z)- R_0^{(1)}(z)\epsilon^{(0)}(z)\,,
\end{equation}
and obtain two towers of charges in terms of the two modes $R_0^{(m)}$ for $m=0,1$. Similar statements hold for the opposite helicty photon and its shadow. 
The associated charges
\be
Q^{m,n}=-\frac{1}{\sqrt{2}e}\frac{(-1)^m}{2\pi i} \oint_\Sigma dz z^n R_0^{(1-m)}\, \qquad m=0,1
\ee
act on generic operators as follows~\cite{Banerjee:2020vnt}
\be
\begin{aligned}
[Q^{0,n},\Ocal_{\Delta,\ell}(z,\bz)]&=-\eta\mathcal Q z^n \bar\p \Ocal_{\Delta-1,\ell}(z,\bz)\,,\\
[Q^{1,n},\Ocal_{\Delta,\ell}(z,\bz)]&=-\eta\mathcal Q z^n(2\bar h-1+\bz\bar \p)\Ocal_{\Delta-1,\ell}(z,\bz)\, .
\end{aligned}
\ee

Meanwhile, the positive-helicity subsubleading soft graviton theorem takes the form~\cite{Campiglia:2016jdj}
\be
\badat{2}
 \langle H_{-1}^{+}(z,\bar z) \Ocal_{\Delta_1,\ell_1}\cdots  \Ocal_{\Delta_N,\ell_N}\rangle &= -\frac{\kappa}{4}\sum_{i=1}^N \eta_i\frac{\bar z-\bar z_i}{ z- z_i}
 \left(2\bar h_i(2\bar h_i-1)-2(\bar z-\bar z_i)2\bar h_i\partial_{\bar z_i}\right.\\
 & \qquad  \qquad\qquad\quad \left.+(\bz-\bz_i)^2\partial^2_{\bz_i}\right)
 \langle  \Ocal_{\Delta_1,\ell_1}\cdots \Ocal_{\Delta_i-1,\ell_i} \cdots \Ocal_{\Delta_N,\ell_N}\rangle \,.
\eadat
\ee
The spin~$|\ell|=2$ operator $H_{-1}^{+}(z,\bar z)$ is conserved as $\bar \partial^4 H_{-1}^{+}=0$ up to contact terms which corresponds to a type III primary descendant at level~$4$. For $\bar \partial^4 \epsilon=0$ we get the Noether current
\begin{equation}\label{Noethersubsubgraviton}
    \Jcal^\epsilon(z)=6 H_{-1}^{(0)}(z) \epsilon^{(3)}(z)-2 H_{-1}^{(1)}(z)\epsilon^{(2)}(z) +2 H_{-1}^{(2)}(z) \epsilon^{(1)}(z)-6 H_{-1}^{(3)}(z) \epsilon^{(0)}(z)\,,
\end{equation}
and obtain four towers of charges in terms of the four modes $H_{-1}^{(m)}$ for $m=0,1,2,3$. Analogous expressions can be obtained for the opposite helicity graviton and its shadow.
The associated charges
\be
Q^{m,n}=-\frac{4}{\kappa}\frac{(-1)^m}{2\pi i}\oint_\Sigma dz z^n H_{-1}^{(3-m)}\,,\qquad m=0,1,2,3
\ee
act on generic operators as follows~\cite{Banerjee:2021cly} 
\be
\begin{aligned}
[Q^{0,n},\Ocal_{\Delta,\ell}(z,\bz)]&=\eta z^n\bar\p^2 \Ocal_{\Delta-1,\ell}(z,\bz)\,,\\
[Q^{1,n},\Ocal_{\Delta,\ell}(z,\bz)]&=\eta z^n(4\bar h\bar\p+3\bz\bar\p^2)\Ocal_{\Delta-1,\ell}(z,\bz)\,,\\
[Q^{2,n},\Ocal_{\Delta,\ell}(z,\bz)]&=\eta z^n(2\bar h(2\bar h-1)+8\bar h\bz\bar \p+3\bz^2\bar\p^2)\Ocal_{\Delta-1,\ell}(z,\bz)\,,\\
[Q^{3,n},\Ocal_{\Delta,\ell}(z,\bz)]&=\eta z^n( 2\bar h(2\bar h-1)\bz+4\bar h\bz^2\bar \p+\bz^3\bar\p^2)\Ocal_{\Delta-1,\ell}(z,\bz)\,.
\end{aligned}
\ee
Again, $\eta=+1$ if $\Ocal_{\Delta,\ell}(z,\bz)$ is \textit{out} and $\eta=-1$ if it is \textit{in}.
In both gravity and gauge theory cases to take into account all possible generators one again needs to use the Noether currents~\eqref{Noethersubphoton} and~\eqref{Noethersubsubgraviton}, or equivalently, keep track of all relevant modes in the expansion~\eqref{RHexpansion} as was emphasized in~\cite{Banerjee:2020vnt,Banerjee:2021cly}. What sets these more subleading soft theorems apart is that the type III primary descendants of the conformally soft operators $R_0^{+}$ and $H_{-1}^{+}$ are also their shadows\cite{Pasterski:2021fjn}: 
\begin{equation}
{\textstyle \frac{1}{2!}} \bar\partial^2 R_0^{+}=\S[R_0^{+}]\equiv \tilde R_2^-\,, \quad  {\textstyle \frac{1}{4!}}\bar\partial^4 H_{-1}^{+}=\S[H_{-1}^{+}]\equiv\tilde H_{3}^-\,.
\end{equation}
We will uncover a similar relation in higher dimensions in section~\ref{sec:CCFTdSHCWardId}, albeit in a much more subtle form.

\subsection{The Soft, the Hard and the Topological Charge}\label{sec:softhardtopo}

Let us conclude this section by commenting on the relation between the standard CFT approach for computing charges from Ward identities and the charges associated to the asymptotic symmetries of gauge theory and gravity at null infinity in the ``soft theorem = asymptotic symmetry" connection. The latter are constructed using the covariant phase space formalism.
Upon imposing the antipodal matching condition~\cite{Strominger:2013jfa} charges for asymptotic symmetries satisfy the classical conservation law 
\begin{equation}
Q^+=Q^-
\end{equation}
between the past $(-)$ and future $(+)$ null boundary. At the level of the S-matrix the desired conservation law becomes
\begin{equation}\label{ChargeConservation}
\langle out|Q^+ S-S Q^-|in\rangle =0\,.
\end{equation}
To prove this relation\footnote{See~\cite{Strominger:2017zoo} for a review and references therein.} one typically splits the charge on each null boundary into a ``soft charge"~$Q^\pm_S$ that will be associated to zero energy modes and a ``hard charge"~$Q^\pm_H$ that contains the fields carrying energetic excitations.
The role of the soft charge is to define a soft operator $Q^\pm_S=\int d^2z \,\epsilon(z,\bz) \cdot O^\pm_S(z,\bz)$ whose insertion in the S-matrix creates/annihilates soft particles, while $Q^\pm_H$ implements the asymptotic symmetry transformations on the matter fields $[Q^\pm, \,.\,]=\delta_\epsilon(.)$.
The symmetry parameter  $\epsilon(z,\bar z)$ depends on the coordinates on the celestial sphere but not the (null) time coordinates.  Upon a judicious choice of (singular) parameters one finds that the conservation law
\begin{equation}\label{ConservationLawQSH}
\langle out|Q^+_S S-S Q^-_S|in\rangle =-\langle out|Q^+_H S-S Q^-_H|in\rangle
\end{equation}
is equivalent to the associated soft theorem: the integral $\int d^2z$ on the left-hand-side of~\eqref{ConservationLawQSH}, after integration by parts and using $\bar \p\frac{1}{z}=2\pi\delta^{(2)}(z)$, localizes to the insertion of a soft particle while the one on the right-hand-side becomes the soft factor times $\langle out |S|in\rangle$. Conversely, starting from the soft theorem we can ``smear" both sides by integrating over the sphere with some parameter $\epsilon(z,\bar z)$ and recover the soft and hard charges which are then guaranteed to obey the conservation law~\eqref{ChargeConservation}.

The standard way in CFT to construct conserved charges is to identify a conserved operator satisfying a Ward identity and then integrating the associated Noether current over a region of the CFT space. Soft theorems are equivalent to the insertion  of conserved operators in correlation functions  and conformal representation theory tells us the corresponding Ward identities. 
Alternatively, these Ward identities can be obtained in the above language -- without the need of introducing a special parameter $\epsilon(z,\bz)$ -- from an unintegrated version of~\eqref{ConservationLawQSH}. That is we define soft and hard operators, $O_S$ and $O_H$, 
through
\begin{equation}\label{QSQH}
Q_S^\pm=\int_{S^2} d^2z \,\epsilon(z,\bz) \cdot O^\pm_S(z,\bz)\,, \quad Q_H^\pm=\int_{S^2} d^2z \,\epsilon(z,\bz) \cdot O^\pm_H(z,\bz)\,,
\end{equation}
which satisfy the ``unsmeared" conservation law
\begin{equation}\label{ConservationLawOSH}
\langle out|O^+_S S-S O^-_S|in\rangle =-\langle out|O^+_H S-S O^-_H|in\rangle\,.
\end{equation}
In Maxwell theory and Einstein gravity the hard operators $O_H^+$ are given by $u$-integrals of the matter current $j_u$ and stress tensor $T_{uu}$ and $T_{ua}$, while the soft operators $O_S^+$ involve ($x^a$ derivatives of) the field strength $F_{ua}$ and the news tensor $N_{ab}$.
The relation~\eqref{ConservationLawOSH} holds because of the constraint equation of the respective theory and the antipodal matching condition imposed between the {\it in} and {\it out} fields. 
From the embedding-to-CFT-space perspective, since a single light ray passing through the origin of spacetime defines a single point in CFT space it is natural to consider the {\it in} and {\it out} operators as living on the same celestial sphere. The right-hand-side of~\eqref{ConservationLawOSH} computes the variation of the {\it in} and {\it out} operators with distributional support at their location, while the left-hand-side gives the conservation equation for the conformally soft operators discussed in section~\ref{sec:CCFT2CWardId}. 
In the standard CFT approach we would use this conservation equation, construct from it a Noether current and integrate it over some region on the sphere to obtain the surface charge. We can then relate this topological charge to the soft charge which integrates the soft operator over the entire sphere.

Let us illustrate this for the simplest example: the soft photon theorem associated to large gauge symmetry. From the constraint equation of Maxwell theory near future null infinity, $\partial_u F_{ru}^{(2)}+\p_\bz F_{uz}^{(0)}+\p_z F_{u\bz}^{(0)}+e^2 j_u^{(2)}=0$ and the integrand of the global electric charge $\star F=F_{ru}^{(2)}d^2z$, we find that the soft and hard operators are~\cite{Strominger:2017zoo} 
\begin{equation}
\badat{2}
O^+_S(z,\bar{z}) &=-\frac{1}{e^2} \int du  (\partial_z  F^{(0)}_{u \bar z} + \partial_{\bar z}  F^{(0)}_{u z} )\,, \qquad O^+_H(z,\bar{z})&= \int du  j_u^{(2)}\,,
\eadat
\end{equation}
where superscripts denote the inverse power of $r$ in the large radius expansion. These operators are associated with fields on the future null boundary while similar expressions hold for the charges at the past boundary where the retarded time $u$ is replaced by the advanced time $v$.
To compute the right-hand-side of~\eqref{ConservationLawOSH}  we use~\cite{Strominger:2013lka,Strominger:2017zoo} 
\begin{equation}
\badat{2}
[j_u^{(2)}(u',z,\bar z),\Phi_i(u,z_i,\bar z_i)]&=-\mathcal Q_i \Phi_i(u,z_i,\bar z_i) \delta^{(2)}(z-z_i) \delta(u-u')\,,
\eadat
\end{equation}
where $\Phi_i(u,z_i,\bar z_i)$ is a matter field of charge $\mathcal Q_i$. A similar expression holds for $j_v^{(2)}(v,z,\bz)$ albeit with the opposite sign. We thus obtain for the hard operator insertion in the S-matrix
\begin{equation}
\badat{2}
\langle out|O_H^+S -S O_H^-|in\rangle &=\langle out| \int du j_u(u,z,\bz) \,S -S\, \int dv j_v(v,z,\bz) |in\rangle\\
&= \sum_{i=1}^n \delta^{(2)}(z-z_i)\eta_i \mathcal{Q}_i \langle \Ocal_1\dots  \Ocal_i \dots \Ocal_n\rangle\,,
\eadat
\end{equation}
where $\eta_i=\pm$ distinguishes between {\it in} and {\it out} states.
For the soft operator insertion we get 
\begin{equation}
\badat{2}
\langle out|O_S^+S-SO_S^-|in\rangle &= \frac{1}{e^2} \partial_{\bar z}\langle out|\int du F_{uz}^{(0)}\, S-S\,\int dv F_{vz}^{(0)}|in\rangle +(z\leftrightarrow\bz) \\
&=  \partial_{\bar z}\langle out|J^+_zS-SJ_z^-|in\rangle +(z\leftrightarrow\bz) \,,
\eadat
\end{equation}
where we defined the soft photon current
\begin{equation}
J_a^+=\frac{1}{e^2}\int_{-\infty}^{+\infty} du F_{ua}^{(0)}\,,
\end{equation}
and a similar expression for the fields on the past boundary.
The equality of the soft and hard insertions~\eqref{ConservationLawOSH} gives the following relation
\be
 \partial_{\bar z}\langle out|J^+_zS-SJ_z^-|in\rangle +(z\leftrightarrow\bz)=\sum_{i=1}^n \delta^{(2)}(z-z_i)\eta_i \mathcal{Q}_i \langle \Ocal_1\dots  \Ocal_i \dots \Ocal_n\rangle\,.
\ee
This takes the form of a conservation equation for a current
\be
\label{eg:ward_id_AS}
\langle \p^a J_a(x) \Ocal_1\dots  \Ocal_n\rangle
=\sum_{i=1}^n \delta^{(2)}(z-z_i)\eta_i \mathcal{Q}_i \langle \Ocal_1\dots  \Ocal_i \dots \Ocal_n\rangle\,,
\ee
where $x^a$ is written in terms of $z, \bar z$. 
This Ward identity can be used to build a Noether current $\mathcal{J}^{\epsilon}$ by multiplying the current operator $J\equiv J_z$ by a parameter $\epsilon(z,\bz)$. 
For $\mathcal{J}^\epsilon(z)=\epsilon(z,z)J(z)$
such that $\bar \partial \mathcal{J}^\epsilon(z)=0$ (away from contact terms) we find that $\bar \partial \epsilon(z,\bar z) =0$
has to be satisfied and is thus holomorphic $\epsilon(z,\bar z)=\epsilon(z)$. This condition must hold inside the region bounded by $\Sigma=\partial R$ where we are integrating the Noether current to construct the charge
\be
Q^{\pm}_\Sigma=\int_R d^2z \bar \partial \left(\epsilon(z) J^\pm(z) \right) = \oint_\Sigma d z \,\epsilon(z)J^\pm(z)   \, .
\ee
Recall that the label $\pm$ refers to which operators (associated to {\it in} or {\it out} states) it acts on. 
One may want to define $Q_\Sigma=Q^{+}_\Sigma-Q^{-}_\Sigma$ as the total charge.
Expanding 
$\epsilon(z)=\sum_{n=-\infty}^{\infty} c_n z^n $
 in a power series we can write a countable basis for the charges. 
Some of the charges are actual symmetries: this happens for smooth $\epsilon(z)$ when $Q_\Sigma$ is zero for a $\Sigma$ that surrounds all the operators -- these are the global charges. Outside the region containing all the operators $\epsilon(z)$ may have a pole  and such local charges do not vanish, however, they  act in a precise way defined through the Ward identities.

Finally, to connect back to the above discussion note that the soft charge $Q_S^\pm$ is defined as an integral over the entire $S^2$, while the topological charge $Q_\Sigma^\pm$ is defined on a region $R$ bounded by $\Sigma=\partial R$ in $S^2$.
We see that the two quantities are equivalent 
\be\label{QSigmaRbarR}
Q_\Sigma^\pm  =\int_R d^2 z \bar \p (\epsilon J^\pm)\, 
\quad 
\longrightarrow 
\quad 
 Q_S^\pm=\int_{S^2} d^2z \epsilon \bar \p J^\pm \, ,
\ee
by assuming that $\epsilon$ does not have poles in the region $R$ and that this region surrounds all the operators.
Notice that the definition of $Q_\Sigma^\pm$ is more general than $Q_S^\pm$ since it computes the charges contained in a region $\Sigma$ of the celestial sphere, which can be chosen at will. Similarly one can define a more general version of the hard charge by integrating the right-hand-side of \eqref{eg:ward_id_AS} (times the parameter $\epsilon$)  on the same region $R$, which gives as a result the variation of all operators contained in such region. 
As a result, the Ward identity can be rewritten in the usual way as 
\begin{equation}
    [Q
    , \;\cdot \;]=\delta
    (\; \cdot \;)\,,
\end{equation}
where the left hand side arises from the soft operator and the right-hand-side from the hard one. 

Let us wrap up by commenting on the special symmetry parameters $\epsilon=\{\varepsilon, f, Y^z\}$ used in the literature to go from the charge conservation laws for large gauge transformations, BMS supertranslations and superrotations to the factorization theorems for the leading soft photon, leading soft graviton and subleading soft graviton. To localize the sphere integrals one takes $\varepsilon=\frac{1}{z-w}$, $f=\frac{\bz-\bw}{z-w}$, and $Y^z=\frac{(\bz-\bw)^2}{z-w}$.
Their form is readily explained: they satisfy (up to contact terms) $\bar \p \epsilon=0$, $\bar\p^2 \epsilon=0$ and $\bar \p^3 \epsilon=0$ as required by the conservation of the Noether currents.
The choice of a parameter $\epsilon$ that depends on new variables $w,\bar w$ is non-canonical but it simply has the effect of mapping the charge with such $\epsilon$ to the local conserved operator that defines them. In usual CFT manuals it is standard to reconstruct an operator by resumming the Laurent expansion of the charges. E.g. for a current $J(w)=\sum_{n} w^{-n-1} Q^{n}$, where the charges $Q^n=\frac{1}{2\pi i}\oint dz z^{n} J(z)$ are defined by  $Q_\Sigma^{\epsilon(z)=z^n}$. What is done in the CCFT literature is to consider the prescription  $J(w)=Q_\Sigma^{\epsilon(z)=1/(z-w)}$, where $\Sigma$ surrounds the pole at $z=w$, which is ultimately equivalent even if at a first sight it may look like the parameter is fine-tuned. 
While being equivalent we think that the usual CFT-inspired construction presented in this section is less confusing since it does not rely in the introduction of special parameters.

\section{Symmetries in Celestial CFT${}_{d>2}$}
\label{sec:CCFTd}

We now discuss the symmetries in celestial CFTs in $d>2$ dimensions.
We start in section~\ref{sec:SOd_tensors} by reviewing useful technology to deal with $SO(d)$ tensors.
In section~\ref{sec:CelestialNecklaces} we then classify the different types of primary descendant operators associated to CCFT symmetries using conformal representation theory.
By demanding that they satisfy some (higher-derivative) conservation equations we construct in section~\ref{sec:CCFTdCharges} their associated charges. 
In section~\ref{sec:CCFTdCWardId} and~\ref{sec:CCFTdSHCWardId} we identify the types of conserved operators that soft theorems and their shadow transforms give rise to and we  build the associated charges.

\subsection{Technology for $SO(d)$ tensors}
\label{sec:SOd_tensors}
In the rest of the paper it will be convenient to use an efficient technology to deal with $SO(d)$ tensors \cite{Costa:2011mg, Costa:2016hju, Costa:2014rya, Lauria:2018klo}.
In this section we review how this works starting with the traceless and symmetric representations and then generalizing to the mixed-symmetric ones. 
\subsubsection{Traceless and Symmetric Tensors }
We will often encounter CFT$_d$ traceless and symmetric tensors contracted with  polarization vectors ${\mathcal e}^a$ which square to zero. Let us review two ways to recover the indices either using a special differential operator or a projector.
First we introduce the differential operator \cite{PhysRevD.13.887} 
\be 
D^a_\ecal \equiv \left(\frac{d}{2}-1+\ecal\cdot 
\partial_{\ecal}
\right) \partial_{\ecal}^a
 -\frac{1}{2} \ecal^a  \partial_{\ecal}\cdot \partial_{\ecal}\ .
\label{Tod}
\ee
This can be used to differentiate the vectors ${\mathcal e}^a$ and automatically renders the final expression traceless and symmetric.
E.g. given a tensor $t(\ecal)\equiv t^{a_1 \dots a_\ell} \ecal_{a_1} \dots  \ecal_{a_\ell}$, we can recover the indices by
\be
\label{open_index_TST}
t^{a_1 \dots a_\ell}=
\frac{1}{\ell! (\frac{d}{2}-1)_\ell} D^{a_1}_\ecal \cdots D^{a_\ell}_\ecal t(\ecal) \, .
\ee
Indeed by taking derivatives of the polarization vectors we get a projector into traceless and symmetric representations $\pi_{\ell}$,
 \be
 \label{recovertensor}
    \ytableausetup{centertableaux,boxsize=1.8 em}
 \pi_{\ell}
\left(
{ \scriptsize
\begin{ytableau}
a_{1}&\, _{\cdots}&\, a_{\ell}\\
\end{ytableau}
}\ ;
{ \scriptsize
\begin{ytableau}
b_{1}&\, _{\cdots}&\, b_{\ell}\\
\end{ytableau}
}\
\right) = \frac{1}{\ell! (\frac{d}{2}-1)_\ell} D^{a_1}_\ecal \cdots D^{a_\ell}_\ecal \ecal^{b_1}  \cdots \ecal^{b_\ell} \, .
 \ee
This projector $ \pi_{\ell}$ is defined in such a way that if we contract its indices $b_i$ with a tensor $t$ (not necessarily in an irreducible representation) it gives back a tensor $t'$ that depends on the indices $a_i$ which are symmetric and traceless,
 \be
    \ytableausetup{centertableaux,boxsize=1.8 em}
 t'^{a_1 \dots a_\ell}=  \, \pi_{\ell}
\left(
{ \scriptsize
\begin{ytableau}
a_{1}&\, _{\cdots}&\, a_{\ell}\\
\end{ytableau}
}\ ;
{ \scriptsize
\begin{ytableau}
b_{1}&\, _{\cdots}&\, b_{\ell}\\
\end{ytableau}
}\
\right) t^{b_1 \dots b_\ell} \, .
 \ee
Another way to get back the open indices is to use the projector itself. Indeed the projector contracted with unconstrained vectors $\hat \ecal$, $\hat \fcal \in \mathbb{R}^d$ (now $\hat\ecal^2,\hat\fcal^2 \neq 0$) is known in a closed form 
\begin{align}
\label{Pi_l}
\pi_{\ell}(\hat \fcal;\hat \ecal)
&
\equiv 
\pi_{\ell}
\left(
{ \scriptsize
\begin{ytableau}
a_{1}&\, _{\cdots}&\, a_{\ell}\\
\end{ytableau}
}\ ;
{ \scriptsize
\begin{ytableau}
b_{1}&\, _{\cdots}&\, b_{\ell}\\
\end{ytableau}
}
\right)
\hat \fcal^{a_1}\hat \ecal^{b_1} \cdots \hat \fcal^{a_\ell}\hat \ecal^{b_\ell}
\\
&=
\frac{\ell!}{2^{\ell}(d/2-1)_\ell} 
|\hat\ecal|^\ell|\hat\fcal|^\ell
C_{\ell}^{d/2-1}\left(\frac{\hat \ecal \cdot \hat \fcal}{
|\hat\ecal| |\hat\fcal|
}\right) \, ,
\end{align}
where $C_\ell^{n}(x)$ is a Gegenbauer polynomial.
Now, since $\hat \ecal,\hat \fcal$ are unconstrained, by taking simple derivatives  $\partial_{\hat \ecal}, \partial_{\hat \fcal}$ we can open the indices of the projector which can then be contracted with the tensor. E.g. a traceless and symmetric tensor $t^{a_1 \dots a_\ell}$ can be recovered from its contracted form $t(\ecal)$ by\footnote{Notice that since we are in $\mathbb R^d$ we may equivalently use upper or lower indices.} 
\begin{equation}
t^{a_1 \dots a_\ell}=\frac{1}{(\ell!)^2}
\partial_{\hat \fcal}^{a_1} 
\cdots \partial_{\hat \fcal}^{a_\ell} \,
\pi_{\ell}(\hat \fcal ; \partial_\ecal) \, t(\ecal) \, .
\end{equation}
This method is easily generalizable to the  more complicated cases which we will need in the following.

\subsubsection{Mixed-symmetric Tensors }
\label{sec:polarization_vectors}
In general, in $d$ dimensions the representations of $SO(d)$ of the spin of tensor operators can be more complicated with respect to the traceless and symmetric one.
They are labelled by a set of $[d/2]$ numbers $ \ell=(\ell_1, \dots, \ell_{[d/2]-1},  |\ell_{[d/2]}|)$ where $\ell_i \geq \ell_{i+1}$ which can be associated to a Young tableau with $\ell_i$ boxes in the $i$-th row. In odd dimensions all the labels $\ell_i$ are greater or equal than zero while in even $d$ the label $\ell_{[d/2]}$ of the last row can be negative, so the Young tableau has to be labelled by its absolute value $|\ell_{[d/2]}|$.
We can place the indices of the tensor in the boxes of the Young tableau, as follows
\be
\label{mixed_sym_indices}
\bf a\equiv{ \scriptsize
\begin{ytableau}
a^{1}_{\!1}&\, _{\cdots}&\, _{\cdots}&\, _{\cdots}&\, _{\cdots}&\, _{\cdots}&\,a^{1}_{\ell_{1}}\\
a^{2}_{\!1}&\, _{\cdots}&\, _{\cdots}&\, _{\cdots}&a^{2}_{\ell_{2}}\\
\resizebox{!}{0.3 cm}{\bf \vdots}&\resizebox{!}{0.3 cm}{\bf \vdots}& \resizebox{!}{0.3 cm}{\bf \vdots}& \raisebox{0.1 cm}{\resizebox{!}{0.12 cm}{\rotatebox[origin=c]{45}{\dots}}} \\
a^{k}_{\!1}&\, _{\cdots}& a^{k}_{\ell_{ k}}
\end{ytableau}
}\ ,
\ee
with $k\leq [d/2]$.
The indices on the rows are symmetrized while the ones on the columns are antisymmetrized, all traces are then removed. 
Since the indices are not all symmetric or antisymmetric, these representations are sometimes called mixed-symmetric tensor representations.
E.g. a traceless symmetric operator with spin $\ell$ corresponds to $(\ell_1=\ell,\ell_2=0 \dots, \ell_{[d/2]}=0)$.
As before we can obtain an index-free notation by contracting the indices \eqref{mixed_sym_indices} with a set of polarization vectors $\ecal_i$ with $i=1, \dots k$, such that the indices $a^i_{j}$ (for $j=1,\dots, \ell_i$) of the $i$-th row are contracted with $\ell_i$ identical vectors $\ecal_i$, such that
$\ecal_i \cdot \ecal_l=0$ for $i,l=1, \dots, k$.
Using this method any mixed-symmetric tensor $t^{\bf a}$ that depends on the indices $\bf a$ in \eqref{mixed_sym_indices}  can then be encoded by an index-free expression $t(\ecal_1,\dots,\ecal_{k})\equiv t(\ecal)$.

In order to recover the indices it is convenient to introduce projectors into mixed-symmetric representations $ \pi_\ell({\bf a}; {\bf b})$  \cite{Costa:2016hju, Costa:2014rya} which depend on two sets of indices ${\bf a},{\bf b}$ as in \eqref{mixed_sym_indices}, namely 
\be
\label{pi_proj}
\pi_\ell({\bf a}; {\bf b})\equiv  \pi
 _\ell
\left(
{ \scriptsize
\begin{ytableau}
a^{1}_{\!1}&\, _{\cdots}&\, _{\cdots}&\, _{\cdots}&\, _{\cdots}&\, _{\cdots}&\,a^{1}_{\ell_{1}}\\
a^{2}_{\!1}&\, _{\cdots}&\, _{\cdots}&\, _{\cdots}&a^{2}_{\ell_{2}}\\
\resizebox{!}{0.3 cm}{\bf \vdots}&\resizebox{!}{0.3 cm}{\bf \vdots}& \resizebox{!}{0.3 cm}{\bf \vdots}& \raisebox{0.1 cm}{\resizebox{!}{0.12 cm}{\rotatebox[origin=c]{45}{\dots}}} \\
a^{k}_{\!1}&\, _{\cdots}& a^{k}_{\ell_{ k}}
\end{ytableau}
}\ ;
{ \scriptsize
\begin{ytableau}
b^{1}_{\!1}&\, _{\cdots}&\, _{\cdots}&\, _{\cdots}&\, _{\cdots}&\, _{\cdots}&\,b^{1}_{\ell_{1}}\\
b^{2}_{\!1}&\, _{\cdots}&\, _{\cdots}&\, _{\cdots}&b^{2}_{\ell_{2}}\\
\resizebox{!}{0.3 cm}{\bf \vdots}&\resizebox{!}{0.3 cm}{\bf \vdots}& \resizebox{!}{0.3 cm}{\bf \vdots}& \raisebox{0.1 cm}{\resizebox{!}{0.12 cm}{\rotatebox[origin=c]{45}{\dots}}} \\
b^{k}_{\!1}&\, _{\cdots}& b^{k}_{\ell_{ k}}
\end{ytableau}
}\
\right) \, .
\ee
Upon contraction of a tensor with one of the two sets of indices, the projector gives back a tensor dependent on the other set of indices opportunely symmetrized
$ t^{\bf a}=\pi_\ell({\bf a}; {\bf b}) t^{\bf b}$.
In particular, contracting two projectors we obtain the usual idempotence $\pi_\ell({\bf a}; {\bf b}) \pi_\ell({\bf b}; {\bf c})=\pi_\ell({\bf a}; {\bf c})$. 
 It is possible to compute the projectors in a closed form \cite{Costa:2016hju} when contracted with two sets of unconstrained vectors $\hat \fcal \equiv \{\hat\fcal_1,\dots,\hat\fcal_k\} $ and $ \hat\ecal \equiv \{ \hat\ecal_1,\dots, \hat\ecal_k\}$,
\be{}
 \pi_\ell(\hat\ecal;\hat\fcal) \equiv 
\pi_\ell({\bf a}; {\bf b}) \prod_{i=1}^{k}\prod_{j=1}^{\ell_i} \hat\ecal_i^{a^{i}_j}\hat\fcal_i^{b^{i}_j}
  \, ,
 \ee
 the simplest example being \eqref{Pi_l}.
To obtain an uncontracted projector we take derivatives of the vectors $\hat\ecal_i, \hat\fcal_i$.

Given the contracted tensor $t(\ecal)$ we can obtain the uncontracted one $t^{\bf a}$ (opportunely symmetrized) by simply taking derivatives $\partial_{\ecal_i}$ of all $\ecal_i$ and symmetrizing the result with a projector, namely
 \be 
 \label{proj_to_indices}
t(\hat \fcal) = \left(\prod_{i=1}^k \frac{1}{\ell_i!} \right) \pi_\ell(\hat \fcal;\partial_{\ecal}) t(\ecal) \, ,
 \ee
 where $\partial_{\ecal}\equiv \{\partial_{\ecal_1}, \dots, \partial_{\ecal_k}\}$ and the normalization is introduced to compensate for the factorials coming from powers of derivatives (e.g. $\partial_x^n x^n=n!$). In  \eqref{proj_to_indices}, the tensor $t^{\bf a}$ is still contracted with a set of vectors $\hat \fcal$, but since they are unconstrained one can simply recover the indices by taking usual derivatives
 \be
 t^{\bf a}= \left(\prod_{i=1}^k \frac{1}{\ell_i!} \right) \p_{\hat \fcal}^{a^{1}_1} \cdots \p_{\hat \fcal}^{a^{1}_{\ell_1}} \;
 \cdots \; \p_{\hat \fcal}^{a^{k}_1} \cdots \p_{\hat \fcal}^{a^{k}_{\ell_k}} t(\hat \fcal) \, .
 \ee
\subsection{Celestial Necklaces}
\label{sec:CelestialNecklaces}
The soft operators introduced in section~\ref{sec:CCFTWardID} are protected operators with integer dimensions. 
Indeed, they belong to reducible conformal multiplets -- that is multiplets that satisfy shortening conditions where a descendant in the multiplet becomes a primary.
In this section we explain how to determine the symmetries of CCFT$_d$ by classifying primary descendants in CFT$_d$.

The general logic is the same as the one discussed in section~\ref{sec:CelestialDiamonds} for CCFT${}_2$ but there are crucial differences in CCFT$_{d>2}$ which render the classification much more challenging. Let us give a flavor of the classification before going into the detailed discussion in the preceding subsections. 
Acting on a primary $\Ocal_{\Delta, \ell}$ with a combination of derivatives produces a descendant  $\Ocal'_{\Delta', \ell'}$. For this descendant to be itself a primary it has to be annihilated by the generator of special conformal transformations $K^a$, that is it has to satisfy $[K^a,\Ocal'_{\Delta', \ell'}(0)]=0$. 
The first major difference in $d>2$ is that the degeneracy of descendant operators at a given level and spin is typically larger than one, e.g. both $\p^\nu \p_\mu \Ocal^{\mu}_\Delta$ and $\square  \Ocal^{\mu}_\Delta$ are level two descendants of $\Ocal^{\mu}_\Delta$ in the vector representation. This means that there may (and in fact does) exist a special linear combination of such descendants that becomes primary when $\Delta$ takes a special value. The second complication is the much richer structure of $SO(d)$ spin representations. In $d>2$ we write derivatives with tensor indices and we should define the resulting tensor such that it transforms in an irreducible representation of $SO(d)$. This can be conveniently done using the projectors into irreducible $SO(d)$ representations defined in section \ref{sec:SOd_tensors}. Naively, we would now have to build the most general descendant operator in CFT${}_d$ and require it to be a primary. 

Fortunately, our task is much less daunting since the quantum numbers $\Delta,\ell,\Delta',\ell'$ for when a descendant $\mathcal O'_{\Delta',\ell'}$ of a primary $\mathcal O_{\Delta,\ell}$ becomes a primary itself are already classified by representation theory. This imposes a remarkably strong constraint on $\Ocal'$ which often -- in fact every time $\Ocal'$ appears with degeneracy equal to one -- can be used to fix the primary descendant's exact form even before checking that it is a primary.
How to find the set of quantum numbers from representation theory is explained in~\cite{Penedones:2015aga} where many $\Ocal'$ are explicitly constructed.
Here we will review and extend the construction of the different types of descendant operators and when they become primaries.
In even $d$ the independent types of descendant operators are I, II, III (as in CCFT$_2$) while in odd $d$ they are I, II, S, P.\footnote{We use a slightly different nomenclature with respect to~\cite{Penedones:2015aga}:  while the definition of types I, II here match the ones of~\cite{Penedones:2015aga} and~\cite{Pasterski:2021fjn}, in odd $d$ what is called type III in~\cite{Penedones:2015aga} is referred to as type S here (for shadow), type IV in~\cite{Penedones:2015aga} is called type P here (for parity) and in even $d$ type V in~\cite{Penedones:2015aga} is referred to as type III here (in accordance with the nomenclature of \cite{Pasterski:2021fjn} in $d=2$).}
Our main focus will then be on those types that arise from protected CCFT$_d$ operators associated to soft theorems.

The most important  primary descendants, associated to universal soft theorems, correspond to type I and II.
In contrast to CCFT$_2$, we need a more refined definition of type I and II operators for CCFT$_{d>2}$.
A level $\kcal$ descendant of type I has spin $\kcal$ units larger than its parent primary, while a type II descendant at level $\kcal$ has spin $\kcal$ units smaller than its parent primary. 
As discussed in section~\ref{sec:CelestialAmplitudes}, in general dimensions $d$ the representations of $SO(d)$ of the spin of the operator can be more complicated than the traceless and symmetric one and a primary operator $\Ocal_{\Delta,\ell}$ is labeled by a set of $[d/2]$ numbers $\ell=(\ell_1,\dots,\ell_{[d/2]})$ where $\ell_i \geq \ell_{i+1}$ in addition to its conformal dimension $\Delta$. The types I and II are thus refined by a label $k$ corresponding to the $\ell_k$ that is increased or decreased. 
In the following we construct these type $\I_k$ and $\II_k$ descendant operators and review the conditions for them to become primaries. 
They will arise in sections~\ref{sec:CCFTdCWardId} and ~\ref{sec:CCFTdSHCWardId} when we discuss the universal soft theorems and their shadow transforms.

In this context there is another relevant primary descendant which we refer to as type $\S$. Interestingly, we will show in section~\ref{sec:CCFTdSHCWardId} that (an analytic continuation of) the type $\S$ operator can be used to compute shadow transforms of primaries which themselves have respectively type I and type II primary descendants. 

In odd $d$ there exists another type of descendant operator which we refer to as type P. For our purpose, the only relevant type P operators will arise for $d=3$, but we will discuss it for completeness.
A discussion on the role of charges for type I operators responsible for more
subleading soft theorems that receive non-universal contributions is presented in appendix
\ref{app:typeI}.
We leave the discussion of type III operators for future work.

We schematically illustrate in figure~\ref{fig:CelestialNecklace} the structure of a conformal multiplet in CFT$_d$.
The conformal dimension  grows downwards but the rest of the diagram is schematic since an accurate representation would require a $[d/2]+1$ dimensional picture labelled  by the variables $\Delta,\ell_1,\dots,\ell_{[d/2]}$.
Not all cases are presented:
The type III descendants in even $d$ are not depicted, and neither is the type S operator in odd $d$ which would be represented by a separate diagram with just a single arrow -- since it is the only one that relates operators with half integer conformal dimensions it cannot have nested primary descendants.
\tikzset{->-/.style={decoration={
  markings,
  mark=at position #1 with {\arrow{>}}},postaction={decorate}}}
  
\begin{figure}[t]
\centering
\subfloat[]{
\begin{tikzpicture}
\foreach \i in{3,4}{
\draw[thick,->-=0.5]
(0,\i+1)--(0,\i+0.1);
\draw[thick] (0,\i+1.05) circle (1.5pt);
\draw[thick] (0,\i+0.05) circle (1.5pt);
}
\draw[thick,dashed] (0,3)--(0,2.1);
\draw[thick,->-=0.5] (0.05,2.05)--(1,1.1);
\draw[thick,->-=0.5] (-0.05,2.05)--(-1,1.1);
\draw[thick,->-=0.5] (-1,1)--(-0.05,0.05);
\draw[thick,->-=0.5] (1,1)--(0.05,0.05);
\draw[thick,dashed] (0,0)--(0,-0.9);
\draw[thick] (0,2.05) circle (1.5pt);
\draw[thick] (1,1.05) circle (1.5pt);
\draw[thick] (-1,1.05) circle (1.5pt);
\draw[thick] (0,0.05) circle (1.5pt);
\foreach \i in{-2,-3}{
\draw[thick,->-=0.5]
(0,\i+1)--(0,\i+0.1);
\draw[thick] (0,\i+1.05) circle (1.5pt);
\draw[thick] (0,\i+0.05) circle (1.5pt);
}
\node[anchor=west] at (0.25,4.5)  {I$_1$};
\node[anchor=west]  at (0.25,3.5) {I$_2$};
\node  at (1,1.8) {I$_{\left[\frac{d}{2}\right]}$};
\node  at (1.2,0.15) {II$_{\left[\frac{d}{2}\right]}$};
\node[anchor=west]  at (0.25,-1.5) {II$_2$};
\node[anchor=west]  at (0.25,-2.5) {II$_1$};
\end{tikzpicture}
}
$\qquad \qquad \qquad \qquad$
\subfloat[]{
\begin{tikzpicture}
\foreach \i in{3}{
\draw[thick,->-=0.5]
(0,\i+1)--(0,\i+0.1);
\draw[thick] (0,\i+1.05) circle (1.5pt);
\draw[thick] (0,\i+0.05) circle (1.5pt);
}
\draw[thick,dashed] (0,3)--(0,2.1);
\foreach \i in{1,...,-1}{
\draw[thick,,->-=0.5]
(0,\i+1)--(0,\i+0.1);
\draw[thick,dashed] (0,-1)--(0,-1.9);
\draw[thick] (0,\i+1.05) circle (1.5pt);
\draw[thick] (0,\i+0.05) circle (1.5pt);
}
\foreach \i in{-3}{
\draw[thick,,->-=0.5]
(0,\i+1)--(0,\i+0.1);
\draw[thick] (0,\i+1.05) circle (1.5pt);
\draw[thick] (0,\i+0.05) circle (1.5pt);
}
\node[anchor=west] at (0.25,3.5) {I$_1$};
\node[anchor=west] at (0.25,1.5) {I$_{\left[\frac{d}{2}\right]}$};
\node[anchor=west] at (0.25,0.5) {P};
\node[anchor=west] at (0.25,-0.5) {II$_{\left[\frac{d}{2}\right]}$};
\node[anchor=west] at (0.25,-2.5) {II$_1$};
\end{tikzpicture}
}
\caption{Celestial necklaces in even dimensions (a) and odd dimensions (b).}
\label{fig:CelestialNecklace}
\end{figure}
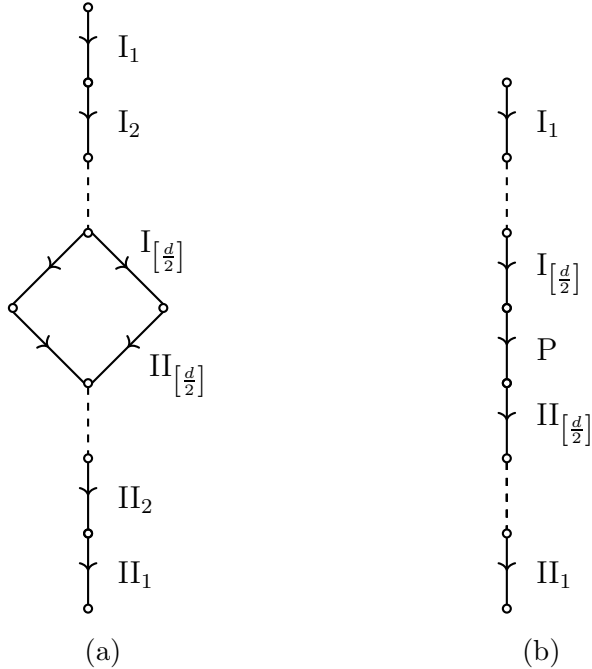

For the classification we will make use of operators contracted with polarization vectors $\Ocal_{\Delta,\ell}(x,\ecal)$ 
where we use the notation   $\ecal\equiv \{ \ecal_1, \dots, \ecal_{[d/2]} \}$ as in section \ref{sec:polarization_vectors}. It is important to stress that these polarization vectors are just a book-keeping device to simplify equations and computations with tensor indices, making all expressions scalar. The logic is that the vectors $\ecal_i$ are contracted with the indices in the  $i$-th row of the Young tableau. The operator $\Ocal_{\Delta,\ell}(x,\ecal)$ is thus a homogeneous function of degree $\ell_i$ in the variable $\ecal_i$. In order to increase the spin $\ell_i$ by one unit we need to introduce one extra vector  $\ecal_i$. Conversely, to reduce the spin $\ell_i$ of the operator by one unit, we need to take one derivative in $\ecal_i$. Finally, in order to opportunely symmetrize the indices we will make use of the projectors of section \ref{sec:polarization_vectors}.

\subsubsection{Type I Operators}
\label{sec:TypeI}
Type I operators have larger spin compared to the parent primary they descend from. In particular given a primary $\Ocal_{\Delta,\ell}$ with dimension $\Delta$ and spin $ \ell=(\ell_1, \dots, \ell_{[d/2]})$, its type $\I_k$ descendant of level~$\kcal$ is defined by the unique operator with spin $\ell'$ such that $\ell'_i=\ell_i+\kcal \delta_{i k}$, namely with the $k$-th spin label increased by $\kcal$ units. Let us consider the primary  $\Ocal_{\Delta,\ell}(x,\fcal)$ contracted with constrained polarization vectors $\fcal$. In order to obtain its $\I_k,\kcal$ descendant  we need to introduce $\kcal$ new vectors $\fcal^a_k$ and $\kcal$ new derivatives $\p^a_x$. Since $\fcal_k^2=0$, there is a single way to do so, 
namely\footnote{This formula generalizes the one of \cite{Penedones:2015aga} to any $k$.
Some examples of $k=2$ primary descendants were already defined in appendix B of \cite{Costa:2016xah}.} 
\begin{equation}\label{O_typeI}
    O_{{\I_k},\kcal}(x,\ecal)\equiv 
    \left(\prod_{i=1}^{[d/2]} \frac{1}{\ell'_i!} \right) \pi_{\ell'}(\ecal;\partial_{\fcal})
    \left[(\fcal_k\cdot \partial_x)^\kcal
    \Ocal_{\Delta,\ell}(x,\fcal)\right]\, .
\end{equation}
Here we use the notation introduced in section \ref{sec:polarization_vectors}, namely    $\ecal\equiv \{ \ecal_1, \dots, \ecal_{[d/2]} \}$, $\fcal\equiv \{ \fcal_1, \dots, \fcal_{[d/2]} \}$ and $\partial_{\fcal}\equiv \{ \partial_{\fcal_1}, \dots, \partial_{\fcal_{[d/2]}}  \}$.
The meaning of this formula is simple: the term $(\fcal_k\cdot \partial_x)^\kcal$ increases the label $\ell_k$ by $\kcal$ units while the projector is only used to properly symmetrize the indices according to the prescription in \eqref{proj_to_indices} (here we used constrained vectors $\ecal$, but the result would have been the same with unconstrained $\hat \ecal$).
Notice that all the variables $\fcal_i$ (the ones contracted with the operator and the $\kcal$ extra vectors $\fcal_k$) are being removed by derivatives coming from the projector and the final expression is independent of $\fcal_i$.
By construction $1\leq \kcal \leq \ell_{k-1}-\ell_{k}$ (the label $\ell_0$ here is defined to be infinite), otherwise the descendant would be labelled by a Young tableau with row $\ell_{k-1}$ shorter that $\ell_k$, which is not allowed.
Type $\I_k$ descendants become primaries when
\begin{equation}
\label{Delta_Ik}
\Delta= \Delta^*_{\I_k, \kcal}\equiv k-\ell_k-\kcal\,  .
\end{equation}
In the following we discuss examples of type $\I_k$  operators where the parent primary $\Ocal_{\Delta, \ell}$ is taken to be traceless and symmetric that will be relevant in section~\ref{sec:CCFTdCWardId}.

\paragraph{Example: Type $\I_{1}$ Operators}
The simplest case corresponds to $k=1$. The type~$\I_{1}$ descendant operators at level~$\kcal$ take the form 
\begin{equation}
    O_{\I_{k=1},\kcal}(x,\ecal)= (\ecal \cdot \partial_x)^\kcal \Ocal_{\Delta, \ell}(x,\ecal)\,,
\end{equation}
where $\ecal$ here is a single polarization vector. Equivalently, in index notation
\begin{equation}
O^{a_1 \dots a_{\ell+\kcal}}_{\I_{k=1},\kcal}(x) =  \partial_x^{\{ a_1} \partial_x^{a_2} \dots \partial_x^{a_\kcal}   \Ocal_{\Delta, \ell}^{a_{\kcal+1} \dots a_{l+\kcal}\}}(x) \, ,
\end{equation}
where the brackets implement symmetrization and subtraction of the traces.
Using \eqref{Delta_Ik} we find that these operators are primaries when $ \Delta= 1-\ell_1-\kcal $.
For a scalar operator $\Ocal_{\Delta}$ (labeled by $\ell=(0,\dots,0)$), its type~$\I_{1}$ descendant at level ${\mathcal n}=1$ takes the simple form $O_{\I_1,1}^a= \partial^{a}\Ocal_{\Delta}$.
It becomes primary when the parent operator $\Ocal_{\Delta}$ has dimension $\Delta=0$, which is the dimension of the identity operator. Indeed $\partial^{a}\Ocal =0$ when $\Ocal$ is the identity.

\paragraph{Example: Type $\I_{2}$ Operators}
Let us now proceed to $k=2$. The simplest example of type~$\I_{2}$ descendants arises when the parent primary is a vector operator $\Ocal^a_{\Delta}$ (i.e. $\ell_1=1$ and $\ell_i=0$ for  $2\leq i\leq[d/2]$). Since $1\leq \kcal\leq \ell_1-\ell_2=1$, this 
operator has a single type~$\I_{k=2}$ primary descendant at level $\kcal=1$
 which transforms in the $(1,1)$ representation, i.e. it is an antisymmetric tensor with two indices. This descendant in index notation takes the simple form $O_{\I_{2},1}^{ab}=\p^{[a}_{\phantom{1}} \Ocal^{b]}_{\Delta}$ and becomes primary when $\Delta=1$.
Another simple example which will play a role in the following is the type~$\I_{k=2}$ primary descendant of a spin two tensor operator $\Ocal^{a_1 a_2}_{\Delta}$ (with $\ell_1=2$ and $\ell_i=0$ for  $2\leq i \leq [d/2]$). This  has two possible type~$\I_{k=2}$ descendants at level $\kcal=1,2$ which respectively increase by one or two units the label $\ell_2$. Restoring the indices (using the projectors \eqref{pi_proj} which are explicitly defined in appendix \ref{app:projectors}) these descendants can be written as
\be
\label{I21_I22}
O^{a^{1}_1 a^{1}_2, a^{2}_1}_{\I_2,\kcal=1}(x)= \pi_{(2,1)}({\bf a},{\bf b}) \p^{b^{2}_1} \Ocal^{b^1_1 b^1_2}_\Delta(x) \, ,
\qquad
O^{a^{1}_1 a^{1}_2, a^{2}_1 a^{2}_2}_{\I_2,\kcal=2}(x)= \pi_{(2,2)}({\bf a},{\bf b}) \p^{b^{2}_1} \p^{b^{2}_2} \Ocal^{b^1_1 b^1_2}_\Delta(x)
\, ,
\ee
and become primaries respectively when $\Delta=1$ and  $\Delta=0$.

In general the level $\kcal$ type $\I_2$ descendants of symmetric and traceless operators (i.e. with $\ell_2=0$) become primary when 
\begin{equation}
\Delta= 2-\kcal \, , \qquad \kcal=1,\dots,\ell_1.
\label{DeltaI2l}
\end{equation}
Operators of type $\I_k$ for  $k=1$ or $k=2$ can appear as descendants of traceless and symmetric primaries. Conversely the cases of $k>2$ can only appear if the parent primary is not traceless and symmetric (due to the condition $1\leq \kcal \leq \ell_{k-1}-\ell_{k}$). We do not consider the latter since  usual Lagrangian fields (such as photons and gravitons) transform in traceless and symmetric representations and thus define primaries $\Ocal_{\Delta,\ell }$ in symmetric and traceless representations. 

\subsubsection{Type II Operators}
\label{subsec:typeII}
Type II operators have smaller spin compared to the parent primary they descend from. The level $\kcal$ descendant of type $\II_k$ of a primary $\mathcal O_{\Delta, \ell}$ with dimension $\Delta$ and spin $ \ell=(\ell_1, \dots, \ell_{[d/2]})$, is defined by the unique operator with spin label $\ell_k$ decreased by $\kcal$ units with respect to the parent primary, and it only exist for $1\leq \kcal \leq \ell_{k}-\ell_{k+1}$ (to avoid $\ell_{k+1}$ larger than $\ell_k$). In order to build this operator in index-free notation we need to introduce $\kcal$ derivatives $\partial_x^a$ and contract them with $\kcal$ derivatives of the $k$-th polarization vector,\footnote{
In even dimensions, the type $\I_{k=d/2}$ would have two distinct realizations like the types Ia and Ib in 
CFT$_2$. The formula~\eqref{typeII_k} gives the analogue of operator Ia in~\eqref{typeIa}.
In this paper the type $\I_{k=d/2}$ operators are not needed, so we do not present the formula for the  analogue of the type Ib operators.
}
\begin{equation}
\label{typeII_k}
     O_{\II_{k},\kcal}(x,\hat \ecal)\equiv  \frac{(\ell_k-\kcal)!}{\ell_k!}    (\partial_{\hat \ecal_k} \cdot \partial_x)^\kcal  \Ocal_{\Delta, \ell}(x,\hat \ecal) \, ,
\end{equation}
where we again make use of the notation $\hat \ecal \equiv \{ \hat \ecal_1, \dots, \hat \ecal_{[d/2]} \}$ introduced in section~\ref{sec:polarization_vectors}.
Here for clarity of the formula, we consider $\Ocal_{\Delta,\ell}$ to be contracted with unconstrained vectors $\hat \ecal_i$, so that simple derivatives in $\hat \ecal_i$ open its indices.\footnote{If we start with an operator contacted with constrained vectors $\fcal_i$, we can recover the unconstrained ones $\hat \ecal_i$ by using a projector as follows
$\Ocal_{\Delta, \ell}(x,\hat \ecal) \equiv  \left(\prod_{i=1}^{[d/2]} \frac{1}{\ell_i!}\right) \pi_{\ell}(\hat \ecal;\partial_{\fcal})\Ocal_{\Delta, \ell}(x,\fcal)$.
After taking the derivatives in $\hat \ecal_i$ in \eqref{typeII_k} we can get back to constrained vectors $\hat \ecal_i \to  \ecal_i$ with $\ecal_i\cdot \ecal_j =0$.}
Type $\II_k$ descendants become primaries when
\be
\label{polesII}
\Delta=\Delta^*_{\II_k, \kcal}\equiv d+\ell_k-k-\kcal\,.
\ee
In the following we discuss examples of type $\II_k$ operators where the parent primary $\Ocal_{\Delta, \ell}$ is taken to be traceless and symmetric that will be relevant in section~\ref{sec:CCFTdSHCWardId}.

\paragraph{Example: Type $\II_{1}$ Operators}
In the simplest case of $k=1$, descendant operators of type $\II_1$ take the form\footnote{
This expression is equivalent to~\eqref{typeII_k} for $k=1$ upon using~\eqref{Tod}. The form~\eqref{II_1_def} is often more convenient in explicit computations since working with constrained vectors $\ecal$ allows one to drop all terms proportional to~$\ecal^2$.
}
\begin{equation}
\label{II_1_def}
    O_{\II_{k=1},\kcal}(x, \ecal)= \frac{(D_\ecal \cdot { \partial_x })^\kcal}{(2-\frac{d}{2}-\ell)_\kcal(-\ell)_\kcal} \Ocal_{\Delta,\ell}(x,\ecal)\,, 
\end{equation}
where $\ecal$ is a single (constrained) vector and the role of $D^a_\ecal$, defined in~\eqref{Tod}, is to recover the tensor indices that are contracted with those constrained vectors. In index notation we can express them as
\begin{equation}
\label{II_1_def1}
 O^{a_1 \dots a_{\ell-\kcal}}_{\II_{k=1},\kcal}(x) =  \partial_{a_{\ell-\kcal+1}}  \dots \partial_{a_\ell}   \Ocal_{\Delta,\ell}^{a_{1} \dots a_{\ell-\kcal} a_{\ell-\kcal+1}  \dots a_{\ell}}(x) \,.
\end{equation}
For a vector (tensor) operator $\mathcal O^{a_1\dots a_{\ell-\kcal}}_{\Delta,\ell}$ labeled by $\ell=(\ell_1,0,\dots,0)$ with $\ell_1=1$ ($\ell_1=2$), its type $\II_1$ descendant at level $\kcal=1$ takes the simple form $O_{\II_1,1}=\partial_{a} \Ocal^{a}_{\Delta, \ell=1} $ ($O^{a}_{\II_1,1}=\partial_{a} \Ocal^{ab}_{\Delta, \ell=2}  $). It becomes a primary for $\Delta=d-1$ ($\Delta=d$) corresponding to the dimension of a conserved current (stress tensor). The primary descendant implements the conservation of such operators. 

When $\Ocal_{\Delta,\ell }$ is traceless and symmetric 
no possible $\II_{k\geq2}$ primary descendant can be constructed since $1\leq \kcal \leq   \ell_{k+1}-\ell_{k} =0$ for any $k\geq2$.
Therefore in this work we will not consider type $\II_{k\geq2}$ primary descendants and  formula \eqref{II_1_def} (or equivalently \eqref{II_1_def1}) will be sufficient.

\subsubsection{Type S Operators}

Type S primary descendants have the same spin as their parent primaries and become primaries themselves when
\begin{equation}
    \Delta=\Delta^*_{\S,\kcal}\equiv \frac{d}{2}-\kcal\,, 
    \qquad 
    \kcal=1,2,\dots \, .
\end{equation}
For this paper it will suffice to consider traceless symmetric parent primary operators $\Ocal_{\Delta,\ell}$  whose level $\kcal$ descendants of type $\S$ are given by 
\be
O_{{\S},\kcal}(x,\ecal) \equiv D_{\S,\kcal}  \Vcal_1 \cdots \Vcal_{n-1}
\Ocal_{\Delta,\ell}  
\ee
where 
\be  
\label{O_typeS0}
D_{\S,\kcal} \equiv \Vcal_0 \cdot \Vcal_1 \cdots \Vcal_{\mathcal n-1} \, ,
\qquad
\Vcal_j\equiv \p_x^2-2 \frac{ (\p_x\cdot \ecal) (\p_x\cdot D_\ecal)}{ (\frac{d}{2}+\ell+j-1) (\frac{d}{2}+\ell-j-2)}  \ .
\ee
It is easy to see that the quantum numbers of the primary descendant $(\Delta=\frac{d}{2}+\kcal, \ell)$ are shadow-related (namely $\Delta \to d- \Delta$) to the ones of its parent primary $(\Delta=\frac{d}{2}-\kcal, \ell)$.
Indeed primary descendant operators of type S have an interesting relation with the shadow transform in CFT (hence the name S) which we discuss in section~\ref{sec:CCFTdCWardId}.

\subsubsection{Type P Operators}
The type P primary descendant exists only in odd $d$.
It appears when the parent primary has dimension 
\be
\Delta=\Delta^*_{\P,\kcal}\equiv\frac{d+1}{2}-\kcal \, ,
\qquad \kcal=1,\dots,\ell_{[\frac{d}{2}]} \, .
\ee
The type P primary descendant appears at level $2\mathcal n-1$ (thus it has dimension $\Delta'=\frac{d-1}{2}+\kcal$, which is the shadow dimension of the parent operator), has the same spin as its parent and has the opposite parity (thus the name P). Indeed it is built using the $\epsilon$ tensor, for this reason this operator has a strong dependence on~$d$ (e.g. we do not know how to analytically continue it in~$d$) and it is so far known only in $d=3$.

For traceless and symmetric parent primaries $\Ocal_{\Delta ,\ell}$ we have $\ell_{[\frac{d}{2}]}=0$ in all $d>3$, and thus their primary descendants will not play a role in our discussion. However when $d=3$ we have $\ell_{[\frac{d}{2}]}=\ell_1$ possible primary descendants of this type.
The type P operator in this case, defined in formula (128) of \cite{Penedones:2015aga}, takes the form
\be
O_{\P,\kcal}(x,\ecal)  %
\equiv \varepsilon(\ecal,\partial_x,D_\ecal)  \Wcal_0 \cdot  \Wcal_1 \cdots \Wcal_{\kcal-2}  \Ocal_{\Delta, \ell}(x,\ecal)
  \ , \label{stateIV}
\ee
where the label $\kcal$  takes the values
 $\kcal=1\dots \ell$ and we introduced $\varepsilon(\ecal,\partial_x,D_\ecal)\equiv \varepsilon_{abc } 
\ecal^{a}\partial^{b}_x
D_\ecal^{c}$   and 
\be
 \Wcal_j=  \partial_x^2-2\frac{ (\partial_x\cdot \ecal) (\partial_x \cdot D_\ecal)}{(\ell+j+1) (\ell-j-1)} \ .
\ee
In CFT${}_3$ these operators appear when $\Delta=2-\kcal$, which is the exact same condition as for type~$\I_2$!
This is not a coincidence:\footnote{See also the discussion in appendix E.6.4 of \cite{Dymarsky:2017xzb}.} indeed
type~$\I_{k=2}$ operators do not exists in $d=3$ since they are only defined when $k>[d/2]$ and type P operators take their place.
This is easily explained, as in $d=3$ an operator with $\ell_2=1$ can be dualized using the epsilon tensor  and rewritten in term of a parity-odd operator with $\ell_2=0$.
This means that in CFT${}_3$ the type $\I_2$ descendants will be replaced by type P.

The easiest instance of such an operator appears for a parent primary of spin $\ell=1$. In this case the  $\kcal=1$ type P operator is given by
\be
\label{TypePspin1}
O^{a}_{\P,\kcal=1}=\varepsilon^{a}_{\phantom{a} b_1 b_2} \partial_x^{b_1} \Ocal^{b_2}_{\Delta} \, ,
\ee 
and becomes a primary when $\Delta=1$.
It is clear that this operator has the exact same properties as the type $\I_{k=2}$ operator, where the antisymmetrization of the indices $b_i$ here is performed by the contraction with the epsilon tensor.
Similarly for spin $\ell=2$ the $\kcal=1$ type P operator takes the form
\be
\label{TypePspin2lev1}
O^{a_1 a_2 }_{\P,\kcal=1}=\varepsilon^{\{a_1|}_{\phantom{\{a_1|} b_1 b_2} \partial_x^{b_1} \Ocal^{b_2 |a_2 \}}_{\Delta} \, ,
\ee 
where the indices $a_i$ are made symmetric and traceless.
This descendant also becomes a primary when $\Delta=1$. 
Finally we have the $\kcal=2$ example 
\be
\label{TypePspin2lev2}
O^{a_1 a_2 }_{\P,\kcal=2}=\varepsilon^{\{a_1|}_{\phantom{\{a_1|} b_1 b_2} \partial_x^{b_1} ( \partial_x^2 \delta^{|a_2\}}_{ b_3} -\partial_{x\,b_3} \partial_x^{|a_2 \}})
\Ocal^{b_2 b_3 }_{\Delta} \, ,
\ee
which becomes primary at $\Delta=0$.

\subsection{Conserved Charges}
\label{sec:CCFTdCharges}
In section~\ref{sec:TopoCharge} we saw that for a given Noether current $\Jcal_a$ with $\partial^a \Jcal_a = 0$ built from conserved operators we can define a topological charge $Q_{\Sigma}$ by integrating $\Jcal_a$. 
Unlike in CFT$_{2}$ where all conservation equations for the operators take the same form, in CFT$_{d>2}$ there are distinct types that gives rise to different expressions for the Noether currents and charges.
In the following we discuss the different types of charges relevant for higher-dimensional soft theorems and CCFT$_d$.

\subsubsection{Charges for Type II}
Let us consider an operator $\Ocal_{\Delta,\ell}$ with a type $\II_k$ primary descendant at level $\kcal$ (thus we are implicitly setting $\Delta=\Delta^*_{\II_k,\kcal}$) with spin $\ell'$ such that $\ell '_i=\ell _i-\delta_{i k} \kcal$.
From $\Ocal_{\Delta,\ell}$ it is possible to construct a current operator, 
\begin{align}
\label{J_TypeIIk}
      \Jcal^{\epsilon \, a}(x)= 
    \frac{\pi_{\ell'}(\partial_{\ecal};\partial_{\hat \fcal})}{\prod_{r=1}^{[d/2]} (\ell'_r!)^2} \;
    \partial_{\hat \fcal_k}^a
    \sum_{j=0}^{\kcal-1}
    (-1)^j
    (\partial_{\hat \fcal_k}\cdot \partial_x)^{\kcal-1-j}\epsilon(x,\ecal)
    (\partial_{\hat\fcal_k} \cdot \partial_x)^j
    \Ocal_{\Delta,\ell}(x,\hat \fcal)\, ,
\end{align}
where $\ecal$ and $\hat \fcal $ are sets of $[d/2]$ vectors, $\partial_x$ acts on the expression immediately to the right while $\partial_{\hat \fcal}$ is understood to act on everything to the right.
Here $\epsilon$ is a function of $x$ with tensor indices transforming in the  representation $\ell '$ with indices contracted with the vectors $\ecal$.

This form of $\Jcal^{\epsilon}_a(x)$ may look complicated at first but it is rigidly fixed to have the most generic vector operator which can be conserved.
Indeed the divergence of the current takes the following simple form
\begin{align}
\label{dJ_IIk}
  \partial^a \Jcal^\epsilon_a 
  &=
   \frac{\pi_{\ell'}(\partial_{\ecal};\partial_{\hat \fcal})}{\prod_{r=1}^{[d/2]} (\ell'_r!)^2}
    \big[
    (\partial_{\hat \fcal_k}\cdot \partial_x)^{\kcal}\epsilon(x,\ecal)   \Ocal_{\Delta,\ell}(x,\hat \fcal)
+ \epsilon(x,\ecal)
     (- \partial_{\hat \fcal_k}\cdot \partial_x)^{\kcal}\Ocal_{\Delta,\ell}(x,\hat \fcal)
    \big] \, .
\end{align}
This result is due to the cancellation of most terms in the sum on \eqref{J_TypeIIk} because of the sign $(-1)^j$: in fact only the terms with $j=0$ and $j=\kcal-1$ survive.
Moreover, the second term in the square bracket in \eqref{dJ_IIk} vanishes because of the type $\II_k$ primary descendant condition.
Therefore conservation $\partial^a \Jcal^\epsilon_a (x)=0$ implies a condition on the parameter $\epsilon$, which, after some massaging,\footnote{
\label{massage_IIk}
We apply the following transformations. 
First we exchange $\hat \fcal \leftrightarrow \p_{\hat \fcal}$ in the following sense: e.g. for vectors $A^a$ and $B^a$ their scalar product can be written as $A^a B_a= (A_a \p^a_{\hat \fcal}) (B_b \hat \fcal^b )=(B_a \p^a_{\hat \fcal}) (A_b \hat \fcal^b )$. We apply the projector $\pi_{\ell'}$ to $\epsilon$ using \eqref{proj_to_indices}. Finally we write the operator $\Ocal$ contracted with a new projector into the representation $\pi_\ell$.
} takes the form
\be
\label{typeIIk_KV}
 \pi_{\ell}(\ecal;\partial_{\fcal})\left[
    (\fcal_k\cdot \partial_x)^\kcal
    \epsilon(x,\fcal)\right] = 0 \, .
\ee
This condition is the same as that of a descendant of type  $\I_k$ at level $\kcal$.
The subscript of the projector does not match with \eqref{O_typeI}  because  here $\ell'$ is the original spin of the tensor $\epsilon$ while $\ell$ is the spin of the new tensor obtained after acting with the derivatives (while in \eqref{O_typeI} it is the opposite).
Thus to build a current with a type $\II_k$ operator we need a parameter $\epsilon$ that satisfies a type $\I_k$ shortening condition.  
In summary, we found that a current of the form \eqref{J_TypeIIk} is conserved if and only if the parameter satisfies \eqref{typeIIk_KV}. Using this current one can write conserved charges in the standard way as in~\eqref{top_op}.

\paragraph{Example: Type $\II_1$ Noether Current}
Of prime interest in celestial CFT$_{d}$ 
are operators of type $\II_1$ and their associated charges which we will focus on in the following. In this case we spell out the form of the current in index notation.
We consider a symmetric traceless spin~$\ell$ operator $\Ocal$ satisfying $\partial_{a_1} \dots \partial_{a_\kcal} \Ocal^{a_1 \dots a_\ell}=0$ with $\kcal\leq \ell$.
We construct a Noether current by combining it with a spin $\ell-\kcal$ symmetric traceless tensor parameter $\epsilon_{a_1,\dots a_{\ell-\kcal}}(x)$. 
The Noether current \eqref{J_TypeIIk} reduces to the form~\cite{Brust:2016gjy}
\be \label{Jcal_gen}
\begin{array}{lll}
\Jcal^{\epsilon\, a}&=&
\Ocal^{a a_1 \dots a_{\ell-1}} \partial_{a_1} \dots \partial_{a_{{\kcal-1}}} \epsilon_{a_\kcal \dots a_{\ell-1}} 
\\
&&
-\partial_{a_1}  \Ocal^{a a_1 \dots a_{\ell-1}} \partial_{a_2} \dots \partial_{a_{{\kcal-1}}} \epsilon_{a_\kcal \dots a_{\ell-1}}
\\
&&
+\partial_{a_1} \partial_{a_2}  \Ocal^{a a_1 \dots a_{\ell-1}} \partial_{a_3} \dots \partial_{a_{{\kcal-1}}} \epsilon_{a_\kcal \dots a_{\ell-1}}
\\
&&
\vdots
\\
&&
+ (-1)^{\kcal-1} \partial_{a_1} \dots \partial_{a_{\kcal-1}}  \Ocal^{a a_1 \dots a_{\ell-1}} \epsilon_{a_\kcal \dots a_{\ell-1}} \, .
\end{array}
\ee
Conservation of~\eqref{Jcal_gen}, i.e. $\partial^a \Jcal^{\epsilon}_a =0$, requires that $\epsilon$ is a generalized conformal Killing tensor  satisfying
\be
\label{generalizedCKT}
\partial_{\{a_1} \dots  \partial_{a_{\kcal}} \epsilon_{a_{\kcal+1} \dots a_\ell\}}= 0 \, ,
\ee
which is a level $\kcal$ type $\I_1$ primary descendant condition.
The general solution to this equation is given by~\cite{Brust:2016gjy} 
\be\label{sympar}
\epsilon_{a_1 \dots a_{\ell-\kcal}}= q^{\mu_1} \dots q^{\mu_{\ell-1}} \partial_{a_1}q^{\nu_1} \dots  \partial_{a_{\ell-\kcal}}q^{\nu_{\ell-\kcal}} c_{\mu_1 \dots \mu_{\ell-1};  \nu_1 \dots \nu_{\ell-\kcal}} \, ,
\ee
where $q$ is defined in the Poincaré section~\eqref{qPoincare}
 and the constant tensor $c_{\mu_1 \dots \mu_{\ell-1};  \nu_1 \dots \nu_{\ell-\kcal}}$ transforms in the irrep $(\ell-1,\ell-\kcal)$. 
To see that \eqref{sympar} is a solution of \eqref{generalizedCKT}, first notice  that $\p_a \p_b q^{\mu}=\delta_{ab}(1,\vec{0},-1)$, thus the result is proportional to $\delta_{ab}$ which is removed when we consider traceless combinations of the indices $a,b$. This means that every time a derivative $\p_{a_i}$ (with $i=1,\dots,\kcal$) in \eqref{generalizedCKT} acts on a $\p_{a_i}q^{\nu_i}$ (with $i=1,\dots,\ell-\kcal$) in \eqref{sympar} the result is zero. All derivatives  in \eqref{generalizedCKT} must then act on the  $q$'s  without derivatives  in \eqref{sympar}. The
result is a product of $\ell$ equal terms $\p_{b_i}q^{\sigma_i}$ symmetrized in the indices $b_i$, which is contracted with a tensor in the representation $(\ell-1,\ell-\kcal)$. This makes the result vanishing because of the  mixed symmetrization  (indeed  a tensor in the $(\ell-1,\ell-\kcal)$ irrep can be at most contracted with $\ell-1$ symmetrized vectors).  
 The charge obtained by integrating~\eqref{Jcal_gen} is given by
\be\label{TypeII_Charge}
Q^\epsilon_\Sigma=\int_\Sigma dS_a \sum_{i=0}^{\kcal-1} (-1)^i
\partial_{a_1} \dots \partial_{a_i}\Ocal^{a a_1 \dots a_{\ell-1}}(x) 
\partial_{a_{i+1}} \dots \partial_{a_{\kcal-1}}
\epsilon_{a_\kcal \dots a_{\ell-1}}(x)
\,,
\ee
with $\epsilon(x)$ given by~\eqref{sympar}. 
Often it will be more convenient to use Gauss' divergence theorem and compute the charge from the volume integral of the divergence of the current.
Indeed using \eqref{generalizedCKT} the divergence can be simply written as $\p^a \mathcal J^{\epsilon}_a=  (-1)^{\kcal-1} \epsilon_{a_\kcal \dots a_{\ell-1}} \partial_{a_1} \dots \partial_{a_{\kcal}}  \Ocal^{a_1 \dots a_\kcal \dots a_{\ell}}  $ and thus the charge takes the simpler form\footnote{Often we will redefine the charge by multiplying it by an overall coefficient to make its action look nicer
which is equivalent to rescaling the constant coefficients in \eqref{sympar}.}
\be
\label{eq:Q_vol}
Q^{\epsilon}_\Sigma=(-1)^{\kcal-1} \int_{R} d^d x \  \epsilon_{a_{\kcal+1} \dots a_{\ell}}(x) \partial_{a_1} \dots \partial_{a_{\kcal}}  \Ocal^{a_1 \dots a_\kcal \dots a_{\ell}}(x) \, ,
\ee
where $\p R=\Sigma$.
We will see examples of type $\II_1$ charges in section~\ref{sec:CCFTdSHCWardId}.

\subsubsection{Charges for Type I}
\label{sec:Charges_I}
Here we explain a construction for charges associated to type I operators. 

Let us consider a primary $\Ocal_{\Delta,\ell}$ 
with a primary descendant of type $\I_k$ at level $\kcal$. As explained earlier the primary descendant has spin labelled by $\ell'_i=\ell_i+\kcal \delta_{i k}$.
It is formally  possible to build an operator $\Jcal^{\epsilon}_a$ such that $\partial^a \Jcal^{\epsilon}_a=0$ out of $\Ocal_{\Delta,\ell}$.
The construction goes as follows.
First we define an operator $\Jcal^{\epsilon}_a$ by an opportune contraction of derivatives with the primary $\Ocal_{\Delta,\ell}$ and with a parameter function $\epsilon$,
\begin{equation}
      \Jcal^{\epsilon\, a}= 
      \frac{\pi_{\ell'}(\partial_{\ecal};\partial_{\fcal})}{\prod_{i=1}^{[d/2]} (\ell'_i!)^2} \;
    \fcal_k^a
    \sum_{j=0}^{\kcal-1}
    (-1)^j
    (\fcal_k\cdot \partial_x)^{\kcal-1-j}\epsilon(x,\ecal)
    (\fcal_k\cdot \partial_x)^j
    \Ocal_{\Delta,\ell}(x,\fcal)\, .
\end{equation}
Here $\epsilon$ is a function of $x$ transforming in the representation $\ell'$ of $SO(d)$ (like the primary descendant of $\Ocal_{\Delta,\ell}$) and it is contracted with polarization vectors $\ecal=\{\ecal_1,\dots, \ecal_{[d/2]}\}$. 
The divergence of  $\Jcal^{\epsilon}_a$ takes the form
\begin{align}
\label{eq:consJ_typeI}
  \partial^a \Jcal^{\epsilon}_a 
  &=
  \frac{\pi_{\ell'}(\partial_{\ecal};\partial_{\fcal})}{\prod_{i=1}^{[d/2]} (\ell'_i!)^2}
    \big[
    (\fcal_k\cdot \partial_x)^{\kcal}\epsilon(x,\ecal)   \Ocal_{\Delta,\ell}(x,\fcal)
+ \epsilon(x,\ecal)
     (- \fcal_k\cdot \partial_x)^{\kcal}\Ocal_{\Delta,\ell}(x,\fcal)
    \big] \, .
\end{align}
 The second term in the square bracket vanishes because of the primary descendant condition \eqref{O_typeI}. 
We thus find 
that $\Jcal^{\epsilon}_a $ is divergenceless
when\footnote{This equation is obtained from a few manipulations of the first term of \eqref{eq:consJ_typeI}: 1)  replacing $\fcal \to \hat \fcal$, 2)  exchanging $\hat \fcal \leftrightarrow \p_{\hat \fcal}$ as explained in footnote \ref{massage_IIk}, 3)  applying the projector to $\epsilon$ using \eqref{proj_to_indices}.
}
\begin{equation}
    (\partial_{\hat \fcal_k} \cdot \partial_x)^{\kcal}
    \epsilon(x,\hat \fcal) = 0 \, .
\end{equation}
This condition on $\epsilon$ is the same as that of a descendant of type $\II_k$ in \eqref{typeII_k}.
So in practice to construct a current from an operator with a primary descendant of type $\I_k$ we find that the associated parameter should
satisfy a shortening condition of type $\II_k$.

Let us give a couple of examples for this construction in some simple $k=1,2$ cases.

\paragraph{Example: Type $\I_1$ Noether Current}
First we consider a vector operator $\Ocal_{\Delta,\ell=1}$, with a primary descendant of the type $\I_{k=1}$ at $\kcal=1$ defined by  $\p^{\{a} \Ocal^{b\}}=0$. We can use this to build a Noether current 
\begin{equation}
   \Jcal^{\epsilon \, a}(x) = \epsilon^{\{ab\}}(x)\Ocal_b(x) \, .
\end{equation}
Conservation of $\Jcal$ can be written as
\be
0=\p^a \Jcal^{\epsilon}_{  a}(x)=[\p_a \epsilon^{\{ab\}}(x)\Ocal_b(x)+ \epsilon^{\{ab\}}(x)\p_a\Ocal_b(x)] \, ,
\ee
but the second term in the square bracket vanishes because of 
 $\partial^{\{a}\Ocal^{b\}}=0$. Therefore $\p^a \Jcal^{\epsilon}_{  a}(x)=0$ implies that the parameter satisfies $\partial_a \epsilon^{\{ab\}}(x)=0$.

\paragraph{Example: Type $\I_2$ Noether Current}
Let us now turn to a vector operator $\Ocal_{\Delta,\ell=1}$with a type $\I_{k=2}$ primary descendant at level $\kcal=1$, defined by $\p^{[a} \Ocal^{b]}$. The associated Noether current can be written as
\be{}
\Jcal^{\epsilon \, a}= \epsilon^{[ab]}(x) \Ocal_{b}(x)\, .
\ee{}
Using $\p^{[a} \Ocal^{b]}=0$, we find that $ \Jcal^{\epsilon}$ is conserved if $\p_a \epsilon^{[ab]} = 0$.
\\

While the construction of type I charges works in principle, in practice these charges take a trivial form and so they do not represent actual symmetries of the theory. This is due to the fact that the Ward identities for type I operators do not contain contact terms as we explain in appendix~\ref{app:typeI}.

\subsection{CCFT$_{d>2}$ Ward Identities}
\label{sec:CCFTdCWardId}
Conformally soft theorems in CCFT$_{d>2}$ and their associated celestial Ward identities were discussed in~\cite{Kapec:2017gsg,Kapec:2021eug}.\footnote{See also~\cite{Banerjee:2019aoy,Banerjee:2019tam}.} Here we make use of our primary descendant classification to explain systematically the origin of conserved operators. 

\subsubsection{Leading Soft Photon Theorem}

In the Poincaré section, using equation \eqref{prods_poinc}, we can easily express the soft photon factor as
\begin{equation}\label{SoftPhotond}
S^{(0)}_p(\omega, x,{\mathcal e})= -2e \sum_{i=1}^N \mathcal Q_i\frac{1}{\omega} \frac{ (x_i-x) \cdot {\mathcal e}  }{ (x_i-x)^2 }\, .
\ee
The soft theorem~\eqref{softtheorem} with the soft factor~\eqref{SoftPhoton} can be mapped to the conformal basis where the power $1/\omega$ selects the operator $R^a_1(x)$ defined in \eqref{conf_soft_ops}. We thus obtain the following conformally soft theorem
\begin{equation}
\label{R1_WI_0}
    \langle R_{1}(x,\mathcal e) \Ocal_{\Delta_1,\ell_1}(x_1,\mathcal e_1) \dots \Ocal_{\Delta_N,\ell_N}(x_N,\mathcal e_N)  \rangle =
   \sum_{i=1}^N \hat S^{(0)}_i(x,{\mathcal e}) 
    \langle \Ocal_{\Delta_1,\ell_1}(x_1,\mathcal e_1) \dots \Ocal_{\Delta_N,\ell_N}(x_N,\mathcal e_N)  \rangle \, ,
\end{equation}
where $R_1(x,\mathcal e)=R^a_1(x) \mathcal e_a$ and the hatted soft factor takes the form  
\begin{equation}
    \hat S^{(0)}_i(x,{\mathcal e}) \equiv -2e  \mathcal Q_i \frac{ (x_i-x) \cdot {\mathcal e}  }{ (x_i-x)^2 }\,.
\end{equation} 
By acting on both sides of the equation with the operator $\partial_{\ecal} \cdot \partial_x$, we can easily compute how the divergence  of $R^a_1(x)$ behaves when inserted in a correlation function, 
\begin{equation}
\label{R1_WI_bad}
\langle \partial_{x}^a  R_{1\, a}(x) \Ocal_{\Delta_1,\ell_1}(x_1,\mathcal e_1) \dots \Ocal_{\Delta_n,\ell_N}(x_N,\mathcal e_N)  \rangle=2e\sum_{i=1}^N \mathcal{Q}_i\frac{d-2}{(x_i-x)^2} \langle  \Ocal_{\Delta_1,\ell_1}(x_1,\mathcal e_1) \dots \Ocal_{\Delta_N,\ell_N}(x_N,\mathcal e_N)  \rangle\, .
\end{equation}
For $d \neq 2$ the dependence on $x$ of the right-hand-side of \eqref{R1_WI_bad} is powerlaw, which is different  from usual Ward identities where $x$ only appears in the argument of a  delta function. 
Thus from \eqref{R1_WI_bad} it is clear that $R^a_1(x)$ is not conserved for $d\neq 2$. 
However, from the classification of section~\ref{sec:CelestialNecklaces} we notice that $R^a_1(x)$ has a type $\I_2$ primary descendant at level $\mathcal n=1$. This descendant has dimensions $\Delta=2$, spin $(1,1)$ (namely has two  antisymmetric  indices) and takes the form defined in section \ref{sec:TypeI},\footnote{See also equation (3.1) of \cite{Banerjee:2019tam}.}
\be
\label{OI21}
 \Ocal_{\I_{2},1}^{a,b}(x) = \frac{1}{2}(\partial^{a}R_1^b(x)-\partial^{b}R^a_1(x)) \, ,
\ee
and it is a primary according to \eqref{DeltaI2l} since the dimension of $R_1$ is equal to one. 
 Using formula \eqref{R1_WI_0} we can  compute the insertion of
 \eqref{OI21}
 in a correlation function. The result is exactly zero and it is also easy to see that  no contact terms arise from this procedure\footnote{Indeed the type I$_2$ operator annihilates  the soft factor even when we introduce a regulator $\epsilon$,
\be{}
   \partial^{[b}   \frac{ (x_i-x)^{a]}  }{ (x_i-x)^2+\epsilon^2 } =    \frac{ (x_i-x)^{[a}  (x_i-x)^{b]} - ((x_i-x)^2+\epsilon^2) \delta^{[a b] }}{ (x_i-x)^2+\epsilon^2 }= 0 \,  .
\ee{}} (as noticed also in  \cite{Kapec:2017gsg}).
This means that we cannot write non-trivial charges for this operator -- see appendix~\ref{app:typeI} for details.
Notice that this primary can be defined in any dimension $d\geq 4$. In $d=3$ this should be replaced by the type P operator defined in \eqref{TypePspin1}.

\subsubsection{Leading Soft Graviton Theorem}
For gravity the soft factor expressed in the Poincaré section is
\begin{equation}\label{SoftGravitond}
S^{(0)}_p(\omega,x,\ecal)=-\kappa\sum_{i=1}^N\eta_i\frac{\omega_i}{\omega}\frac{\left((x_i-x)\cdot\mathcal{e}\right)^2}{(x_i-x)^2}\,.
\end{equation}
In the conformal basis the leading soft graviton theorem becomes
\begin{equation}\label{H1_WI_0}
    \langle H_1(x,\ecal) \Ocal_{\Delta_1,\ell_1} \dots \Ocal_{\Delta_N,\ell_N}  \rangle=\sum_{i=1}^N \hat S^{(0)}_i(x,{\ecal}) \langle  \Ocal_{\Delta_1,\ell_1} \dots \Ocal_{\Delta_i+1,\ell_i}\dots \Ocal_{\Delta_N,\ell_N}  \rangle\, ,
\end{equation}
where $H_1(x,\ecal)=H^{ab}_1(x) \ecal_a \ecal_b$, the hatted soft factor takes the form 
\begin{equation}
    \hat S^{(0)}_i(x,{\ecal}) \equiv -\kappa\eta_i \frac{\left((x_i-x)\cdot\mathcal{e}\right)^2}{(x_i-x)^2}\,,
\end{equation}
and the shift in $\Delta_i$ of the $i$th operator is due to the factor of $\omega_i$ in~\eqref{SoftGravitond}.
From the divergence 
\begin{equation}\label{H1_WI_1}
     \langle \p^b_x\partial^a_x H_{1\,ab}(x) \Ocal_{\Delta_1,\ell_1} \dots \Ocal_{\Delta_N,\ell_N}  \rangle=-\kappa\sum_{i=1}^N \eta_i \frac{(d-1)(d-2)}{(x_i-x)^2} \langle  \Ocal_{\Delta_1,\ell_1} \dots \Ocal_{\Delta_i+1,\ell_i}\dots \Ocal_{\Delta_N,\ell_N}  \rangle\, ,
\end{equation}
we see that $H_{1}^{ab}(x)$ is not a conserved operator in $d>2$.

As in the soft photon case, one can nevertheless use the classification of section \ref{sec:TypeI} to show that $H_{1}^{ab}(x)$ has a primary descendant. It is of type $\I_2$ and level $\kcal=1$ with dimensions $\Delta=2$ and spin $(2,1)$  and takes the form
$\pi_{(2,1)}({\bf a},{\bf b}) \p^{b^{2}_1} H^{b^1_1 b^1_2}_1(x) $
as defined in equation \eqref{I21_I22}.
By substituting the expression of the projector written in \eqref{pi21_indices} the primary descendant can be written as follows
 \begin{align}
O^{a^{1}_1 a^{1}_2, a^{2}_1}_{\I_2,\kcal=1}(x)
=&\frac{1}{3}(2 \p^{a_1^2} H_1^{a_1^1,a_2^1}- \p^{a_2^1} H_1^{a_1^1,a_1^2}- \p^{a_1^1} H_1^{a_1^2,a_2^1})
\nonumber
\\
&
+\frac{1}{3 (d-1)}(2 \delta ^{a_1^1,a_2^1} \p_c H_1^{c a_1^2}-\delta ^{a_1^1,a_1^2} \p_c H_1^{c a_2^1}-\delta ^{a_1^2,a_2^1} \p_c H_1^{ca_1^1}) \, .
\end{align}
Inserting this operator in correlation functions gives again zero without contact terms,
so $H_1$ does not give rise to non-trivial charges (see appendix \ref{app:typeI}).
This type of primary descendant operator was considered in equation (3.6)  of~\cite{Banerjee:2019tam} but there the full symmetrization of the indices was not performed, thus the resulting operator was not transforming in an irreducible representation of $SO(d)$.
Again this primary descendant is only defined for $d\geq4$, while in $d=3$ this should  be replaced  by the type P in~\eqref{TypePspin2lev1}.

\subsubsection{Subleading Soft Graviton Theorem}
The subleading soft graviton factor can be expressed in the Poincaré section as
{\small
\begin{equation}\label{SubSoftGravitond}
\begin{aligned}
S^{(1)}_p(\omega,x,\ecal)=-\kappa \sum_{i=1}^N\frac{(x_i-x)\cdot \mathcal e}{(x_i-x)^2}&\bigg[\frac{(x_i-x)^2}{2}\mathcal e \cdot \p_{x_i} +(x_i-x)\cdot \mathcal e \big(-(x_i-x)\cdot \p_{x_i} +\omega_i \p_{\omega_i}\big)\\
&+i(x_i-x)\cdot M_i\cdot \mathcal e\bigg]\, ,
\end{aligned}
\end{equation}}
where we denote $\p_{x_i}\equiv \frac{\p}{\p {x_i}}$, $\p_{\omega_i}\equiv\frac{\p}{\p \omega_i}$ and $M^{ab}_i$ is antisymmetric in its indices and implements rotations of the tensor indices of the $i$-th operator.  The subleading conformally soft graviton theorem is given by
\begin{equation}\label{H0_WI_0}
    \langle H_0(x,\ecal) \Ocal_{\Delta_1,\ell_1} \dots \Ocal_{\Delta_N,\ell_N}  \rangle=\sum_{i=1}^N \hat S^{(1)}_i(x,{\ecal})  \langle  \Ocal_{\Delta_1,\ell_1} \dots \Ocal_{\Delta_i,\ell_i}\dots \Ocal_{\Delta_N,\ell_N}  \rangle\, ,
\end{equation}
where $H_0(x,\ecal)=H^{ab}_0(x) \ecal_a \ecal_b$ and since~\eqref{SubSoftGravitond} is already of $O(\omega^0)$ the hatted soft factor takes the same form
\begin{equation}
    \hat S^{(1)}_i(x,{\ecal}) = S^{(1)}_i(\omega,x,{\ecal})\,.
\end{equation}

Unlike in $d=2$ where insertions of $H_0$ with $\ell=-2$ ($\ell=+2$) acted on by three (anti) holomorphic derivatives can be shown to vanish up to contact terms, in $d>2$ we cannot act with more than $\ell$ derivatives $\partial_x$ while  decreasing the spin of the operator. Instead, in order to identify the conserved operators we make use of the primary descendant classification of section~\ref{sec:CelestialNecklaces}.

We find that $H_{0}^{ab}(x)$ has a primary descendant of type $\I_2$ and level $\kcal=2$ with dimensions $2$ and spin $(2,2)$ which takes the form $\pi_{(2,2)}({\bf a},{\bf b}) \p^{b^{2}_1} \p^{b^{2}_2} H^{b^1_1 b^1_2}_0(x)$ defined in \eqref{I21_I22}.
Substituting the projector of appendix \ref{app:projectors}  the operator can be written as\footnote{Again this type of operator was considered in equations (3.35) of \cite{Banerjee:2019tam} but there the full symmetrization of the indices was not performed.}
\begin{align}
&O^{a^{1}_1 a^{1}_2, a^{2}_1 a^{2}_2}_{\I_2,\kcal=2}(x)= \frac{1}{6}\bigg\{
-2  \left(\frac{ \p^2 \delta ^{a_1^2,a_2^2}}{d-2}- \p^{a_1^2} \p^{a_2^2}\right) H_0^{a_1^1\,a_2^1}
-2  \left(\frac{ \p^2 \delta ^{a_1^1,a_2^1}}{d-2} -\p^{a_1^1} \p^{a_2^1}\right) H_0^{a_1^2\,a_2^2}
 \\
&
\quad
+\left(\frac{\p^2 \delta ^{a_1^1, a_2^2}}{d-2}-\p^{a_1^1} \p^{a_2^2}\right)  H_0^{a_1^2\,a_2^1} 
+  \left(\frac{\p^2 \delta ^{a_1^2,a_2^1}}{d-2}-\p^{a_1^2} \p^{a_2^1}\right) H_0^{a_1^1\,a_2^2}
\nonumber \\
&
\quad
+\left(\frac{\p^2 \delta ^{a_1^1,a_1^2}}{d-2}-\p^{a_1^1} \p^{a_1^2}\right)  H_0^{a_2^1\,a_2^2} 
+  \left(\frac{\p^2 \delta ^{a_2^1,a_2^2}}{d-2}-\p^{a_2^1} \p^{a_2^2}\right) H_0^{a_1^1\,a_1^2}
\nonumber \\
&
\quad
-\left(\p^{a_2^1} \delta ^{a_1^1,a_1^2}+\p^{a_1^1} \delta ^{a_1^2,a_2^1}-2 \p^{a_1^2} \delta ^{a_1^1,a_2^1}\right) \frac{\p_c  H_0^{c \, a_2^2}}{d-2}
+
\left(\p^{a_2^2} \delta ^{a_1^1,a_1^2}+\p^{a_1^2} \delta ^{a_1^1,a_2^2}-2 \p^{a_1^1} \delta ^{a_1^2,a_2^2}\right)\frac{ \p_c  H_0^{c \, a_2^1}}{d-2}
\nonumber \\
&
\quad
-\left(\p^{a_2^2} \delta ^{a_1^2,a_2^1}+\p^{a_1^2} \delta ^{a_2^1,a_2^2}-2 \p^{a_2^1} \delta ^{a_1^2,a_2^2}\right) \frac{\p_c  H_0^{c\, a_1^1}}{d-2}
-\left(\p^{a_2^1} \delta ^{a_1^1,a_2^2}+\p^{a_1^1} \delta ^{a_2^1,a_2^2}-2 \p^{a_2^2} \delta ^{a_1^1,a_2^1} \right) \frac{\p_c  H_0^{c\, a_1^2}}{d-2}
\nonumber \\
&
\quad
+2  \left(\delta ^{a_1^1,a_2^2} \delta ^{a_1^2,a_2^1} -2 \delta ^{a_1^1,a_2^1} \delta ^{a_1^2,a_2^2}+\delta ^{a_1^1,a_1^2} \delta ^{a_2^1,a_2^2}\right)
\frac{\p_b \p_c   H_0^{b \,c}}
{(d-2) (d-1)}
\bigg\} \, .
\nonumber 
\end{align}
This operator is exactly zero in correlation functions and thus does not yield non-trivial charges.
As in the previous cases, this primary descendant operator exactly vanishes in correlation functions.
This operator exists in $d\geq4$, while in $d=3$ it should  be  replaced  by    \eqref{TypePspin2lev2}.

\subsection{CCFT$_{d>2}$ Shadow Ward Identities}
\label{sec:CCFTdSHCWardId}

In the previous section we showed that the conformally soft operators that arise from recasting soft theorems in a boost eigenbasis are operators of type $\I_2$. These have shortening conditions which do not give rise to contact terms and thus we cannot build non-trivial charges from them.
Moreover type $\I_2$ operators cannot define CFT$_d$ currents or stress tensors which are of type $\II_1$. 
The latter however can be obtained from the shadow transformed Ward identities. 
Indeed the quantum labels of type $\I_2$ operators is related to  the ones of type $\II_1$ by shadow transform (see figure \ref{fig:CelestialNecklaceShadow}).
\begin{figure}[H]
\centering
\subfloat[]{
\begin{tikzpicture}
\foreach \i in{3,4}{
\draw[thick,->-=0.5]
(0,\i+1)--(0,\i+0.1);
\draw[thick] (0,\i+1.05) circle (1.5pt);
\draw[thick] (0,\i+0.05) circle (1.5pt);
}
\draw[thick,dashed] (0,3)--(0,2.1);
\draw[thick,->-=0.5] (0.05,2.05)--(1,1.1);
\draw[thick,->-=0.5] (-0.05,2.05)--(-1,1.1);
\draw[thick,->-=0.5] (-1,1)--(-0.05,0.05);
\draw[thick,->-=0.5] (1,1)--(0.05,0.05);
\draw[thick,dashed] (0,0)--(0,-0.9);
\draw[thick] (0,2.05) circle (1.5pt);
\draw[thick] (1,1.05) circle (1.5pt);
\draw[thick] (-1,1.05) circle (1.5pt);
\draw[thick] (0,0.05) circle (1.5pt);
\foreach \i in{-2,-3}{
\draw[thick,->-=0.5]
(0,\i+1)--(0,\i+0.1);
\draw[thick] (0,\i+1.05) circle (1.5pt);
\draw[thick] (0,\i+0.05) circle (1.5pt);
}
\filldraw[thick,violet!90!black] (0,-3+0.05) circle (1.5pt);
\filldraw[thick,purple] (0,-2+0.05) circle (1.5pt);
\filldraw[thick,red!70!orange] (0,-1+0.05) circle (1.5pt);
\filldraw[thick,orange!90!lightgray] (0,0+0.05) circle (1.5pt);
\filldraw[thick,violet!90!black] (0,5+0.05) circle (1.5pt);
\filldraw[thick,purple] (0,4+0.05) circle (1.5pt);
\filldraw[thick,red!70!orange] (0,3+0.05) circle (1.5pt);
\filldraw[thick,orange!90!lightgray] (0,2+0.05) circle (1.5pt);
\draw[gray,thick,->-=0.5] (-0.2,4+0.05) to[out=-145,in=145,distance=2.5cm] (-0.2,-2+0.05);
\node[anchor=west] at (0.25,4.5)  {I$_1$};
\node[anchor=west]  at (0.25,3.5) {I$_2$};
\node  at (1,1.8) {I$_{\left[\frac{d}{2}\right]}$};
\node  at (1.2,0.15) {II$_{\left[\frac{d}{2}\right]}$};
\node[anchor=west]  at (0.25,-1.5) {II$_2$};
\node[anchor=west]  at (0.25,-2.5) {II$_1$};
\end{tikzpicture}
}
$\qquad \qquad \qquad \qquad$
\subfloat[]{
\begin{tikzpicture}
\foreach \i in{3}{
\draw[thick,->-=0.5]
(0,\i+1)--(0,\i+0.1);
\draw[thick] (0,\i+1.05) circle (1.5pt);
\draw[thick] (0,\i+0.05) circle (1.5pt);
}
\draw[thick,dashed] (0,3)--(0,2.1);
\foreach \i in{1,...,-1}{
\draw[thick,,->-=0.5]
(0,\i+1)--(0,\i+0.1);
\draw[thick,dashed] (0,-1)--(0,-1.9);
\draw[thick] (0,\i+1.05) circle (1.5pt);
\draw[thick] (0,\i+0.05) circle (1.5pt);
}
\foreach \i in{-3}{
\draw[thick,,->-=0.5]
(0,\i+1)--(0,\i+0.1);
\draw[thick] (0,\i+1.05) circle (1.5pt);
\draw[thick] (0,\i+0.05) circle (1.5pt);
}
\filldraw[black!65!teal,thick] (0,-3+0.05) circle (1.5pt);
\filldraw[black!65!teal,thick] (0,4+0.05) circle (1.5pt);
\filldraw[blue!45!teal,thick] (0,-2+0.05) circle (1.5pt);
\filldraw[blue!45!teal,thick] (0,3+0.05) circle (1.5pt);
\filldraw[teal,thick] (0,-1+0.05) circle (1.5pt);
\filldraw[teal,thick] (0,2+0.05) circle (1.5pt);
\filldraw[cyan!65!teal,thick] (0,0+0.05) circle (1.5pt);
\filldraw[cyan!65!teal,thick] (0,1+0.05) circle (1.5pt);
\draw[gray,thick,->-=0.5] (-0.2,3+0.05) to[out=-145,in=145,distance=2cm] (-0.2,-2+0.05);
\node[anchor=west] at (0.25,3.5) {I$_1$};
\node[anchor=west] at (0.25,1.5) {I$_{\left[\frac{d}{2}\right]}$};
\node[anchor=west] at (0.25,0.5) {P};
\node[anchor=west] at (0.25,-0.5) {II$_{\left[\frac{d}{2}\right]}$};
\node[anchor=west] at (0.25,-2.5) {II$_1$};
\end{tikzpicture}
}
\caption{Shadow transform relating primary operators (indicated by the same color) in celestial
necklaces in even dimensions (a) and odd dimensions (b). In our examples, the relation between the soft operators (type I$_2$) and their shadows (type II$_1$) is depicted by the curved arrow. }
\label{fig:CelestialNecklaceShadow}
\end{figure}
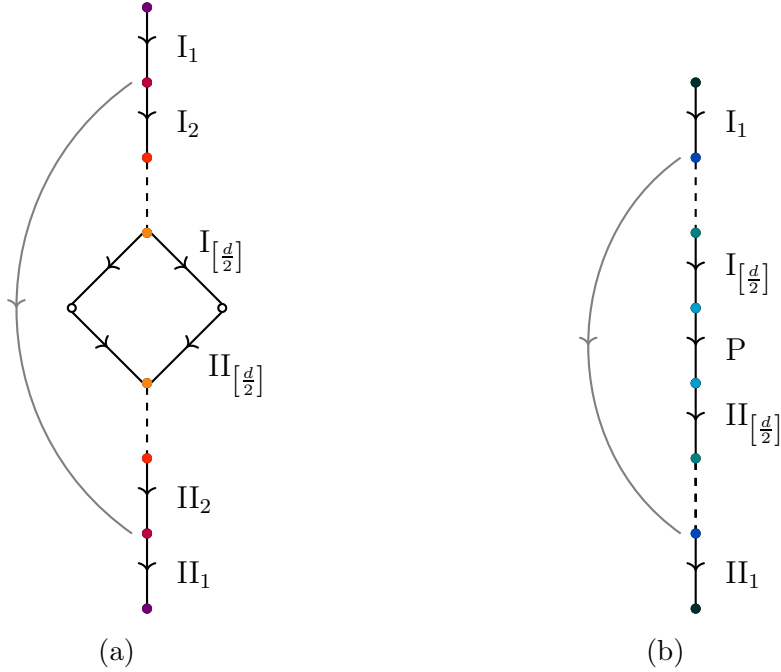

To show that confromally soft operators can be shadowed to obtain type $\II_1$ operators is rather subtle because the shadow integrals naively
annihilate type $\I$ operators.
The shadow is thus obtained by a regularization  of the integral, which gives a finite result.
In the following we will show that, very surprisingly, in even dimensions this regularization procedure gives rise to  analytically continued type $\S$ operators (this relation generalizes to  all values of $\ell$ and $k$ of the conformally soft operators \eqref{conf_soft_ops}).
Alternatively, we can directly compute the shadow transform of the soft factors in any dimensions which we detail in appendix~\ref{app:Shadow}.
We will then use these results to compute the Ward identities for the shadow transform of the leading soft photon and the leading and subleading soft graviton operators.

\subsubsection{Shadows and Type S operators}
 The shadow transform of a symmetric and traceless operator $\Ocal_{\Delta,\ell}$ corresponds to the following integral transform  
\begin{equation}\label{I_Shadow_indexfree}
\S[\Ocal_{\Delta,\ell}](x,\ecal)=\frac{N_{\Delta,\ell}}{\ell! (\frac{d}{2}-1)_\ell}\int d^dx' \frac{I(x'-x,\ecal,D_{\ecal'})^\ell}{[(x'-x)^2]^{d-\Delta}} \Ocal_{\Delta,\ell}(x',\ecal') \, ,
\end{equation}
where $I(x,y,z)\equiv y\cdot z- 2(y\cdot x)(z\cdot x)/x^2$ is the inversion tensor. The choice of normalization of the shadow transform 
\be\label{C}
N_{\Delta,\ell}=\frac{\pi^{-\hd}\Gamma(d-\Delta+\ell)\Gamma(\Delta-1)}{\Gamma(\Delta-1+\ell)\Gamma\left(\Delta-\hd\right)}
\ee
yields $\S[\S[\Ocal_{\Delta,\ell}]]=\Ocal_{\Delta,\ell}$.
In appendix~\ref{app:Shadow} we compute a regulated version of~\eqref{I_Shadow_indexfree} which we apply to the universal soft theorems. 

We now want to take a different route.
Indeed in~\cite{Kapec:2017gsg} it was shown that in even dimensions the shadow transform  can be written in terms of local differential operators when acting on conformally soft operators. Here we want to generalize this result to any soft operator $\Ocal_{k,\ell}$ with dimensions $k=1,0,-1,\dots$ and spin $\ell$. At the same time we will show that the local differential operators are actually written in terms of an analytic continuation of the type S operator.

To show the relation between the type S operator and the shadow transform, it is convenient  to use an alternative definition (see \cite{Penedones:2015aga}) for the differential operator $D_{\S,\kcal}$ given in \eqref{O_typeS0},
\be{} \label{D_typeS}
D_{\S,\kcal}\equiv \square^{\kcal-\ell} \widehat D_{\S,\kcal} \, ,
\qquad 
\widehat D_{\S,\kcal}\equiv\sum_{j=0}^\ell a_{j,\kcal} \square^{\ell-j}(\ecal\cdot \partial_x)^j(D_\ecal\cdot \partial_x)^j \, ,
\ee{}
 where the coefficients $a_{j,\kcal}$ are written as
\be
\label{eq:coeff_a}
a_{j,\kcal}= \frac{(-2)^j \ell! \kcal!}{j! (\ell-j)! (\kcal-j)! (-\ell)_j (-\frac{d}{2}-\ell+2)_j (\frac{d}{2}-j+\ell+\kcal-1)_j} \, .
\ee
Since  $D_{\S,\kcal}$ contains 
$\square^{\kcal-\ell}$, this  operator makes sense only for $\kcal\geq \ell$.\footnote{
It is possible to extend the definition of $D_{\S,\kcal}$ to $\kcal < \ell$ by replacing $\square^{\kcal-\ell} \square^{\ell-j} \to \square^{\kcal-j}  $ and replacing the upper limit of the sum $\ell \to \kcal$. The resulting formula is written in (173) of \cite{Penedones:2015aga}. Formula \eqref{D_typeS} however will be more convenient for our purposes.} 
On the other hand, this representation is convenient since $\widehat D_{\S,\kcal}$ can be analytically continued to complex values of $\kcal$.

For $\kcal=\frac{d}{2}-\Delta$ we can thus use $\widehat D_{\S,\kcal}$ to rewrite the shadow kernel in~\eqref{I_Shadow_indexfree} as follows
\begin{equation}
\label{DhatI}
\frac{I(x,\ecal,v)^\ell}{(x^2)^{d-\Delta}} = c_{\Delta, \ell} \widehat D_{\S, \frac{d}{2}-\Delta}\frac{(\ecal \cdot v)^\ell}{(x^2)^{d-\Delta-\ell}} \, ,
\end{equation}
for vectors $v$ in CFT$_d$ with the property $v^2=0$ and where
\be{}
c_{\Delta,\ell}=-\frac{1}{2^{2 \ell} (-\Delta +d+\ell-1) (-\frac{d}{2}+\Delta )_\ell (-d+\Delta +2)_{\ell-1}}\,.
\end{equation}

So far we have not restricted the conformal dimension of the operator. In the following we are interested in the conformally soft values $\Delta\in 1- \mathbb Z_{\geq 0}$.
Performing the shadow integral~\eqref{I_Shadow_indexfree} is subtle and
to make the result finite we introduce an infinitesimal regulator $\epsilon$.

In the rest of this section we set   $\Delta = k+\epsilon$ where $k=1,0,-1,\dots$ and we take $\epsilon\to 0$ at the end. To proceed we specify to even dimensions $d$ where we use the following identity

\begin{equation}
\label{boxx2}
\frac{1}{(x^2)^{d-k-\epsilon-\ell}}=d^\epsilon_{k,\ell}
\square^{\frac{d}{2}-\ell-k} \frac{1}{(x^2)^{\frac{d}{2}-\epsilon}}
\end{equation}
for $\frac{d}{2}-\ell-k\in \mathbb Z_{\geq 0}$
where 
\begin{equation}
d^\epsilon_{k,\ell}=(-1)^{\ell+k+1} 4^{-\frac{d}{2}+\ell+k}\frac{ \left(-\frac{d}{2}+\epsilon +1\right)_{\ell+k-1}}{(1-\epsilon )_{d-k-\ell-1}} \, .
\end{equation}
Using these relations we first study what happens to the integral~\eqref{I_Shadow_indexfree} for  $\epsilon=0$ exactly. 
Integrating by parts in~\eqref{I_Shadow_indexfree} using~\eqref{DhatI} with  $v=D_{\ecal'}$ such that $ \frac{1}{\ell! (\frac{d}{2}-1)_\ell}(\ecal \cdot D_{\ecal'})^\ell \Ocal_{\Delta,\ell}(x,\ecal')=\Ocal_{\Delta,\ell}(x,\ecal)$ and the identity~\eqref{boxx2} 
gives 
\begin{align}
S[\Ocal_{k,\ell}](x,\ecal)]={N_{k,\ell}}c_{k,\ell} d^{\epsilon=0}_{k,\ell} \int d^dx' \frac{1}{|x'-x|^{d}}   D_{\S, \frac{d}{2}-k} \Ocal_{k,\ell}(x',\ecal) \, .
\end{align}
Now one can argue that $\widehat D_{\S, \frac{d}{2}-k} \Ocal_{k,\ell}$ is exactly zero (notice also that the integral \mbox{$\int d^{d}x' |x-x'|^{-d}$} by itself is divergent). The rough idea is that $\widehat D_{\S, \frac{d}{2}-k}$ in even dimensions acts like a type I differential operator which  annihilates the primary, for more details see the discussion in appendix \ref{app:typeSeven}. This is also checked directly for the examples in sections \ref{sec:CCFTdShadowPhoton}~-~\ref{sec:CCFTdShadowSubGraviton} by explicitly acting with $D_{\S, \frac{d}{2}-k}$ on the soft factors associated to the insertion of $\Ocal_{k,\ell}$. 

In\cite{Kapec:2017gsg,Kapec:2021eug} the shadow of soft modes is then regulated by keeping $\epsilon$ infinitesimal and taking the limit $\epsilon \to 0$ at the end. This gives
\begin{align}\label{Shadow_infeps}
S[\Ocal_{k+\epsilon,\ell}](x,\ecal)
&={N_{k+\epsilon,\ell}}c_{k+\epsilon,\ell} d^\epsilon_{k,\ell} \int d^dx' \frac{1}{|x'-x|^{d-2\epsilon}}  \square^{\frac{d}{2}-\ell-k} \widehat D_{\S, \frac{d}{2}-k-\epsilon} \Ocal_{k+\epsilon,\ell}(x',\ecal)\,.
\end{align}
We expand the action of the differential operator to the first non-vanishing order in $\epsilon$,
\be
\widehat D_{\S, \frac{d}{2}-k-\epsilon}   \Ocal_{k,\ell}(x,\ecal) = \epsilon  \left(\partial_{\epsilon} \widehat D_{\S, \frac{d}{2}-k-\epsilon}\right)_{\epsilon=0} \Ocal_{k,\ell}(x,\ecal)
+O(\epsilon^2)\, .
\ee
We further discard the order $\epsilon$ term in $\Ocal_{k+\epsilon,\ell}$ because the soft operators are actually defined by taking the residue of the operator $\Ocal_{\Delta,\ell}$ for $\Delta=k$, and if we shift $k$ by $\epsilon$, the residue vanishes. 
We then notice that the remaining integral kernel  in~\eqref{Shadow_infeps} is proportional to
\be
\lim_{\epsilon \to 0}  \frac{\epsilon}{|x'-x|^{d-2\epsilon }} =\frac{S_d}{2} \delta^{(d)}(x'-x) \, ,
\ee
where $S_d\equiv \frac{2 \pi ^{d/2}}{\Gamma \left(\frac{d}{2}\right)}$. The integral in $x'$ can now be trivially performed giving the result 
\begin{equation}\label{Shadow_typeS}
    \boxed{
    S[\Ocal_{k,\ell}](x,\ecal)=n_{k,\ell} \,
    \square^{\frac{d}{2}-\ell-k}
    \left(\partial_{\epsilon}  \widehat D_{\S, \frac{d}{2}-k-\epsilon} \right)_{\epsilon=0} 
    \Ocal_{k,\ell}(x,\ecal)
    }
\end{equation}
where $n_{k,\ell}\equiv  \frac{S_d}{2} \lim_{\epsilon \to 0}{N_{k+\epsilon,\ell}} c_{k+\epsilon,\ell} d^\epsilon_{k,\ell} $. 
Formula \eqref{Shadow_typeS} is very explicit since 
the derivative of $\widehat D$ 
is obtained in a closed form by acting on the coefficients $a_{j,\kcal}$ in \eqref{eq:coeff_a}, namely 
\be
(\partial_\epsilon a_{j,\frac{d}{2}-k-\epsilon})_{\epsilon=0}=
-\frac{ \ell! \left(\frac{d}{2}-j-k+1\right)_j \left(H_{\frac{d}{2}-k}-H_{\frac{d}{2}-k-j}-H_{d-k+\ell-2}+H_{d-k+\ell-2-j} \right)}{ (-2)^{-j} j! (\ell-j)! (-\ell)_j \left(-\frac{d}{2}-\ell+2\right)_j (d-j-k+\ell-1)_j} \, ,
\ee
where $H_k$ are harmonic numbers.
Formula \eqref{Shadow_typeS} not only gives a nice interpretation of the differential operators in \cite{Kapec:2017gsg}, but also extends those results to all $k=1,0,-1,\dots$ and $\ell$ (with the only constraint that $d/2-\ell - k\geq0$), furnishing  a  very efficient tool to compute the shadow transform of
bosonic soft operators.  

Let us give the explicit expressions of \eqref{Shadow_typeS}  for $\ell=1$ with $k=1$ and $\ell=2$ with $k=1,0$, which correspond to the shadow transforms of respectively the leading soft photon and the leading and subleading soft graviton,
\begin{align}
&\!S[\Ocal_{1,1}]
=
n_{1,1} \frac{2 \square{}^{\frac{d}{2}-2}}{(d-2)^2}  (\ecal\cdot \partial_x) (D_{\ecal} \cdot \partial_x)\Ocal_{1,1}(x,\mathcal e)
 \, , \\
&\!S[\Ocal_{1,2}]
= n_{1,2} 
\frac{2\square{}^{\frac{d}{2}-3}}{(d-1)^2}
\bigg(\square  (\ecal\cdot \partial_x )(D_{\ecal}\cdot  \partial _x)
-\frac{2 d-5}{(d-2)^2} (\ecal\cdot \partial _x){}^2 \left(D_{\ecal} \cdot \partial _x \right){}^2 \bigg)\Ocal_{1,2}(x,\mathcal e)
\, , \\
&\!S[\Ocal_{0,2}]
=n_{0,2}
\frac{2\square^{\frac{d}{2}-2}}{d^2}
\bigg(\square (\ecal\cdot \partial _x)  (D_\ecal \cdot \partial _x)
-\frac{(2 d^2-3 d+2)}{(d-2) (d-1)^2}
(\ecal \cdot \partial _x){}^2 \left(D_\ecal \cdot \partial _x\right){}^2 \bigg)\Ocal_{0,2}(x,\mathcal e)
\, .
\end{align}
These results match the ones of \cite{Kapec:2017gsg} and slightly generalize them  since in \cite{Kapec:2017gsg} it is only described how to get  $ \partial_a S[\Ocal_{0,2}]^{ab}$ and $ \partial_a S[\Ocal_{1,2}]^{ab}$ as local operators.

To conclude let us mention that the relation \eqref{Shadow_typeS} can also be expressed in a more compact but less transparent\footnote{ 
\label{footnote:Deps}
Notice that  \eqref{Shadow_S} contains a
seemingly problematic
term where $\square$ is elevated to the non-integer power $\frac{d}{2}-\ell-k-\epsilon$. However the order $O(\epsilon)$ contribution coming from expanding this term vanishes because, as we argued above, $\widehat D_{\S, \frac{d}{2}-k} \Ocal_{k,\ell}=0$. So for any practical  purposes one can (and should) use  $\square^{\frac{d}{2}-\ell-k}$ as in \eqref{Shadow_typeS}, which is well defined.
} fashion as,
\begin{equation}
\label{Shadow_S}
    \boxed{ 
     S[\Ocal_{k,\ell}](x,\ecal)= n_{k,\ell} \lim_{\epsilon \to 0} \frac{1}{\epsilon}  D_{\S, \frac{d}{2}-k-\epsilon}  \Ocal_{k+\epsilon,\ell}(x,\ecal) 
    }\, .
\end{equation}
Here we did not assume that $\Ocal_{k+\epsilon,\ell}$ can be replaced by $\Ocal_{k,\ell}$ so \eqref{Shadow_S} works not only for soft operators but also for generic primaries.
This formula shows that the shadow transform of primary $\Ocal_{k, \ell}$ corresponds to the action of an analytically continued type $\S$ operator on the analytically continued primary $\Ocal_{k+\epsilon, \ell}$. 
To be precise, we find a type S operator at level $\kcal=\frac{d}{2}-k-\epsilon$ but since $\kcal$  is not integer for non-vanishing $\epsilon$ (and likewise the dimension $\Delta=k+\epsilon$ of the operator), it is not a priori  clear if this descendant is also a primary. 
Notice that if the shadow \eqref{Shadow_S} were a (non-primary) descendant it would be extremely problematic, firstly because the shadow of a primary should be a primary, and secondly because
 in the following we want to argue that these shadowed operators define the stress tensor and currents in celestial CFT, which must be primaries. 
 Fortunately we find that \eqref{Shadow_S} defines a primary in the following sense
\be \label{PrimarySO}
\lim_{\epsilon \to 0} \frac{1}{\epsilon} [K^a ,  D_{\S, \frac{d}{2}-k-\epsilon} \Ocal_{k+\epsilon,\ell}(0,\ecal)] = 0 \, ,
\ee
where $K^a$ is the generator of special conformal transformations (and we consider the remark of footnote \ref{footnote:Deps}).
To show that this vanishes we take the $\epsilon\to 0$ limit after computing the commutator and we
further need to mod out by the module defined by the primary descendant of $\Ocal_{k,\ell}$. Namely we have to require that  the shortening conditions  for the operator $\Ocal_{k,\ell}$ (derived in section~\ref{sec:CelestialNecklaces}) are satisfied. 
We checked this explicitly in several cases for various values of $d,\ell,k$ corresponding to the leading soft photon and leading and subleading soft graviton operators. In all cases the result of computing~\eqref{PrimarySO} is proportional to a shortening condition for a type $\I_2$ operator, which we set to zero by modding out by the module defined by this primary descendant.
In the simplest example of $\ell=1$ and $k=1$ in $d=4$ we have $\widehat D_{\S, 1-\epsilon} \Ocal_{1+\epsilon,1}(x,\ecal) \propto \ecal_b (  \delta^b_{\,\, c} \square-\frac{
2 (\epsilon -1) }{\epsilon -2} \p^b \p_c  ) \Ocal^c_{1+\epsilon,1}(x)$  and thus it is very easy to perform the  commutation by simply replacing the derivative with the generator of translations and using the conformal algebra, 
\begin{align}
\lim_{\epsilon \to 0} \frac{1}{\epsilon}
\left[K^a,\left( \delta^b_{\,\, c} \square -\frac{
2 (\epsilon -1) }{\epsilon -2} \p^b \p_c  \right) \Ocal^c_{1+\epsilon,1}(0)\right] \propto
\p^a \Ocal^b_{1,1}(0)-\p^b \Ocal^a_{1,1}(0) = 0 \, .
\end{align}
The result is zero since it is  proportional to the type $\I_2,\kcal=1$ shortening condition which the operator $\Ocal^a_{1,1}$  satisfies.
Let us stress that \eqref{Shadow_S} is not a canonical type of primary descendant and for this reason it escapes the conventional classification. Indeed the limiting prescription in \eqref{Shadow_S} extracts  the order $O(\epsilon)$ term which appears by either expanding the differential operator or the primary operator:
$D_{\S, \frac{d}{2}-k-\epsilon} \Ocal_{k,\ell}+D_{\S, \frac{d}{2}-k} \Ocal_{k+\epsilon,\ell}$. These two terms however define descendants of two different primaries. 
Thus by opportunely summing descendants of two different conformal multiplets (with arbitrarily close conformal dimensions) it is possible to define a new primary operator given by~\eqref{Shadow_S}.
This mechanism is very different from the one in standard CFTs in which the  primary  descendant arises because  a single multiplet  becomes degenerate. Moreover, the result of the commutator only vanishes when the shortening conditions are implemented, which is another feature that does not arise for usual primary descendants, where the result of the commutator is instead exactly zero.

\subsubsection{Leading Soft Photon Theorem}\label{sec:CCFTdShadowPhoton}
The divergence of the conformally soft photon operator $R^a_{1}(x)$ inside a correlation function~\eqref{R1_WI_bad}  does not give a standard CFT$_{d>2}$ Ward identity.
Besides the fact that $R^a_{1}(x)$ is not divergenceless, 
it does not have the correct conformal dimension to be a current. Indeed a current should have dimensions $d-1$, while $R^a_1(x)$ has dimension $1$. 
It is thus natural to consider its shadow $\tilde R^a_{d-1}(x)$ which instead has the correct conformal dimension.
Shadowing the soft factor we obtain
\be\label{R1t_WI}
\tilde{S}^{(0)}(x,\mathcal e)=-2e
{\textstyle \left[\Gamma\left(\frac{d}{2}\right)\right]^2}
\sum_{i=1}^N\mathcal Q_i\frac{1}{\omega}\frac{(x_i-x) \cdot {\mathcal e}}{(x_i-x)^{d}}\, . 
\ee
This computation can be performed either using ~\eqref{Shadow_typeS} (which in principle requires even $d$) or by the method detailed in appendix~\ref{app:Shadow} (which works in any $d$).

To determine the divergence of $\tilde R^a_{d-1}(x)$ in a correlation function we act with $\partial_{\ecal} \cdot \partial_x$ on \eqref{R1t_WI} and Mellin transform. The result is 
\be
\label{R1_WI_1}
\langle\p_a \tilde R^a_{d-1}(x)\Ocal_{\Delta_1,\ell_1}\dots \Ocal_{\Delta_N,\ell_N}\rangle=4e\pi^{\hd}{\textstyle \Gamma\left(\hd\right)}
\sum_{i=1}^N \mathcal Q_i \delta^{(d)}(x_i-x)\langle\Ocal_{\Delta_1,\ell_1}\dots \Ocal_{\Delta_N,\ell_N}\rangle\, ,
\ee
where the appearance of the delta function follows from appendix~\ref{app:deltadist}.
Clearly, $\tilde R^a_{d-1}(x)\equiv J^a(x)$ is a conserved $U(1)$ current, i.e. it satisfies 
$\partial_a J^a(x)=0$ away from contact points. 
Its primary descendant $\partial_a \tilde R^a_{d-1}(x)$ at level $\kcal=1$ is the simplest example of a spin~$\ell=1$ operator of type $\II_{1}$. The associated Noether current~\eqref{Jcal_gen} can be expressed as
\be 
\Jcal^{\epsilon \, a}(x)=\tilde R^a_{d-1}(x) \epsilon(x),
\ee 
whose conservation equation $\partial^a \Jcal^{\epsilon}_a=0$ implies that the associated symmetry parameter
\be 
\epsilon(x)=c 
\ee 
must be a constant in $d>2$. 
Thus, following \eqref{eq:Q_vol}, we obtain a single charge $Q^{\epsilon}_\Sigma=c Q$ defined as  
\be \label{ChargeJ}
Q=\frac{1}{4e\pi^\hd\Gamma\left(\hd\right)}\int_R d^d x \, \partial_a \tilde R^a_{d-1}(x)\, .
\ee
 for a region  $R$ in $ \mathbb R^d$ with boundary given by $\Sigma$.
From the identity \eqref{R1_WI_1}, we can deduce the charge action on an operator $\mathcal O_{\Delta,\ell}(x)$ 
\be
\left[Q,\Ocal_{\Delta,\ell}(x)\right]=\mathcal Q \Ocal_{\Delta,\ell}(x)\, .
\ee
For a non-abelian symmetry the current transforms in the adjoint representation of the symmetry group and so it has more components to each of which one can associate a conserved charge.

\subsubsection{Leading Soft Graviton Theorem}\label{sec:CCFTdShadowGraviton}

The conformally soft graviton operator $H^{ab}_1(x)$ is likewise not conserved in correlation functions~\eqref{H1_WI_1} in $d>2$ and we consider instead its shadow $\tilde H^{ab}_{d-1}(x)$ which has the correct dimension of a type $\II_1$ operator at level $\kcal=2$, according to equation \eqref{polesII}. 
The shadow of the soft graviton factor is given by
\be\label{ShadowGravd}
\tilde{S}^{(0)}(x,\ecal)=-\frac{\kappa}{2}d{\textstyle \left[\Gamma\left(\hd\right)\right]^2}\sum_{i=1}^N\eta_i\frac{\omega_i}{\omega}\frac{[(x_i-x)\cdot \ecal]^2 }{(x_i-x)^d}\, .
\ee
Computing the divergence of $\tilde H^{ab}_{d-1}(x)$ in correlation functions using  appendix~\ref{app:deltadist} 
yields 
\be\label{H1_WI}
\langle\p_a\p_b\tilde H^{ab}_{d-1}(x)\Ocal_{\Delta_1,\ell_1}\dots \Ocal_{\Delta_N,\ell_N}\rangle={-2\kappa}\pi^\hd (d-1){\textstyle \Gamma\left(\hd\right)}
\sum_{i=1}^N \eta_i\delta^{(d)}(x_i-x)\langle\Ocal_{\Delta_1,\ell_1}\dots \Ocal_{\Delta_i+1,\ell_i}\dots \Ocal_{\Delta_N,\ell_N}\rangle\, .
\ee
This higher-derivative Ward identity is associated to  the  level $\kcal=2$ primary descendant operator of type II$_{1}$ defined by  
$\p_a\p_b\tilde H^{ab}_{d-1}(x)$. 
The Noether current~\eqref{Jcal_gen} can be written as
\be 
\mathcal J^{\epsilon \, a}(x)= \tilde H^{ab}_{d-1}(x)\partial_b \epsilon(x)-\partial_b  \tilde H^{ab}_{d-1}(x)\, \epsilon(x)\, ,
\ee
and the conservation of $\mathcal J^{\epsilon}_a$ requires the parameter to take the form~\eqref{sympar}, namely
\begin{equation}
    \epsilon(x)=c_\mu q^\mu(x) \,,
\end{equation}
where $q^\mu$ is defined  in \eqref{qPoincare} and $c_\mu$ is a constant vector $c_\mu=\left(c_0,c_a,c_{d+1}\right)$.
Using \eqref{eq:Q_vol} the charge takes the form
\begin{equation}
Q^{\epsilon}_\Sigma={-\frac{1}{2\kappa}\frac{1}{\pi^\hd (d-1)\Gamma\left(\hd\right)}}\int_{R}d^dx  \
\epsilon(x) \partial_a \partial_b  \tilde H^{ab}_{d-1}(x)\, ,
\end{equation}
where $\p R=\Sigma$.
This can be thus written as a sum of $d+2$ independent charges $Q^{\epsilon}_\Sigma=c_{\mu} Q^{\mu}$. 
From the Ward identity \eqref{H1_WI}, we can deduce the charge action on an operator $\Ocal_{\Delta,\ell}(x)$,
\be
\left[Q^{\mu},\Ocal_{\Delta,\ell}(x)\right]=\eta  q^\mu(x) \Ocal_{\Delta+1,\ell}(x)\, ,
\ee
where the shift in the conformal dimension   and the multiplication by $\eta q^{\mu}$ is exactly what produces the translations in physical $d+2$-dimensional space.\footnote{Given a momentum state $|p^{\mu}\rangle$ with $p^{\mu}=\eta \omega q^{\mu}$ and its Mellin transform  $|\Delta,\eta q^{\mu}\rangle$, bulk translations $\Pcal^{\nu}$ act as
$$
\Pcal^{\nu} |p^{\mu}\rangle=p^{\nu} |p^{\mu}\rangle 
\quad \to \quad 
\Pcal^{\nu} |\Delta,\eta q^{\mu}\rangle =\eta q^{\nu} |\Delta+1,\eta q^{\mu}\rangle \, .
$$
}

\subsubsection{Subleading Soft Graviton Theorem}\label{sec:CCFTdShadowSubGraviton}
Among the conserved operators of spin $\ell=2$ we expect the appearance of the stress tensor which has conformal dimension $\Delta=d$. 
Indeed, the shadow transform of the subleading soft graviton operator $H_0^{ab}$, denoted by $\tilde H^{ab}_d(x)$, has the right conformal dimension to be the stress tensor. Shadowing the subleading soft graviton factor gives 
\begin{align}
\tilde S^{(1)}(x,\mathcal e)&={\kappa} {\textstyle \left[\Gamma\left(\hd+1\right)\right]^2}\sum_{i=1}^N\frac{(x-x_i)\cdot \mathcal e}{[(x-x_i)^2]^{\hd+1}}\bigg[\frac{1}{2}(x-x_i)\cdot \mathcal e\left(-2\omega_i \p_{\omega_i}+(d-2)(x-x_i)\cdot \p_{x_i}\right) \nonumber \\
&\qquad \qquad\qquad\qquad\qquad \qquad +(x-x_i)^2\mathcal e \cdot \p_{x_i}-i(d-1)\mathcal e\cdot M_i \cdot (x-x_i)\bigg]\,.
\label{ShadowSubGravd}
\end{align}
We again use appendix \ref{app:deltadist} to land, after Mellin transforming over the energies, on the following Ward identity 
\be\label{H2_WI}
\begin{aligned}
&\langle \p_a \tilde H^{ab}_{d}(x)\Ocal_{\Delta_1,\ell_1}\dots \Ocal_{\Delta_N,\ell_N}\rangle=\frac{\kappa}{2}\pi^\hd(d-1){\textstyle \Gamma\left(\hd+1\right)}
\sum_{i=1}^N \bigg[\delta^{(d)}(x-x_i)\p_{x_i}^b
\\
& \qquad \qquad  \qquad \qquad 
-\frac{\Delta_i}{d}\p^b\delta^{(d)}(x-x_i)+\frac{i}{2}M_i^{bc} \p_{c} \delta^{(d)}(x-x_i)\bigg]
\langle\Ocal_{\Delta_1,\ell_1}\dots \Ocal_{\Delta_N,\ell_N}\rangle \, .\\
\end{aligned}
\ee
The first of the three terms in the  square brackets corresponds to the usual Ward identity for a $d$-dimensional stress tensor $T^{ab}$ which is conserved, traceless and symmetric   away  from other operator insertions. Instead our primary operator
$\tilde H^{ab}_d(x)\equiv -T^{\{ab\}}=-T^{ab}+T^{[ab]}+\frac{1}{d}\delta^{ab}T^{c}_c$
is {\it exactly} traceless and symmetric 
because we explicitly project it
into the traceless and symmetric representation. 
This explains the two additional terms in the second line of~\eqref{H2_WI}.
Here the square brackets denote anti-symmetrization, while the curly brackets make the indices symmetric and traceless.
The primary descendant $\p_a \tilde H^{ab}_{d}(x)$ is the prime example of a spin $\ell=2$ operator of type II$_{1}$ at level $\kcal=1$.
The Noether current~\eqref{Jcal_gen} takes the form
\be \label{Jcal_T}
\Jcal^{\epsilon\, a}(x) = \tilde H^{ab}_{d}(x) \epsilon_{b}(x)
\ee 
for some vector $\epsilon_{b}(x)$ such that $\partial^a \Jcal_a^{\epsilon}=0$. Using the properties of the stress tensor, we find that $\epsilon_a$ must satisfy the following equation
\be
\label{CKE}
\frac{\partial_a  \epsilon_{b} + \partial_b  \epsilon_{a}}{2} -  \frac{\eta_{a b } }{d}  \partial_c  \epsilon^{c} =0 \, .
\ee
This is the conformal Killing equation, which can also be obtained by studying the transformations that preserve the flat metric up to a conformal factor.
In $d>2$ this equation has a finite number of solutions
\be
\label{CKV}
 \epsilon_{a}(x) =c^P_a+c^R_{[ a\, b]} x^b+   c^D x_a +c^K_{b } (2 x_a x^b-\eta_{a}^{\phantom{a} b} x^2) \,,
\ee
which are the conformal Killing vectors that are parametrized by $
 \frac{(d+2)(d+1)}{2}$ coefficients $c^P_a, c^R_{[a \,b]},  c^D, c^K_a$ with the brackets denoting anti-symmetrization.
Notice that the equation for $\epsilon_a(x)$ obtained in~\eqref{sympar}, which for the present case reduces to 
\begin{equation}
\label{eq:eps_cmunu}
    \epsilon_a(x)=c_{[\mu\,\nu]} q^{\mu}  \partial_{a}q^{\nu} \, , 
\end{equation}
is equivalent to~\eqref{CKV} upon identifying
\be \label{}
c^D \equiv c_{[d+1\,0]} \, ,
\quad 
c^P_a \equiv \frac{1}{2}  (c_{[0\,a]}+c_{[d+1\,a]})\, ,
\quad 
c^K_a\equiv \frac{1}{2}(c_{[a\,0]} -c_{[a\,d+1]}) \, ,
\quad 
c^R_{[a\,b]} \equiv c_{[a\,b]}   \, .
\ee
From \eqref{eq:Q_vol}, the charge takes the form  
 \be
 Q^{\epsilon}_{\Sigma}={\frac{2}{\kappa}\frac{1}{\pi^\hd(d-1){\textstyle \Gamma\left(\hd+1\right)}}}\int_R d^d x  \ \epsilon_a(x) \p_ b \tilde H^{ab}_{d}(x)\, , \ee
 where $\p R= \Sigma$.
Using $\epsilon_a$ as defined in \eqref{CKV} and \eqref{eq:eps_cmunu} we shall  expand the charge in terms of the independent constant coefficients 
as
$Q^{\epsilon}_{\Sigma}\equiv c_{[\mu \nu]}Q^{\mu \nu}\equiv c_D Q^D+c_a^P Q^{P \, a}+c_a^K Q^{K \, a}+c_{[ab]}^{R} Q^{R \, ab}$. The action of the charges $Q^{\mu \nu}$  on a primary operator  $\Ocal_{\Delta,\ell}(x)$ can then be computed from \eqref{H2_WI} and takes the following form
\be
[Q^{\mu \nu},\Ocal_{\Delta,\ell}(x)]=\bigg[ q^{\mu} (\p^a q^\nu) \p_a  +\frac{\Delta}{d}  [\p^a(q^\mu \p_a q^\nu )] 
-\frac{i}{2} (\p_b q^\mu) (\p_a q^\nu )  M^{ab}  \bigg] \Ocal_{\Delta,\ell}(x) \, .
\ee
 This compact expression  can then be rewritten -- replacing $q^{\mu}$ as in \eqref{qPoincare} and using the definition above for the charges -- in terms of the more familiar CFT$_d$ charges 
 associated to translations~$P_a$, rotations~$R_{a b}$, dilations~$D$ and special conformal transformations~$K_a$,
\be
\begin{aligned}
[Q^{P\,a},\Ocal_{\Delta,\ell}(x)]&=\p^a \Ocal_{\Delta,\ell}(x)\,,\\
[Q^{R\, ab},\Ocal_{\Delta,\ell}(x)]&=\left((x^a \p^b-x^b \p^a)+i M^{ab}\right)\Ocal_{\Delta,\ell}(x)\,,\\
[Q^{D},\Ocal_{\Delta,\ell}(x)]&=\left(x^a\p_a+\Delta\right)\Ocal_{\Delta,\ell}(x)\,,\\
[Q^{K\, a},\Ocal_{\Delta,\ell}(x)]&=\left((2x^a x_b-\delta_b^a x^2)\p^b+2 \Delta x^a-2ix_b M^{ba}\right)\Ocal_{\Delta,\ell}(x)\, .
\end{aligned}
\ee
We thus recovered  
the generators of the conformal algebra, which simply correspond to the Lorentz transformations in the bulk.

\section{Conclusions and Outlook}
\label{sec:Conclusions}
In this work we have focused on the universal soft theorems for gauge theory and gravity in $d+2$ spacetime dimensions and the classification of soft symmetries in the dual celestial CFT${}_d$.

In the conformal basis, the universal soft theorems in $d+2>4$ dimensions take the form of $d$ dimensional correlation functions with the insertion of a (conformally) soft operator with special integer dimension~$\Delta$ that transforms in a short representation.
Soft operators have primary descendants of type~I$_2$ at level $\kcal=2-\Delta$ which have the spin label $\ell_2$ increased by $\kcal$ units and thus do not transform in traceless and symmetric representations. 
The conservation equations defined by these primary descendants 
do not give rise to contact terms and so the associated charges are trivial.
Non-trivial conserved charges can instead be built from their shadow transforms.

The shadow transform of the  soft operators -- which in even $d$ acts as an analytic continuation of the type S differential operator -- maps them to operators with 
type~II$_1$ primary descendants whose spin label $\ell_1$ is decreased by $\kcal=\Delta+\ell_1-1$ units.
Conformally soft shadow operators define familiar conserved CFT$_d$ operators such as currents and stress tensor as well as operators satisfying higher-derivative conservation equations such as those generating translations in $d+2$ spacetime dimensions. For all such operators we explained how to construct the full set of associated conserved charges.
Importantly, these shadow operators generate the symmetries corresponding to $d+2$ dimensional Poincaré and global $U(1)$ transformations -- which are finite-dimensional groups.
This is in contrast to the infinite local enhancement in $d+2=4$ spacetime dimensions by BMS and large gauge transformations for which we have constructed Noether currents and infinite towers of charges. 

Let us conclude with some comments and open questions.

\paragraph{The top of the necklace}
In CCFT${}_2$ the conformal multiplets containing the universal soft operators were completed into a diamond by adding a primary at the top: the universal soft operators in figure~\ref{fig:CelestialDiamond} are at the left and right corners, with their type II primary descendant at the bottom corresponding to their conservation equation, and can themselves be understood as type I primary descendants of a new operator at the top. (For degenerate multiplets the soft operator, whose type III primary descendant corresponds to its conservation equation, already coincides with the top operator of a zero-area diamond and so no additional operator needs to be added.) 
This was a useful thing to do as it defines the Goldstone modes of spontaneously broken asymptotic symmetries which are used to dress celestial amplitudes to make them infrared finite~\cite{Arkani-Hamed:2020gyp}. The top operators have logarithmic correlation functions (similar to the ones of a free boson in  $d=2$ which is the simplest example of a top operator in a celestial diamond) and should not be considered as part of the spectrum of the original theory (it may be thought as a primary of a logarithmic extension of the theory or just as an auxiliary operator); because of the logarithmic behaviour the arguments of appendix \ref{app:typeI} do not apply to this operator. 

In a similar vein, we can complete the celestial necklaces in CCFT${}_{d>2}$ by adding a top primary operator - this is the one already shown in figure~\ref{fig:CelestialNecklace}. Indeed, all universal soft operators have type $\I_2$ primary descendants and can themselves be understood as type $\I_1$ primary descendants of an operator at the top. As in $d=2$ this top operator has logarithmic correlation functions. 
Its quantum numbers are fixed by representation theory and the form of its primary descendant follows the classification in section~\ref{sec:TypeI}. 
For the leading soft photon the top-of-the-necklace operator must be an operator $\Ocal_{\Delta=0,\ell=0}$ with type $\I_1,\kcal=1$ primary descendant  $\p^a \Ocal_{0,0}$.
This is set by the fact that its type $\I_1$ primary descendant has $(\Delta,\ell)=(1,1)$. 
Similarly for the leading soft graviton the top-of-the-necklace operator is $\Ocal_{\Delta=-1,\ell=0}$ with a primary descendant of type $\I_1,\kcal=2$ that takes the form $ \p^{\{a} \p^{b\}}\Ocal_{\Delta=-1,\ell=0}$.
For the subleading soft graviton we find $\Ocal_{\Delta=-1,\ell=1}$ with type $\I_1,\kcal=1$ primary descendant $ \p^{\{a}_{\phantom{\big|}} \Ocal^{b\}}_{\Delta=-1,\ell=1}$.
Note that the equations for the type $\I_1$ primary descendants appeared in the literature (e.g.~\cite{Kapec:2021eug} and \cite{Banerjee:2019tam}). In~\cite{Banerjee:2019tam} they were computed (see formulae (3.4), (3.26), (3.38)) as solutions of the type $\I_2$ shortening condition thought of as a ``classical field equation''.\footnote{In our language 
 the type $\I_1$ primary descendants trivially solve the type $\I_2$ shortening conditions.} For us they descend automatically from the necklace structure.

\paragraph{Subleading soft photon and subsubleading soft graviton}
In $d+2=4$ dimensions, the subleading soft photon ($\Delta=0$) and the subsubleading soft graviton ($\Delta=-1$) theorems  correspond to type III primary descendants in CCFT$_2$.
In even $d>2$ there are descendants of type $\III_k$ which become primary for $\Delta= k-\ell_k$~\cite{Penedones:2015aga}. We thus expect these subleading soft photon and subsubleading soft graviton theorems in even $d+2>4$ dimensions to map to correlation functions of conserved operators with primary descendants of type III$_1$ with $\ell_1=1$ and $\ell_1=2$, respectively, in CCFT$_{d>2}$. The explicit form of these primary descendants is not known and we leave their construction for future work.
The situation in odd~$d$ is even more tricky. There appears to be no primary descendant whose parent has the correct conformal dimension to match the conformally soft operators with $\Delta=0$ (for spin one) and $\Delta=-1$ (for spin two). We leave the resolution of this puzzle for the future.

\paragraph{Towers of ever more subleading soft theorems}
Conformally soft operators beyond the subleading soft theorem in gauge theory and the subsubleading soft theorem in gravity have type I$_1$ primary descendant operators at level $\kcal$.
Given a spin $\ell$ particle they appear at $\Delta=1-\ell-\kcal$ for $\kcal=1,2,\dots, \infty$, while in terms of the power expansion in $\omega$ they arise at $\omega^{k}$ for $k=\ell+\kcal-1$.
In appendix \ref{app:typeI} we describe why from such operators we cannot construct non-trivial charges.
However it would be interesting to see if the complete list of conformally soft operators (or their shadows) in $d>2$ forms an interesting algebra.

\paragraph{Fermionic symmetries}
While we have focused on bosonic symmetries here, it would be interesting to extend our $d>2$ discussion to fermionic symmetries. 
Infinite-dimensional fermionic symmetries were found  in~\cite{Dumitrescu:2015fej,Avery:2015iix,Lysov:2015jrs} to be implied by the soft gluino and gravitino theorems in $d+2=4$ spacetime dimensions. These fermionic symmetries were shown to be generated by CCFT$_2$ operators in~\cite{Fotopoulos:2020bqj,Pano:2021ewd}. In particular, large supersymmetry transformations are generated by the shadow transformed conformally soft gravitino primary operator. In higher dimensions we expect a classification of fermionic primary descendants to yield a $\Delta=\frac{3}{2}$ (shadow) operator that generates global supersymmetry transformations.

\paragraph{
Generalization to higher spin
}
As a proof of concept we can see how our technology can be used to study  soft particles with generic spin $\ell$ on the same footing.
Given a spin $\ell$ particle, we obtain a primary operator with dimension $\Delta$ and spin $\ell$.
Without doing any computations we expect $\ell$ soft theorems associated to type $\I_2$ primary descendants at level $\kcal=1,\dots,\ell$ which appear at dimensions $\Delta=2-\kcal$. 
The shortening condition of these operators is of the form
\be
\label{spinltypeI2}
O^{a^{1}_1 \dots a^{1}_\ell, a^{2}_1 \dots a^{2}_\kcal}_{\I_2,\kcal}(x)= \pi_{(\ell,\kcal)}({\bf a},{\bf b}) \p^{b^{2}_1} \dots \p^{b^{2}_\kcal} \Ocal^{b^1_1 \dots b^1_\ell}_{\Delta, \ell}(x) \, .
\ee
The first $\ell$ universal conformally soft theorems  are  exactly annihilated by the differential operators above and thus do not give rise to non-trivial conserved charges.
However the shadow transform of these $\I_2,\kcal$ operators gives rise to type $\II_1$ operators at level $\kcal'=\ell-\kcal+1$ which instead can be used to build non-trivial charges.

As an example, the leading soft factor, which generalizes the ones of photons  \eqref{SoftPhotond} and gravitons  \eqref{SoftGravitond} to spin $\ell$ particles,
is proportional to
\begin{equation}
\label{soft_spin_l}
\sum_{i=1}^N g_i  \frac{(\eta_i\omega_i)^{\ell-1}}{\omega}\frac{\left((x_i-x)\cdot\mathcal{e}\right)^\ell}{(x_i-x)^2}\, ,
\end{equation}
for some coefficients $g_i$.\footnote{
When $\ell=1$, $g_i=e \mathcal{Q}_i$ as in equation \eqref{SoftPhotond}. Because of Weinberg's soft theorem $g_i$ must be all equal for $\ell=2$, giving  $g_i=\kappa$ as  in  \eqref{SoftGravitond}.  Finally $g_i$ should vanish for $\ell>2$. Still it is an instructive exercise to show what the  soft operators and charges look like in the celestial basis in a closed form in $\ell$.}
We associate this contribution to the insertion of a conformally soft operator $\Ocal_{\Delta=1,\ell}$.
One can indeed check that~\eqref{soft_spin_l} is  annihilated by the differential operator in~\eqref{spinltypeI2} for $\kcal=1$. The resulting Ward identity does not produce contact terms and cannot be used to build non-trivial charges.
The shadow transform of $\Ocal_{\Delta=1,\ell}$ is the operator $\tilde \Ocal_{\Delta=d-1,\ell}$ which following the classification of section \ref{subsec:typeII} is of type  $\II_1,\kcal=\ell$. 
Indeed by taking the shadow of expression \eqref{soft_spin_l} using formula~\eqref{Shadow_typeS}, we obtain
\be
S\left[\frac{\left(x\cdot\mathcal{e}\right)^\ell}{x^2}\right]=\frac{2 (-1)^d \Gamma \left(\frac{d}{2}\right) \Gamma \left(\frac{d}{2}+\ell-1\right) }{(\ell-1)!} \frac{(\ecal \cdot x)^\ell}{|x|^{d}} \, .
\ee
It is easy to check that by taking a descendant of type $\II_1,\kcal=\ell$, this equation gives rise to a contact term (see equation \eqref{delta_gen_spin}),
\be
\label{Shortening_spin_ell}
(D_\ecal \cdot \p_x)^\ell  \frac{(\ecal \cdot x)^\ell}{|x|^{d}} \propto \delta^{(d)}(x) \, .
\ee
The resulting Ward identity for $\tilde \Ocal_{\Delta=d-1,\ell}$ can thus be  used to define non-trivial charges using equation  \eqref{TypeII_Charge}.
From formula \eqref{sympar}, without doing any computations, we can also predict that the number of such charges is finite and equals the dimension of the $SO(d+2)$ spin $\ell-1$ representation, namely $\frac{(d+2 \ell-2) \Gamma (d+\ell-1)}{\Gamma (d+1) \Gamma (\ell)}$. 
Finally, using \eqref{sympar}, \eqref{eq:Q_vol} and \eqref{Shortening_spin_ell} it is straightforward to see that the action of the (opportunely normalized) charges on a primary $\Ocal_{\Delta,\ell}$ takes the form
\be
[Q^{\mu_1 \dots \mu_{\ell-1}},\Ocal_{\Delta,\ell}(x)]=
g \, \eta^{\ell-1} q^{\{ \mu_1} \dots q^{\mu_{\ell-1} \}} \Ocal_{\Delta+\ell-1,\ell}(x) \, .
\ee
This can be considered as an example of the power of the classification and techniques that we introduced in this paper. 
Because of the structure of primary descendants we know exactly which computations we should perform and often we can also predict their result.

\section*{Acknowledgements}

We would like to thank Daniel Kapec, Alok Laddha and Prahar Mitra for useful discussions. AP and ET are supported by the European Research Council (ERC) under the European Union’s Horizon 2020 research and innovation programme (grant agreement No 852386). The work of YP is supported by the PhD track fellowship of Ecole Polytechnique.

\appendix{

\section{Projectors}
\label{app:projectors}
In this appendix we exemplify the form of the projectors. We will use the results of \cite{Costa:2016hju} where a number of projectors were computed. These were found in the contracted form $\pi_{(\ell_1,\ell_2,\ell_3,0,\dots,0)}(\ecal,\fcal)$ for  generic $\ell_1$ and for all $\ell_2,\ell_3$ such that $\ell_2+\ell_3\leq 4$. Following the notation of section \ref{sec:polarization_vectors}, the polarization vectors are defined by the sets $\ecal=\{\ecal_1,\dots,\ecal_{[d/2]}\}$, $\fcal=\{\fcal_1,\dots,\fcal_{[d/2]}\}$, however in this appendix we consider these vectors as unconstrained, dropping the hats to avoid clutter.
All projectors $\pi_{\ell}(\ecal,\fcal)$ are polynomials in the scalar products $(\ecal_i\cdot \ecal_j),(\ecal_i\cdot \fcal_j),(\fcal_i\cdot \fcal_j)$.
The results are in general quite lengthy but they can be readily used in computations in Mathematica.
The easiest example of such projectors is the symmetric and traceless one defined in \eqref{Pi_l}. The next to easiest example is the projector in the hook representation $\pi_{(\ell,1)}$,
\begin{align}
&\pi_{(\ell,1)}(\ecal,\fcal)=
\frac{(d-2) 2^{-\ell } \ell ! }{(\ell +1) (d+\ell -3) \left(\frac{d}{2}-1\right)_{\ell }}
\left| \fcal_1\right| {}^{\ell -2} \left| \ecal_1\right| {}^{\ell -2}
\times
\\
&\qquad\qquad \times \Bigg\{-(d-2) \left| \fcal_1\right|  \left| \ecal_1\right|  \left(\ecal_1\!\cdot\! \fcal_2 \ecal_2\!\cdot\! \fcal_1-\ecal_1\!\cdot\! \fcal_1 \ecal_2\!\cdot\! \fcal_2\right) C_{\ell -1}^{\left(\frac{d}{2}\right)}\left(\frac{\ecal_1\cdot \fcal_1}{\left| \ecal_1\right|  \left| \fcal_1\right| }\right) 
\nonumber \\
& \qquad\qquad \qquad  + d \bigg[
\fcal_1^2 \left(\ecal_1\!\cdot\! \ecal_2 \ecal_1\!\cdot\! \fcal_2-\ecal^2_1 \ecal_2\!\cdot\! \fcal_2\right)+\ecal^2_1 \fcal_1\!\cdot\! \fcal_2 \ecal_2\!\cdot\! \fcal_1+\ecal_2\!\cdot\! \fcal_2 \left(\ecal_1\!\cdot\! \fcal_1\right)^2
\nonumber \\
& \qquad\qquad \qquad\qquad \qquad
-\left(\ecal_1\!\cdot\! \fcal_2 \ecal_2\!\cdot\! \fcal_1+\fcal_1\!\cdot\! \fcal_2 \ecal_1\!\cdot\! \ecal_2\right) \ecal_1\!\cdot\! \fcal_1 \, .
\bigg]
 C_{\ell -2}^{\left(\frac{d}{2}+1\right)}\left(\frac{\ecal_1\cdot \fcal_1}{\left| \ecal_1\right|  \left| \fcal_1\right| }\right)\Bigg\}
 \, .
\nonumber 
\end{align}
For this paper it is enough to consider the projector $\pi_{(\ell,1)}$ in the cases $\ell=1,2$. The case $\pi_{(1,1)}$ is trivial and just gives antisymmetrization. In particular the contracted projector takes the form
\be
\pi_{(1,1)}(\ecal,\fcal)=\frac{1}{2} \left(\ecal_1\!\cdot\! \mathcal{f}_1 \ecal_2\!\cdot\! \mathcal{f}_2-\ecal_1\!\cdot\! \mathcal{f}_2 \ecal_2\!\cdot\! \mathcal{f}_1\right) \, ,
\ee
and similarly one recovers the indices by taking derivatives of the vectors as follows
\be
\pi_{(1,1)}({\bf a},{\bf b})=\partial_{\ecal_1}^{a^1_1}\partial_{\ecal_2}^{a^2_1}\partial_{\fcal_1}^{b^1_1}\partial_{\fcal_2}^{b^2_1}
\pi_{(1,1)}(\ecal_i,\fcal_i)
=\frac{1}{2} \left(\delta _{a_1^1,b_1^1} \delta _{a_1^2,b_1^2}-\delta _{a_1^1,b_1^2} \delta _{a_1^2,b_1^1}\right) \, .
\ee
For $\pi_{(2,1)}$ the situation  is more non-trivial. The contracted projector takes the form
\begin{align}
\pi_{(2,1)}(\ecal,\fcal)=&\frac{2}{3 (d-1)} \bigg((d-1) \ecal_2\!\cdot\! \mathcal{f}_2 \left(\ecal_1\!\cdot\! \mathcal{f}_1\right)^2+\mathcal{f}_1^2 \left(\ecal_1\!\cdot\! \ecal_2 \ecal_1\!\cdot\! \mathcal{f}_2-\ecal_1^2 \ecal_2\!\cdot\! \mathcal{f}_2\right)
\\
&
-\ecal_1\!\cdot\! \mathcal{f}_1 \big((d-1) \ecal_1\!\cdot\! \mathcal{f}_2 \ecal_2\!\cdot\! \mathcal{f}_1+\mathcal{f}_1\!\cdot\! \mathcal{f}_2 \ecal_1\!\cdot\! \ecal_2\big)+\ecal_1^2 \mathcal{f}_1\!\cdot\! \mathcal{f}_2 \ecal_2\!\cdot\! \mathcal{f}_1\bigg) \, .
\end{align}
The projector with open indices can be easily obtained by taking derivatives of all vectors $\ecal_i,\fcal_i$, namely
\be
\pi_{(2,1)}({\bf a},{\bf b})=\frac{1}{4}\partial_{\ecal_1}^{a^1_1}\partial_{\ecal_1}^{a^1_2}\partial_{\ecal_2}^{a^2_1}\partial_{\fcal_1}^{b^1_1}\partial_{\fcal_1}^{b^1_2}\partial_{\fcal_2}^{b^2_1}\pi_{(2,2)}(\ecal_i,\fcal_i) \, ,
\ee
where the factor of $1/4$ is included because there are two derivatives of the two vectors $\fcal_1$ which produce a factor of $2$ in the numerator, and similarly for $\ecal_1$.
The result can be easily obtained in Mathematica and it reads
\begin{align}
\label{pi21_indices}
\pi_{(2,1)}({\bf a},{\bf b})= &
\frac{-1}{6 (d-1)} \bigg[-2 (d-1) \delta _{a_1^2,b_1^2} \left(\delta _{a_1^1,b_2^1} \delta _{a_2^1,b_1^1}+\delta _{a_1^1,b_1^1} \delta _{a_2^1,b_2^1}\right)
\\
&
+\delta _{a_2^1,b_2^1} \left((d-1) \delta _{a_1^1,b_1^2} \delta _{a_1^2,b_1^1}+\delta _{a_1^1,a_1^2} \delta _{b_1^1,b_1^2}\right)
+\delta _{a_1^1,b_2^1} \left((d-1) \delta _{a_1^2,b_1^1} \delta _{a_2^1,b_1^2}+\delta _{a_1^2,a_2^1} \delta _{b_1^1,b_1^2}\right)
\nonumber \\
&
+\delta _{a_2^1,b_1^1} \left((d-1) \delta _{a_1^1,b_1^2} \delta _{a_1^2,b_2^1}+\delta _{a_1^1,a_1^2} \delta _{b_1^2,b_2^1}\right)
+\delta _{a_1^1,b_1^1} \left((d-1) \delta _{a_1^2,b_2^1} \delta _{a_2^1,b_1^2}+\delta _{a_1^2,a_2^1} \delta _{b_1^2,b_2^1}\right)
\nonumber \\
&
-2 \delta _{b_1^1,b_2^1} (\delta _{a_1^2,a_2^1} \delta _{a_1^1,b_1^2}
+ \delta _{a_1^1,a_1^2}  \delta _{a_2^1,b_1^2})
-2 \delta _{a_1^1,a_2^1} \left(\delta _{b_1^1,b_1^2} \delta _{a_1^2,b_2^1}-2 \delta _{b_1^1,b_2^1} \delta _{a_1^2,b_1^2}+\delta _{b_1^2,b_2^1} \delta _{a_1^2,b_1^1}\right)
\bigg]
\nonumber \, .
\end{align}
In this form one can easily check that taking traces over the indices $a$ or over the indices $b$ gives zero. Also one can check that $\pi_{(2,1)}({\bf a},{\bf b})\pi_{(2,1)}({\bf b},{\bf c})=\pi_{(2,1)}({\bf a},{\bf c})$.

In \cite{Costa:2016hju} a result for the projector $\pi_{(\ell,2)}$ is also presented, but we do not report it here since it is too lengthy.
For this work we only need the simplest of such projectors, namely $\pi_{(2,2)}$, which in its contracted form reads
\begin{align}
&\pi_{(2,2)}(\ecal,\fcal)=
\frac{(d-1)}{3 (d-2) (d-1)}\bigg[ \left(\ecal_1\!\cdot\! \mathcal{f}_1\right)^2 \left((d-2) \left(\ecal_2\!\cdot\! \mathcal{f}_2\right)^2-\mathcal{f}_2^2 \ecal_2^2\right)
\nonumber \\
&
-2 (d-1) \ecal_1\!\cdot\! \ecal_2 \left(\mathcal{f}_1\!\cdot\! \mathcal{f}_2 \ecal_2\!\cdot\! \mathcal{f}_2-\mathcal{f}_2^2 \ecal_2\!\cdot\! \mathcal{f}_1\right) \ecal_1\!\cdot\! \mathcal{f}_1
+(d-1) \left(\ecal_1\!\cdot\! \mathcal{f}_2\right)^2 \left((d-2) \left(\ecal_2\!\cdot\! \mathcal{f}_1\right)^2-\mathcal{f}_1^2 \ecal_2^2\right)
\nonumber \\
&
+2 d \ecal_1^2 \mathcal{f}_1\!\cdot\! \mathcal{f}_2 \ecal_2\!\cdot\! \mathcal{f}_1 \ecal_2\!\cdot\! \mathcal{f}_2
-d \mathcal{f}_1^2 \ecal_1^2 \left(\ecal_2\!\cdot\! \mathcal{f}_2\right)^2
+\mathcal{f}_2^2 \left(-(d-1) \ecal_1^2 \left(\ecal_2\!\cdot\! \mathcal{f}_1\right)^2
-2 \mathcal{f}_1^2 \left(\left(\ecal_1\!\cdot\! \ecal_2\right)^2-\ecal_1^2 \ecal_2^2\right)\right)
\nonumber \\
&
-2 (d-1) \ecal_1\!\cdot\! \mathcal{f}_2 \left(\ecal_1\!\cdot\! \mathcal{f}_1 \left((d-2) \ecal_2\!\cdot\! \mathcal{f}_1 \ecal_2\!\cdot\! \mathcal{f}_2
-\ecal_2^2 \mathcal{f}_1\!\cdot\! \mathcal{f}_2\right)+\ecal_1\!\cdot\! \ecal_2 \left(\mathcal{f}_1\!\cdot\! \mathcal{f}_2 \ecal_2\!\cdot\! \mathcal{f}_1-\mathcal{f}_1^2 \ecal_2\!\cdot\! \mathcal{f}_2\right)\right)
\nonumber \\
&
+2 \left(\mathcal{f}_1\!\cdot\! \mathcal{f}_2\right)^2 \left(\ecal_1\!\cdot\! \ecal_2\right)^2-2 \ecal_1^2 \mathcal{f}_1\!\cdot\! \mathcal{f}_2 \ecal_2\!\cdot\! \mathcal{f}_1 \ecal_2\!\cdot\! \mathcal{f}_2
-2 \ecal_1^2 \ecal_2^2 \left(\mathcal{f}_1\!\cdot\! \mathcal{f}_2\right)^2
+\mathcal{f}_1^2 \ecal_1^2 \left(\ecal_2\!\cdot\! \mathcal{f}_2\right)^2
\bigg] \, .
\end{align}
To define the projector with open indices it again suffices to take derivatives of all the vectors,
\be
\pi_{(2,2)}({\bf a},{\bf b})=\frac{1}{16}\partial_{\ecal_1}^{a^1_1}\partial_{\ecal_1}^{a^1_2}\partial_{\ecal_2}^{a^2_1}\partial_{\ecal_2}^{a^2_2}\partial_{\fcal_1}^{b^1_1}\partial_{\fcal_1}^{b^1_2}\partial_{\fcal_2}^{b^2_1}\partial_{\fcal_2}^{b^2_2}\pi_{(2,2)}(\ecal_i,\fcal_i) \, .
\ee
The result can be easily obtained in Mathematica but it is too lengthy to present here.

While for this paper we restricted to $\ell_1\leq 2$ because we are interested in photons and gravitons, for higher spin particles one would 
also need the projectors with $\ell_1> 2$.

\section{Type I Operators and Trivial Charges}
\label{app:typeI}
In this appendix we want to show that type I operators are associated to trivial charges. 

Let us first explain what we mean by trivial charges. 
Let us consider an operator $\Ocal(x)$ with a primary descendant $D \Ocal(x)$, where $D$ is some differential operator that creates the descendant. Now let us assume that 
\be
\label{trivial WI}
\langle (D \Ocal(x)) \Ocal_1(x_1) \dots  \Ocal_N(x_N) \rangle =0\,,
\ee
exactly without any contact term. 
We call \eqref{trivial WI} trivial Ward identity in contrast  with the usual Ward identity \eqref{Ward_id}, where delta functions are present on the right-hand-side.
Now because of the absence of contact terms it is easy to see that no non-trivial charges can be defined. Indeed the interesting feature of the charges is that their insertion in a correlation function equals the variation of the operators which is computed through the integral of the delta functions on the right-hand-side of \eqref{Ward_id}. When these are absent, the variations become trivial and so do the charges themselves. 
We thus conclud that \eqref{trivial WI} gives rise to trivial charges.

To clarify the discussion let us present the simplest example of a trivial charge, the one associated to the identity operator. 
The identity operator is a primary with a level-one primary descendant of type $\I_1$, i.e. which takes the form $\partial_a \mathbb{1}$. We want to show that from this operator, as expected, it is not possible to build non-trivial conserved charges.

Using our construction of section \ref{sec:Charges_I}, one can in principle define a Noether current 
$
    \Jcal^a(x) = \epsilon^{a}(x) \mathbb{1}
$, with $\p_a \epsilon^{a}(x)=0$, which has a constant solution $\epsilon^{a}(x)=c^{a}$.\footnote{One could also consider the case in which the parameter is not smooth $\epsilon^{a}(x)\propto x^a |x|^{-d}$ which satisfies $\p_a \epsilon^{a}(x) \propto \delta^{(d)}(x) $. In this case $Q^a_\Sigma$ is equal to a constant if $\Sigma$  contains the origin and vanishes otherwise. Thus the result is still trivial. } One can then proceed in  defining
a set of topological charges $Q^\epsilon_\Sigma=c_{a} Q^{a}_\Sigma$ with
\be{}
Q^a_\Sigma \equiv \int_\Sigma dS^a  \mathbb{1}  \, .
\ee{}
It is easy to see that $Q^a_\Sigma=0$ because of rotation invariance (e.g. by choosing $\Sigma$ to be the surface of a sphere). In this case the triviality of the charge is obtained by saying that $Q^a_\Sigma$ vanishes.

Let us also see how this can be obtained from the Ward identity \eqref{trivial WI}.
Of course since $\mathbb{1}$ does not depend on an insertion point,
\be{}
\langle (\partial_a \mathbb{1}) \Ocal_1(x_1) \dots  \Ocal_N(x_N) \rangle =0\,.
\ee{}
Now we integrate this  Ward identity over the volume of a sphere and we also find that the insertion of the charge $Q^a_\Sigma$ (defined by integrating the left hand side) acts like zero (by integrating the right-hand-side). Thus, as expected, we obtain the same result as above. 

In this case it is automatic to know that $\partial_a \mathbb{1}=0$. Conversely for other operators the fact that $D \Ocal(x)=0$ inside a correlation function must be proven. This is what we plan to do in the rest of the appendix.
There are two strategies.
For (standard) CFTs we will  study the $N$-point functions with the insertion of $\Ocal(x)$ and prove that they are polynomial in $x$, and thus they are annihilated by $D$ (since derivatives of a polynomial cannot give contact terms).
For celestial CFTs we will make use of the wavefunction formalism to define the operators and show that, since the corresponding wavefunction vanishes, then also $D \Ocal(x)$ must vanish when inserted in a correlation function.
 
\subsection*{Trivial Type I Ward Identities 
in CFTs 
}
In this subsection we show that equation \eqref{trivial WI} holds in CFTs, by proving that correlation functions with the insertion of type I operators are polynomial  as a function of their insertion point.

Let us start with the case of $d=2$. The idea of the proof is simple. We consider an $N$-point function with a type I operator inserted at  $z,\bar z$. This function must satisfy conservation equations in both $z$ and $\bar z$ (type Ia and Ib) which imply that the function is polynomial in these variables (and thus it cannot give rise to contact terms from the conservation equations). Let us however give a more concrete  argument, by considering the OPE of type I operators which is fixed by the three-point function and can be explicitly written down. 
We consider a generic three-point function in a usual CFT$_2$ where the operator $\Ocal$ has $h=\frac{1-k}{2}$ and $\bar h=\frac{1- \bar k}{2}$. For simplicity, since the holomorphic and antiholomorphic behaviours factorize and are of the same form we consider only the holomorphic part (but we should remember in the end to include the antiholomorphic one). The holomorphic dependence of the three-point function on the primaries $\Ocal,\Ocal_2,\Ocal_3$ is (here we consider usual three-point functions which are not distributions in $z_i$) 
\be
\langle \Ocal(z_1)\Ocal_2(z_2)\Ocal_3(z_3)\rangle= c_{123}\frac{1}{z_{12}^{h+h_2-h_3}z_{13}^{h+h_3-h_2}z_{23}^{h_2+h_3-h}} \, .
\ee
Now we consider $h=\frac{1-k}{2}$ and we ask for $\partial^{k} \Ocal=0$ (keeping all the $z_i$ distinct, which means that the result is zero away from contact points). It is easy to see that this shortening condition implies that $h_3-h_2=\frac{1-k}{2}+i$ with $i=0,\dots k-1$. 
We thus get
\be
\langle \Ocal(z_1)\Ocal_2(z_2)\Ocal_3(z_3)\rangle= c_{123} \delta_{h_3-h_2-\frac{1-k}{2}-i}
z_{12}^{i}z_{13}^{k-1-i}z_{23}^{-2h_3+i+1-k}
\, .
\ee
The result is clearly polynomial in $z_1$ for any allowed value of $i$.
We can then expand this expression e.g. for $z_1\to z_2$ using the binomial $z_{13}^{a}=(z_{12}+z_{23})^{a}=
\sum_{b=0}^{a} \binom{a}{b} z_{12}^{b} z_{23}^{a-b}
$.
The result is 
\be
\langle \Ocal(z_1)\Ocal_2(z_2)\Ocal_3(z_3)\rangle= c_{123} \delta_{h_3-h_2-\frac{1-k}{2}-i}
\sum_{j=0}^{k-1-i} \binom{k-1-i}{j} z_{12}^{i+j} 
z_{23}^{-2h_3-j}\, .
\ee
The OPE can thus be recast as
\be
\Ocal(z)\Ocal_2(0) \sim c_{123} \delta_{h_3-h_2-\frac{1-k}{2}-i}
\sum_{j=0}^{k-1-i} \binom{k-1-i}{j}  \frac{\Gamma(1-2h_3)}{\Gamma(1-2h_3-j)} z^{i+j} \p^j \Ocal_3(0) \, .
\ee
This OPE is polynomial in $z$ of order $k-1$, thus by applying $\p_z^k$ to this expression we find exactly zero. No contact term can be generated. 
Notice that this would also happen when the operators depend on the antiholomorphic coordinates. Starting with $\Ocal$ with 
$h=\frac{1-k}{2}$ and $\bar h=\frac{1- \bar k}{2}$ and conservations $\p^k\Ocal=0={\bar \p}^{\bar k}\Ocal $
we find
\begin{align}
\Ocal(z,\bz)\Ocal_2(0,0) \sim 
&
c_{123} \delta_{h_3-h_2-\frac{1-k}{2}-i}\delta_{\bar h_3-\bar h_2-\frac{1-\bar k}{2}-\bar i}
\sum_{j=0}^{k-1-i}\sum_{\bar j=0}^{\bar k-1-\bar i} \binom{k-1-i}{j} 
\binom{\bar k-1-\bar i}{\bar j}
\times
\nonumber \\
&
\qquad \times \frac{\Gamma(1-2h_3)}{\Gamma(1-2h_3-j)}
\frac{\Gamma(1-2 \bar h_3)}{\Gamma(1-2 \bar h_3-\bar j)}
z^{i+j}\bz^{\bar i+\bar j} \p^j \bar\p^{\bar j}  \Ocal_3(0,0) \, ,
\end{align}
where $\bar i=0,\dots \bar k -1$.
The crucial observation is that no negative powers of $z$ or $\bz$ are present in this OPE thus there is no possible way to get a delta function by applying derivatives in $\p_z$ and $\p_\bz$.
We therefore conclude that when we insert the operator $\Ocal(z,\bz)$ in any $N$-point function we still get that the dependence in $z$ and $\bz$ is polynomial. Therefore
\be
\langle (\p^k \Ocal(z,\bz)) \Ocal_1 \dots \Ocal_{N}  \rangle = 0 \, ,
\ee
without contact terms and similarly for ${\bar \p}^{\bar k} \Ocal(z,\bz)$.

For this computation we required minimal assumptions like the usual form of the three-point function which is fixed by symmetry, and the fact that the $\Ocal$ is a type I operator which thus satisfies the type I shortening conditions. Notice that for operators of type II and type III the same logic does not work because these have shortening conditions which only involve either $\p_z$ or  $\p_\bz$, but not both of them. To ensure a polynomial behavior it is instead necessary to have shortening conditions in both variables.

Let us now turn to $d>2$.
As for $d=2$, the idea behind the demonstration is that an $N$-point correlation function with a type I operator inserted  at a position $x^a$ should have a polynomial dependence in $x^a$ because it has to satisfies the type I conservation equation (one can project the indices of the $d>2$ type I conservation equation into a plane, e.g. using coordinates $x^1\pm i x^2$, and use a similar argument as in $d=2$). Again we find it useful to rephrase this argument in terms of three-point functions and OPE, where all computations can be explicitly performed. 
To exemplify why the OPE of type I operators is polynomial let us consider the next to trivial example after the identity. We take a vector operator with a type  $\I_1,\kcal=1$ primary descendant. This operator must have $\Delta=-1$. 
The OPE of such an operator is fixed by its three-point functions with all other two operators. For simplicity let us consider the example of the three-point function with two scalar operators, 
\be
\langle \Ocal^{a}_{\Delta=-1}(x_1) \Ocal_{\Delta_2}(x_2) \Ocal_{\Delta_3}(x_3) \rangle \propto \frac{x_{12}^{a} x_{13}^2-x_{13}^{a} x_{12}^2}{|x_{12}|^{\Delta_{2}-\Delta_{3}} 
|x_{13}|^{\Delta_{3}-\Delta_{2}} |x_{23}|^{\Delta_{2}+\Delta_{3}+2}
}\, .
\ee
We require that $\p^{\{b}_{\phantom{\Delta}}
\Ocal^{a\}}_{\Delta=-1}(x_1)=0$ away from possible contact terms.
It is easy to see that this requirement
implies that the three-point function vanishes unless $\Delta_{2}=\Delta_{3}$, and therefore
that the three-point function is a polynomial in the distances $x_{12}$ and $x_{13}$.
This means that the OPE takes the simple form
\be
\Ocal^{a}_{\Delta=-1}(x)\Ocal_{\Delta_2}(0) \sim \delta_{\Delta_2, \Delta_3 }\big[x^a-\frac{1}{2\Delta_{2}}(2x^a x^b-x^2 \delta^{a b})\p_b\big] \Ocal_{\Delta_3 }(0) \, ,
\ee
where the square bracket contains all the possible descendants. We clearly see that the square bracket is a polynomial in $x$.
Of course if we now take 
$\p^{\{b}_{\phantom{\Delta}}
\Ocal^{a\}}_{\Delta=-1}(x)$ on the OPE we get exactly zero and no possible contact terms can be generated. By considering spinning operators $\Ocal_2$ and $\Ocal_3$ it is easy to see that similar results are obtained. 
Therefore we proved that 
\be
\left \langle \left(\p^{\{b}_{\phantom{\Delta}}
\Ocal^{a\}}_{\Delta=-1}(x)\right) \Ocal_1(x_1) \dots \Ocal_N(x_N) \right\rangle =0 \, ,
\ee
without any contact term on the right-hand-side.

This demonstration can be repeated for different spins $\ell$ of the operator $\Ocal$ and 
different levels $\kcal$ of the type I shortening condition.

\subsection*{Trivial Type I Ward Identities from CCFT Wavefunctions }
In this subsection we want to show that equation \eqref{trivial WI} can be obtained in CCFTs by studying the wavefunctions associated to the operators of type I.

To start we quickly  review the wavefunction formalism in CCFTs.
The CCFT operators $\Ocal_{\Delta \ell}(x) $ can be obtained \cite{Donnay:2020guq} as a convolution of usual QFT$_{d+2}$ operators $O(X)$ in position space with so called ``primary'' wavefunctions
$f_{\Delta,\ell}(X;q)$ \cite{Pasterski:2017kqt}, 
\be
\label{Ops_BulkToBdy}
\Ocal_{\Delta \ell}(q) = i \left(O(X),f_{\Delta,\ell}(X;q) \right) \, ,
\ee
where the inner product $(\cdot,\cdot)$ is given by a standard integral in $X^{\mu}$, e.g. for spin zero this is related to the Klein Gordon inner product.
The wavefunctions $f_{\Delta,\ell}$ must satisfy a set of conditions (e.g. they should  transform as $d$ dimensional primary operators in CCFT) which allows one to classify them.
For this appendix we consider the most standard (``radiative'') wavefunctions $f_{\Delta,\ell}$ which are defined when $\ell$ is equal to the spin of the bulk primary $O(X)$ (generalized wavefunctions can also be defined \cite{Pasterski:2020pdk, Pasterski:2021fjn}, but will not play a role here).

Equation \eqref{Ops_BulkToBdy} generates operators inserted at a point $q^{\mu}$ in the embedding space which can then be projected to any given section, e.g. the Poincaré section $q^{\mu} \to x^a$ \eqref{qPoincare}.
In this formulation the only dependence on the CFT$_d$ coordinate $x^a$ is encoded in the wavefunction. 
The idea of \cite{Pasterski:2021fjn} is that in order to classify the types of operators one can  simply classify the respective wavefunctions. 
E.g. for a bulk scalar operator, the correspondent wavefunction takes the form $f_{\Delta,\ell=0}(X,q)\propto  (X \cdot q)^{-\Delta}$. 
It is very easy to classify the shortening conditions of such wavefunctions. E.g. when $\Delta=0$ the wavefunction $f_{\Delta,\ell=0}$  becomes constant and is annihilated by $\p^a$. This means that the operator $\p^a \Ocal_{\Delta=0,\ell=0}$ inserted in a celestial correlator is computed by the convolution of a vanishing wavefunction, and thus it is zero. 

In \cite{Pasterski:2021fjn} it was shown that in $d=2$ all wavefunctions of type I, are  annihilated by the differential operators that create the primary descendants. This is in contrast with the wavefunctions of type II and III where the differential operators give new (generalized) wavefunctions which are not vanishing.
This is in perfect agreement with the fact that type I operators have trivial Ward identities while type II and III have non-trivial Ward identities with delta functions on the right-hand-side.

Here we want to show using the same CCFT wavefunction approach
that also in $d>2$ the
Ward identities for type $\I_1,\kcal$ operators are trivial.
To this end we introduce the massless radiative wavefunctions in $d>2$. These take the form of spin $\ell$ bulk-to-boundary propagators\cite{Pasterski:2017kqt, Costa:2014kfa},
\be
\label{BulkBdy_Prop}
f_{\Delta,\ell}(X,Z;q,\varepsilon) \propto \frac{[(X\cdot q)(Z\cdot \varepsilon)-(X \cdot \varepsilon)(q \cdot Z)]^\ell}{(X\cdot q)^{\Delta+\ell}}  \, ,
\ee
where $Z$ is a polarization vector that contracts the indices of the bulk operator $O(X)$, while $\varepsilon$ is a polarization vector for the  CCFT operator in embedding space $\Ocal_{\Delta,\ell}(q,\varepsilon)$.

It is easy to see that, after setting $q$ and $\varepsilon$ to the Poincaré section \eqref{qPoincare} and \eqref{epsPoincare},  the following relation holds
\be\label{shorteningWF}
(\ecal \cdot \p_x)^{\kcal} f_{\Delta=1-\ell-\kcal,\ell}(X,Z;q,\varepsilon) = 0 \, .
\ee
To prove \eqref{shorteningWF} we use the relations $(\ecal\cdot \partial_x) q^\mu =\varepsilon^{\mu}$ and $(\ecal\cdot \partial_x) \varepsilon^\mu =0$ which imply that $(\ecal\cdot \partial_x)$ annihilates the term inside the square parentheses in \eqref{BulkBdy_Prop}. The rest of the proof is basically that $(\ecal\cdot \partial_x)^\kcal (X\cdot q)^{\kcal-1}=0$  because we are taking $\kcal$ derivative of a polynomial of order $\kcal-1$.
Therefore we conclude that the type $\I_{k=1},\kcal$ shortening condition annihilates the wavefunction.
This implies that the associated CCFT operators constructed from \eqref{Ops_BulkToBdy} must satisfy a trivial Ward identity of the form \eqref{trivial WI}.
For completeness let us also mention that \eqref{shorteningWF} is special to type I  and that we checked that for other types (like II or S) the shortening of the wavefunction gives rise to non vanishing contributions proportional to  (generalized) primary wavefunctions.

Finally let us mention that in section \ref{sec:CCFTdCWardId} we have various examples of type $\I_2$ operators inserted in a correlation function (these are the primary descendants of soft operators) and we can explicitly see that also these  operators  have trivial Ward identities. 
For all such operators with trivial Ward identities we expect  the associated charges to be trivial.

\section{Ward Identities and Dirac Delta Distribution}
\label{app:deltadist}
A more careful treatment of the conservation equations for conformally soft shadow operators shows that the statements 
\begin{equation}
    \p_a \tilde S^{(0),a}=0\,, \quad \p_a \p_b \tilde S^{(0),ab}=0\,, \quad \p_a \tilde S^{(1),ab}=0
\end{equation}
only hold up to contact terms. Indeed accounting for the Dirac  delta distributions is necessary to obtain standard Ward identities. This is the purpose of this appendix.

The leading soft photon and graviton cases can be treated in one go since $\p_a \tilde S^{(0),ab}$ and  $\tilde S^{(0),b}$ have the same behaviour.
We need to show that $\p_a\left(\frac{x^a}{x^d}\right)$ has distributional support. To do so we define the regulated expression
\be\label{der}
\p_a \left(\frac{x^a}{(x^2)^\hd}\right)=\lim_{\varepsilon\to 0}\p_a\left(\frac{x^a}{(x^2+\varepsilon^2)^{\hd}}\right)=\lim_{\varepsilon\to 0}d\frac{\varepsilon^2}{(x^2+\varepsilon^2)^{\hd+1}}\,,
\ee
Integrating \eqref{der} against a test function $f$, we obtain
\be
\begin{aligned}
\int_{\mathbb R ^d}d^d x\, \lim_{\varepsilon\to 0}d\frac{\varepsilon^2}{(x^2+\varepsilon^2)^{\hd+1}}f(x)=d\int_{S^{d-1}}d\Omega\int_0^\infty dr \, \frac{r^{d-1}}{(r^2+1)^{\hd+1}}\lim_{\varepsilon\to 0}f(\varepsilon x)\, ,
\end{aligned}
\ee
where $\Omega$ denotes the solid angle in $d$ dimensions, $r$ is the radial direction and we have performed the change of variable $x^a\to \varepsilon x^a$. Taking the limit $\varepsilon \to 0$, the above integral yields 
\be
\int_{S^{d-1}}d\Omega f(0)=\frac{2\pi^{\hd}}{\Gamma\left(\hd\right)}f(0)\, .
\ee
Therefore, we find
\be\label{del1}
\p_a \left(\frac{x^a}{(x^2)^\hd}\right)=\frac{2\pi^{\hd}}{\Gamma\left(\hd\right)} \delta^{(d)}(x)\, .
\ee
Following the same logic we can generalize this formula to any spin as 
\be 
\label{delta_gen_spin}
\frac{1}{(-1)^\ell \ell! \left(-\frac{d }{2}-\ell+2\right)_\ell}(\p_x \cdot D_{\ecal})^\ell \frac{(\ecal \cdot x)^\ell}{(x^2)^{\frac{d}{2}}} = \frac{\pi ^{d/2} (-1)^\ell  (\ell-1)!  (d-2)_\ell \left(\frac{d}{2}\right)_\ell}{
\Gamma \left(\frac{d}{2}+\ell\right)
\left(-\frac{d}{2}-\ell+2\right)_\ell 
} \delta^{(d)} (x) \, .
\ee

For the subleading soft graviton theorem, the situation is more tricky.  Taking a derivative of the shadow transformed subleading soft graviton factor \eqref{ShadowSubGravd} we define
\be\label{T}
\begin{aligned}
V^b(x)\equiv \p_a \bigg[&
\frac{\p_{x_i}^{(a}x^{b)}}{(x^2)^{\hd}}+\frac{1}{2}(d-2)x\cdot \p_{x_i}\frac{x^a x^b}{(x^2)^{\hd+1}}-\frac{1}{2}x\cdot \p_{x_i}\frac{\delta^{ab}}{(x^2)^\hd}\\
&-\omega_i\p_{\omega_i}\left(\frac{x^a x^b-\frac{\delta^{ab}}{d}x^2 }{(x^2)^{\hd+1}}\right)
+i(d-1) \frac{x_c M_i^{c \, (a} x^{b)}  }{(x^2)^{\hd+1}}
\bigg]\, ,
\end{aligned}
\ee
where we dropped various prefactors which are irrelevant for this calculation and we set $x_i=0$ for simplicity. In the above formula the round brackets denote a symmetrization of the indices.
The terms in the second line of \eqref{T} are proportional to the derivative of the Dirac distribution. We can see this as follows.
For the $\omega_i \p_{\omega_i}$ term we have
\be
\begin{aligned}
\p_a \left(\frac{x^a x^b-\frac{\delta^{ab}}{d}x^2 }{(x^2)^{\hd+1}}\right)&=\lim_{\varepsilon\to 0}\p_a \left(\frac{x^a x^b-\frac{\delta^{ab}}{d}x^2 }{(x^2+\varepsilon^2)^{\hd+1}}\right)
\\&=\lim_{\varepsilon\to 0}\frac{(d+2)(d-1)}{d}\frac{x^b \varepsilon^2}{(x^2+\varepsilon^2)^{\hd+2}}\,.
\end{aligned}
\ee
Noticing that $(d+2)\frac{x^b}{(x^2+\varepsilon^2)^{\hd+2}}=-\partial^b\frac{1}{(x^2+\varepsilon^2)^{\hd+1}}$ and integrating against a test function we obtain
\be
\begin{aligned}
-\frac{d-1}{d}\int_{\mathbb R^d}d^d x\, \lim_{\varepsilon\to 0}\p^b\frac{\varepsilon^2}{(x^2+\varepsilon^2)^{\hd+1}}f(x)&=\frac{d-1}{d}\int_{\mathbb R^d}d^dx\, \frac{1}{(x^2+1)^{\hd+1}}\lim_{\varepsilon\to 0} \frac{\p}{\p(\varepsilon x_b)}f(\varepsilon x)
\\&=\frac{d-1}{d^2}\frac{2\pi^\hd}{\Gamma\left(\hd\right)}\left(\p^b f\right)(0)\, .
\end{aligned}
\ee
Therefore we have the following relation
\be\label{del2}
\p_a\left(\frac{x^a x^b-\frac{\delta^{ab}}{d}x^2}{x^{d+2}}\right)=-\frac{d-1}{d^2}\frac{2\pi^\hd}{\Gamma\left(\hd\right)}\p^b \delta^{(d)}(x)\, .
\ee
For the $M_i$ term we have
\be
\begin{aligned}
\lim_{\varepsilon\to 0}\p_a \frac{-x_c M_i^{c(a}  x^{b)}}{(x^2+\varepsilon^2)^{\hd+1}}&=\frac{1}{2}(d+2){M_i}^{b c}\lim_{\varepsilon\to 0}\frac{\varepsilon^2 x_c}{(x^2+\varepsilon^2)^{\hd+2}}\\
&=-\frac{1}{2d} M_i^{bc}\lim_{\varepsilon\to 0}\p_c\frac{d\varepsilon^2}{(x^2+\varepsilon^2)^{\hd+1}}\, .
\end{aligned}
\ee
Then using \eqref{der}, we obtain
\be
\lim_{\varepsilon\to 0}\p_a \frac{-x_c M_i^{c(a} x^{b)}}{(x^2+\varepsilon^2)^{\hd+1}}=-\frac{\pi^\hd}{d\Gamma\left(\hd\right)}M_i^{bc}\p_c \delta^{(d)}(x)\, .
\ee
Having shown that the second line of \eqref{T} gives a derivative of the Dirac  delta distribution, it remains to show that the first line of \eqref{T} gives a  delta distribution (not a derivative thereof). The first line of \eqref{T} can be written as 
\be
\lim_{\varepsilon\to 0} \frac{1}{2}\left[ \frac{d \varepsilon^2}{(x^2+\varepsilon^2)^{\hd+1}}\p_{x_i}^b+(d^2-4)\frac{\varepsilon^2 x^b x^c}{(x^2+\varepsilon^2)^{\hd+2}}\p_{x_i\, c}\right]\, .
\ee
The first term can be evaluated using~\eqref{der}. For the second term, we have that
\be
\frac{\varepsilon^2 x^b x^c}{(x^2+\varepsilon^2)^{\hd+2}}=-\frac{1}{d+2}\left[-\frac{1}{d}\p^b\p^c\frac{\varepsilon^2}{(x^2+\varepsilon^2)^\hd}-\frac{\varepsilon^2 \delta^{bc}}{(x^2+\varepsilon^2)^{\hd+1}}\right]\, .
\ee
The second term in the above equation can be clearly evaluated using equation \eqref{der}. The first term vanishes identically in the limit $\varepsilon\to 0$. 

Putting everything together we land on the result 
\be\label{Tfinal}
V^b(x)=(d-1)\frac{\pi^{\hd}}{\Gamma\left(\hd+1\right)}\left(\delta^{(d)}(x)\p_{x_i}^b-\frac{1}{d}\omega_i \p_{\omega_i}\p^b \delta^{(d)}(x)+\frac{i}{2}M_i^{bc}\p_c \delta^{(d)}(x)\right)\, .
\ee

\section{Two-Point Integral in CFT$_d$ Space}\label{app:2pt}

The computations of the shadow transformed soft factors in section~\ref{sec:CCFTdSHCWardId} and appendix~\ref{app:Shadow} make use of the following result for two-point integrals in CFT$_d$ space
\be\label{twoptint}
\int_{\mathbb{R}^d}\frac{d^d x_1}{(x_{12}^2)^\alpha(x_{13}^2)^\beta}=\frac{\pi^{\hd}\Gamma\left(\a+\b-\hd\right)\Gamma\left(\hd-\a\right)\Gamma\left(\hd-\b\right)}{\Gamma(\a)\Gamma(\b)\Gamma(d-\a-\b)}\frac{1}{(x_{23}^2)^{\a+\b-\hd}}\, ,
\ee
where $x_{12}^a\equiv(x_1-x_2)^a$. To obtain this result note that the shift $x_1\rightarrow x_1+x_2$ in~\eqref{twoptint} is inconsequential and we may as well compute 
\begin{equation}
    I(x_2,x_3)\equiv \int_{\mathbb{R}^d}\frac{d^d x_1}{(x_1^2)^\a \left((x_1+x_{23})^2\right)^\b}\,.
\end{equation}
Using the Feynman-Schwinger parametrization and going to spherical coordinates we find
\be
\begin{aligned}
I(x_2,x_3)&=\frac{\Gamma(\a+\b)}{\Gamma(\a)\Gamma(\b)}\int_0^1 dt\, t^{\a-1}(1-t)^{\b-1}\int_{\mathbb{R}^d}d^d x_1\, \frac{1}{\left[\left(x_1+(1-t)x_{23}\right)^2+t(1-t)x_{23}^2\right]^{\a+\b}}\\
&=\frac{\Gamma(\a+\b)}{\Gamma(\a)\Gamma(\b)}\int_0^1 dt\, t^{\a-1}(1-t)^{\b-1}\int d\Omega_d \int_0^\infty dr \frac{r^{d-1}}{\left[r^2+t(1-t)x_{23}^2\right]^{\a+\b}}\\
&=\frac{\pi^{\hd}\Gamma(\a+\b-\hd)}{\Gamma(\a)\Gamma(\b)}(x_{23}^2)^{\hd-\a-\b}\int_0^1 dt\, t^{\hd-\b-1}(1-t)^{\hd-\a-1}\\
&=\frac{\pi^{\hd}\Gamma(\a+\b-\hd)\Gamma\left(\hd-\a\right)\Gamma\left(\hd-\b\right)}{\Gamma(\a)\Gamma(\b)\Gamma(d-\a-\b)}\frac{1}{(x_{23}^2)^{\a+\b-\hd}}
\end{aligned}
\ee
which is the result on the right-hand-side of~\eqref{twoptint}.

\section{Shadow Transforms of Soft Factors}
\label{app:Shadow}

In this appendix we give a complementary derivation of the shadow transformed soft theorems that applies for any spacetime dimension $d$ -- that is for both even and odd. While in section~\ref{sec:CCFTdSHCWardId} we have worked in the index-free notation, here we give the shadow transformed soft photon and graviton factors in their alternative form with the indices made explicit.

\subsection*{Leading Soft Photon}\label{LSP}
The shadow transform of the soft photon operator expressed in index form is
\be
\tilde R_1^a(x)=\lim_{\Delta\to 1}N_{\Delta,1}\int_{\mathbb R^d}d^d y \frac{I^{ab}(x-y)}{\left[(x-y)^2\right]^{d-\Delta}}R_{1\,b}(y)\,,
\ee
where $I^{ab}(x)=\delta^{ab}-2\frac{x^a x^b}{x^2}$. 
To obtain the shadow transformed soft photon theorem we need to compute the shadow of the soft photon factor \eqref{SoftPhotond} which is given by
\be\label{photonshadow}
\tilde S_p^{(0)\,a}(\omega,x)=-2e\sum_{i=1}^N \mathcal Q_i \frac{1}{\omega} \lim_{\Delta\to 1}N_{\Delta,1}\mathcal S^a_{\Delta,1}\, ,
\ee
where 
\be
\mathcal S^a_{\Delta,1}=\int_{\mathbb R^d}d^d y \frac{I^{ab}(x-y)}{\left[(x-y)^2\right]^{d-\Delta}}\frac{(x_i-x)_b}{(x_i-x)^2}\, .
\ee
The shadow kernel can be rewritten as follows
\be
\frac{I^{ab}(x-y)}{\left[(x-y)^2\right]^{d-\Delta}}=\widehat D_S^{ab}\frac{1}{\left[(x-y)^2\right]^{d-\Delta-1}}\, ,
\ee
where we defined the differential operator $c_{\Delta,1}\widehat D_{S,\,\frac{d}{2}-\Delta}$ in index form
\be
\widehat D_S^{ab}=\frac{1}{2(1+\Delta-d)(d-\Delta)}\left(\p^a\p^b-\frac{1+\Delta-d}{2\Delta-d}\delta^{ab}\square\right)\, .
\ee
Then, integrating by parts and using equation \eqref{twoptint}, we obtain
\be\label{k}
\mathcal S^a_{\Delta,1}=\frac{\pi^\hd\Gamma\left(\hd\right)\Gamma\left(\hd-\Delta+1\right)\Gamma\left(\Delta-\hd\right)}{\Gamma(\Delta)\Gamma(d-\Delta+1)}(\Delta-1)\frac{(x_i-x)^a}{\left[(x_i-x)^2\right]^{\hd-\Delta+1}}\, .
\ee
Plugging this result into~\eqref{photonshadow} we obtain in the limit $\Delta\to 1$ the shadow transformed soft photon factor~\eqref{R1t_WI}.\footnote{\label{evenodddim}Notice that in even dimensions both $\mathcal S^a_{\Delta,1}$ and $N_{\Delta,1}$ are finite in the limit $\Delta\to 1$. 
In odd dimensions, $\mathcal S^a_{\Delta,1}$ goes to zero and $N_{\Delta,1}$ diverges as $\Delta\to 1$ but their product is still finite.
A similar subtlety occurs for the shadow transform of the leading and subleading conformally soft graviton.}
Taking the divergence of~\eqref{photonshadow} and using equation~\eqref{del1}, we obtain the Ward identity~\eqref{R1_WI_1}.

\subsection*{Leading Soft Graviton}
The shadow transform on the leading soft graviton operator is
\be
\tilde H_1^{ab}(x)=\lim_{\Delta\to 1}N_{\Delta,2}\int_{\mathbb R^d}d^d y \frac{I^{aa'}(x-y)I^{bb'}(x-y)}{\left[(x-y)^2\right]^{d-\Delta}}H_{1\,a'b'}(y)\,.
\ee
Applying this transform to the leading soft graviton factor~\eqref{SoftGravitond} yields
\be
\tilde S^{(0)\, ab}_p(\omega,x)=-\kappa \sum_{i=1}^N \frac{\omega_i}{\omega}\lim_{\Delta\to 1}N_{\Delta,2}\mathcal S^{ab}_{\Delta,2}\, ,
\ee
where
\be
 \mathcal S^{ab}_{\Delta,2}=\int_{\mathbb R^d} d^d y\, \frac{I^{a a'}(x-y)I^{b b'}(x-y)}{\left[(x-y)^2\right]^{d-\Delta}}\left(\frac{(x_i-y)_{a'}(x_i-y)_{b'}}{(x_i-y)^2}-\frac{\delta_{a' b'}}{d}\right)\, .
\ee
The shadow kernel can be expressed as
\be
\frac{I^{aa'}(x-y)I^{bb'}(x-y)}{[(x-y)^2]^{d-\Delta}}=\widehat D_S^{a a' b b'}\frac{1}{[(x-y)^2]^{d-\Delta-2}}\, ,
\ee
where we defined the differential operator $c_{\Delta,2}\widehat D_{S,\frac{d}{2}-\Delta}$ in index form
{\small
\be\label{diffop}
\begin{aligned}
\widehat D_S^{a a' b b'}=&\alpha_1\bigg(\p^a\p^{a'}\p^b\p^{b'}\\
-&\alpha_2\left(\delta^{a'b'}\p^{a}\p^{b}+\delta^{a'b}\p^{a}\p^{b'}+\delta^{ab}\p^{a'}\p^{b'}+\delta^{ab'}\p^{a'}\p^{b}+(\Delta-d)\left(\delta^{aa'}\p^{b}\p^{b'}+\delta^{bb'}\p^{a}\p^{a'}\right)\right)\square\\
+&\alpha_3 \left(\delta^{a'b'}\delta^{ab}+\delta^{a'b}\delta^{ab'}+[(\Delta-d)(\Delta-d+1)-1]\delta^{aa'}\delta^{bb'}\right)\square^2\bigg)\, ,
\end{aligned}
\ee
}
with
\be
\begin{aligned}
\alpha_1&=\frac{1}{4(\Delta-d)(\Delta-d-1)(1+\Delta-d)(2+\Delta-d)}\, ,\\[10pt]
\alpha_2&=\frac{1}{2+2\Delta-d}\, ,\\[10pt]
\alpha_3&=\frac{1}{(2\Delta-d)(2+2\Delta-d)}\, .
\end{aligned}
\ee
Using integration by parts, the result \eqref{twoptint} and taking the limit $\Delta\to 1$, the shadow leading soft graviton factor becomes equation~\eqref{ShadowGravd}. The result is finite with a similar subtlety mentioned in footnote~\ref{evenodddim} regarding odd dimensions.
Using equation~\eqref{der}, we can show the Ward identity~\eqref{H1_WI}.

\subsection*{Subleading Soft Graviton}
The shadow transformed subleading soft gravtion operator and the soft graviton factor~\eqref{SubSoftGravitond} can also be computed in index notation but the expressions are much more cumbersome to write down. Let us outline the necessary steps starting from the index-free expression for the shadow transform of the subleading soft graviton factor, equation~\eqref{SubSoftGravitond}, which is given by
\be
\begin{aligned}
\tilde S^{(1)}_p(\omega,x,\mathcal e)=\lim_{\Delta\to 0}N_{\Delta,2}\int_{\mathbb R^d}d^d y\, \frac{1}{[(x-y)^2]^{d-\Delta-2}}\widehat D_S^{aa'bb'}S^{(1)}_{p\, ab}(\omega,y)e_{a'} e_{b'}\,.
\end{aligned}
\ee
Note that we used integration by parts to move the differential operator $D^{aa'bb'}_S$ on the soft factor. This operator was defined in \eqref{diffop}  and acting with it on the soft factor using \eqref{twoptint} to evaluate the integrals and taking the limit $\Delta\to 0$, we obtain the shadow subleading soft graviton factor~\eqref{ShadowSubGravd}. We obtain a finite result in both even and odd dimensions (see footnote~\ref{evenodddim}). 
Using equation \eqref{Tfinal}, we can show the Ward identity~\eqref{H2_WI}.

\section{Type S Operators in Even Dimensions} \label{app:typeSeven}
There is a puzzle about type S primary descendants. They can be defined in any dimensions but from representation theory one knows that this type of operator should exist only in odd $d$. How is this possible? What happens to these operators in even $d$?
The most plausible answer is that in even $d$ a type S operator can be written as a composite of other primary descendants. 
We now show in a few examples that indeed this is what happens.

In the following we will focus on spin $\ell$ primary operators $\Ocal_{\Delta, \ell}$ with dimensions $\Delta=2-\ell$  with type S primary descendants at level 
 $\kcal=\frac{d}{2}-2+\ell$. These primary descendants are only defined when $ \kcal$ is integer, and therefore  for even $d$. 
 First we show that the type S operators can be written in a very simple form in term of projectors which makes manifest many of their properties.
For the sake of clarity we  exemplify the form of $D_{\S, \frac{d}{2}-2+\ell}$ for $\ell=0,1,2$ (up to overall normalization factors),
\begin{align}
& \square^{\frac{d}{2}-2} \, , \quad (\ell = 0) \, , \\
& \square^{\frac{d}{2}-2}[\ecal \cdot \partial_x \, \partial_{\hat \fcal} \cdot \partial_x -\square (\ecal\cdot \partial _{\hat \fcal}) ] \, ,\quad  (\ell = 1) \, , \\
&\square^{\frac{d}{2}-2}[ (d-2) (\ecal \cdot \partial _x){}^2 (\partial _x\cdot \partial _{\hat \fcal}){}^2 -2 (d-1)\square  (\ecal \cdot \partial _x)  (\ecal\cdot \partial _{\hat \fcal}) (\partial _x\cdot \partial _{\hat \fcal})  \nonumber \\
&\qquad\qquad\qquad\qquad\qquad\qquad\qquad +(d-1) \square^2 (\ecal\cdot \partial _{\hat \fcal}){}^2+ \square (\ecal \cdot \partial _x)^2   (\partial _{\hat \fcal}\cdot \partial _{\hat \fcal}) ] \, ,\quad  (\ell = 2) \, .
\end{align}
The differential operators $D_{\S, \frac{d}{2}-2+\ell} $ above are defined such that they act on a spin $\ell$ primary $\Ocal_{\Delta,\ell}(x,\hat \fcal)$ whose tensor indices are contracted with unconstrained vectors $\hat \fcal$. The resulting descendants are spin $\ell$ operators with indices contracted with constrained vectors $\ecal$ (such that $\ecal^2=0$). As stated above, they become primaries when $\Delta= 2-\ell$. 
For any even $d>2$ these differential operators can be written in terms of projectors as follows\footnote{
In the following we will use a notation in which a vector contracted with an index inside a box is represented as the vector inside the box, e.g. 
\be{}
{
   \ytableausetup{centertableaux,boxsize=1.3 em}
\scriptsize
\begin{ytableau}
a&b&c&\, _{\cdots}
\end{ytableau}
}
\;
\ecal^b \equiv 
{ \scriptsize
\begin{ytableau}
a & \ecal&c& \, _{\cdots}
\end{ytableau}
}
   \ytableausetup{centertableaux,boxsize=1.8 em}
\, .
\ee{}}
\begin{align}
D_{\S, \frac{d}{2}-2+\ell}
&
\propto \pi_{\frac{d}{2}-2+\ell,\ell}
\left(
{ \scriptsize
\begin{ytableau}
\partial_x&\, _{\cdots}&\, _{\cdots}& \,\partial_x \\
\ecal&\, _{\cdots}&\ecal \\
\end{ytableau}
}\ ;
{ \scriptsize
\begin{ytableau}
\partial_{x}
&\, _{\cdots} 
&\, _{\cdots}
& \partial_{x} \\
\partial_{\hat \fcal}&\, _{\cdots}&\,\partial_{\hat \fcal}
\\
\end{ytableau}
}\
\right)
 \, 
 \\
&  \propto 
  \square^{\frac{d}{2}-2} 
  \;
  \pi_{\ell,\ell}
\left(
{ \scriptsize
\begin{ytableau}
\partial_x&\, _{\cdots}& \,\partial_x \\
\ecal&\, _{\cdots}&\ecal \\
\end{ytableau}
}\ ;
{ \scriptsize
\begin{ytableau}
\partial_{x}
&\, _{\cdots} 
& \partial_{x} \\
\partial_{\hat \fcal}&\, _{\cdots}&\,\partial_{\hat \fcal}
\\
\end{ytableau}
}\
\right)
 \, .
 \label{DS_proj}
\end{align}
From this form it is straightforward to see that the resulting operators are conserved. Indeed, conservation is obtained by acting with $(\partial_x \cdot D_\ecal)$ on the projector which is ensured by the symmetry properties of the projector.

One can also show that the operator in \eqref{DS_proj} is related to other types of  primary descendants of  $\Ocal_{\Delta=2-\ell,\ell}$. Indeed the projector $\pi_{\ell, \ell}$ can be written as the consecutive action of two other operators that create primary descendants, namely
\be{}
 \pi_{\ell,\ell}
\left(
{ \scriptsize
\begin{ytableau}
\partial_x&\, _{\cdots}& \,\partial_x \\
\ecal&\, _{\cdots}&\ecal \\
\end{ytableau}
}\ ;
{ \scriptsize
\begin{ytableau}
\partial_{x}
&\, _{\cdots} 
& \partial_{x} \\
\partial_{\hat \fcal}&\, _{\cdots}&\,\partial_{\hat \fcal}
\\
\end{ytableau}
}\
\right) \propto D_{\II_{k=2},\kcal=\ell} D_{\I_{k=2},\kcal=\ell} \, .
\ee{}
Therefore we obtain that 
\be
\label{DStoDIIBoxDI}
D_{\S, \frac{d}{2}-2+\ell} \Ocal_{\Delta=2-\ell,\ell}  \propto     D_{\II_{k=2},\kcal=\ell} \; \square^{\frac{d}{2}-2} \; D_{\I_{k=2},\kcal=\ell}  \Ocal_{\Delta=2-\ell,\ell} \, .
\ee
The type $\I_2$ operator at level $\kcal=\ell$ is acting on a primary with dimensions $\Delta=2-\ell=\Delta^*_{\I_2,\kcal=\ell}$, which thus has the correct dimension to define a primary descendant according to equation~\eqref{Delta_Ik}.
When $d=4$, the $\square$ disappears and the type S operator becomes equal to the action of type $\I_2$ and $\II_2$ primary descendants. Moreover the type $\I_2,\kcal=\ell$ primary descendant has dimension $\Delta=2$ which is the correct dimension for a primary with a type   $\II_2,\kcal=\ell$ primary descendant according to~\eqref{polesII}. In $d=4$ we can thus prove that the action  of $D_{\S, \frac{d}{2}-2+\ell} $ can be decomposed in terms of more basic primary descendants. 
For even $d>4$ one needs to write the powers of the $\square$ in terms of other primary descendants, but we did not attempt this here.
However let us mention that in any even $d\geq 4$ the operator \eqref{DStoDIIBoxDI} always starts with the action $D_{\I_{k=2},\kcal=\ell} \Ocal_{\Delta=2-\ell,\ell}$. Since $D_{\I_{k=2},\kcal=\ell}$ is of type I, we expect it to annihilate the primary without generating contact terms. If this is taken into account, then the action of the rest of the terms in \eqref{DStoDIIBoxDI} would be trivial. 

Finally let us  mention that the values $\Delta^*_{\S,\kcal}$ for which the type S operator becomes a primary descendant in even dimensions always coincides  with the ones of either type~$\I_k$ or $\III_k$, but never of type~$\II_k$.
We thus expect that the composite primary descendant of type~S,$\kcal$ will start with the action of differential operators of either type $\I_k$ or $\III_k$. For all type $\I_k$ we expect the differential operator to annihilate the primary without generating contact terms. In these cases we thus conclude that the full action of $D_{\S,\kcal}$ should annihilate the primary.

In summary, in this appendix we exemplified why the type S operator is not considered a new type in even dimensions, by showing that in explicit cases it can be understood as the action of more fundamental types such as $\I_{k=2}$ and $\II_{k=2}$.

\bibliographystyle{utphys}
\bibliography{CCFTd_Draft}

\end{document}